\documentclass{aa}  

\usepackage{graphicx}
\usepackage{txfonts}
\usepackage{amsmath,amsfonts,amssymb}
\usepackage{textcomp,gensymb}
\usepackage[T1]{fontenc}
\usepackage{aecompl}
\usepackage{ulem}
\usepackage{aas_macros}
\usepackage[dvipsnames]{xcolor}
\usepackage{color}
\usepackage{times}
\usepackage{subfig}
\usepackage{stfloats}
\usepackage[utf8]{inputenc}
\usepackage{hyperref}
\usepackage{multirow}
\usepackage{natbib}
\usepackage{placeins}

\makeatletter
\newcommand*{\hyperlinkcite}[1]{\hyper@link{cite}{cite.#1}} 
\makeatother

\newcommand{\lon}[1]{\hyperlinkcite{Long+2021}{L21}#1}

\newcommand{\mdy}[1]{#1}

\newcommand{\rmvd}[1]{}

\newcommand{\lang}[1]{#1}

\begin{document}

   \title{V892~Tau: A tidally perturbed circumbinary disc in a triple stellar system \thanks{Outputs of the hydrodynamical simulations underlying this article are available at \url{https://zenodo.org/records/11350568}}}

   \author{Antoine Alaguero \inst{1}
          \and
          Nicol\'as Cuello \inst{1}
          \and
          Fran\c cois Ménard \inst{1}
          \and
          Simone Ceppi \inst{2}
          \and
          \'Alvaro Ribas \inst{3}
          \and
          Rebecca Nealon \inst{4,5}
          \and\\
          Miguel Vioque \inst{6,7}
          \and
          Andr\'es Izquierdo \inst{8,9}$^{,14}$
          \and
          James Miley \inst{10,11,12}
          \and 
          Enrique Mac\'ias \inst{6}
          \and
          Daniel J. Price \inst{13}
          }

   \institute{\mdy{Univ. Grenoble Alpes, CNRS, IPAG, F-38000 Grenoble, France}\\
              \email{antoine.alaguero@univ-grenoble-alpes.fr}  
         \and
              Dipartimento di Fisica, Universit\'a degli Studi di Milano, via Celoria 16, 20133 Milano, Italy 
         \and
             Institute of Astronomy, University of Cambridge, Madingley Road, Cambridge, CB3 0HA, UK 
         \and
             Department of Physics, University of Warwick, Coventry CV4 7AL, UK 
         \and   
             Centre for Exoplanets and Habitability, University of Warwick, Coventry CV4 7AL, UK 
         \and
             European Southern Observatory, Karl-Schwarzschild-Str. 2, 85748 Garching bei München, Germany 
         \and
             Joint ALMA Observatory, Alonso de C\'ordova 3107, Vitacura, Santiago 763-0355, Chile 
         \and
             Leiden Observatory, Leiden University, P.O. Box 9513, \mdy{NL-}2300 RA Lei- den, The Netherlands 
         \and
             \mdy{Department of Astronomy, University of Florida, Gainesville, FL 32611, USA} 
         \and
            Departamento de Física, Universidad de Santiago de Chile, Av. Victor Jara 3659, Santiago, Chile  
         \and
            Millennium Nucleus on Young Exoplanets and their Moons (YEMS), Chile 
        \and
            Center for Interdisciplinary Research in Astrophysics and Space Exploration (CIRAS), Universidad de Santiago, Chile 
        \and
            School of Physics and Astronomy, Monash University, Vic 3800, Australia 
             }

   \date{Received 21 February 2024; Accepted 19 May 2024}

 
  \abstract
   {V892~Tau is a young binary star surrounded by a circumbinary disc \lang{that} 
   \lang{shows hints of an} interaction with the low-mass nearby star V892~Tau~NE. }
   { The goal of this paper is to constrain the orbit of V892~Tau~NE and to determine the resulting circumbinary disc dynamics.}
   {We present new ALMA observations of the V892~Tau circumbinary disc at a twice higher angular and spectral resolution. 
   \lang{We modelled} the data with V892~Tau as a triple system and \lang{performed} a grid of hydrodynamical simulations testing several orbits of the companion. The simulation outputs \lang{were} then post-processed to build synthetic maps that we \lang{compared} to the observations.}
   {The $^{12}$CO emission of the disc shows clear non-Keplerian features such as spiral arms. When comparing the data with our synthetic observations, we \lang{interpreted} these features as ongoing interactions with the companion. Our simulations indicate that an eccentricity of $e\sim0.5$ of the companion is needed to reproduce the observed disc extent and that a mutual inclination of $\Delta i \sim 60\degree$ with the inner binary reproduces the measured disc tilt.}
   {In order to explain most of the features of the circumbinary disc, we propose that V892~Tau~NE follows an orbit with a mild eccentricity $0.2 < e < 0.5$ and a mutual inclination of $30\degree < \Delta i < 60\degree$. Such a misaligned companion suggests the disc is oscillating and precessing with time, stabilising in an intermediate plane with a non-zero mutual inclination with the inner binary. Given \lang{the} orbital configuration, we show that the stability of future planets is compromised in the second half of the disc once the gas has dissipated.}


   \maketitle
%
\footnotetext[14]{\mdy{NASA Hubble Fellowship Program Sagan Fellow}}

\section{Introduction}
\label{sec:intro}

During the early stages of star formation, a significant fraction of stars are found to be part of multiple systems \citep{Reipurth+2014, Offner+2022}. Multiple systems naturally evolve to hierarchical configurations \citep{DucheneKraus2013}, but allow a large diversity of possible configurations for protoplanetary discs. Circumstellar discs can in principle form around any hierarchical level of the system. As a consequence, discs in multiple systems are shaped by the disc-disc and star-disc interactions that depend on the geometrical configuration and physical properties of the system \citep{Bate2018}. By means of those interactions, a large cavity is carved in circumbinary discs up to several times the binary semi-major axis, while the outer edge of the disc is set by tidal interactions with the outer stars of the system \citep{Artymowicz1994, MirandaLai2015}. Spiral arms are usually launched in the process \citep{Rafikov2002}. Gravitational torques from misaligned stars tend to incline discs and to make them precess \citep{PapaloizouTerquem1995}, resulting in misaligned geometries of the disc plane with respect to the stellar orbital plane and even leading to a disc warp or break in some cases (e.g. \cite{Nixon+2013, Facchini+2013, Rabago+2023}). The investigation of these dynamical behaviours is key to understand disc dynamics in multiple systems and their resulting planetary architectures.

Hydrodynamical simulations have proven to be a powerful tool to infer orbits from disc morphology and kinematics in multiple systems. \lang{Multiplicity-induced} substructures depend on the orbital parameters of companions. By running a grid of hydrodynamical models and comparing them to disc observations, one can now constrain orbits in multiple systems \citep{Price+2018, Gonzalez+2020, Nowak+2024}. Even if this method is limited by observations in snapshots in time and the cost of hydrodynamical simulations, it provides constrains independent from astrometric measurements, and it can help to discriminate between available orbits. 

\begin{figure}
   \centering
   \includegraphics[width=\columnwidth]{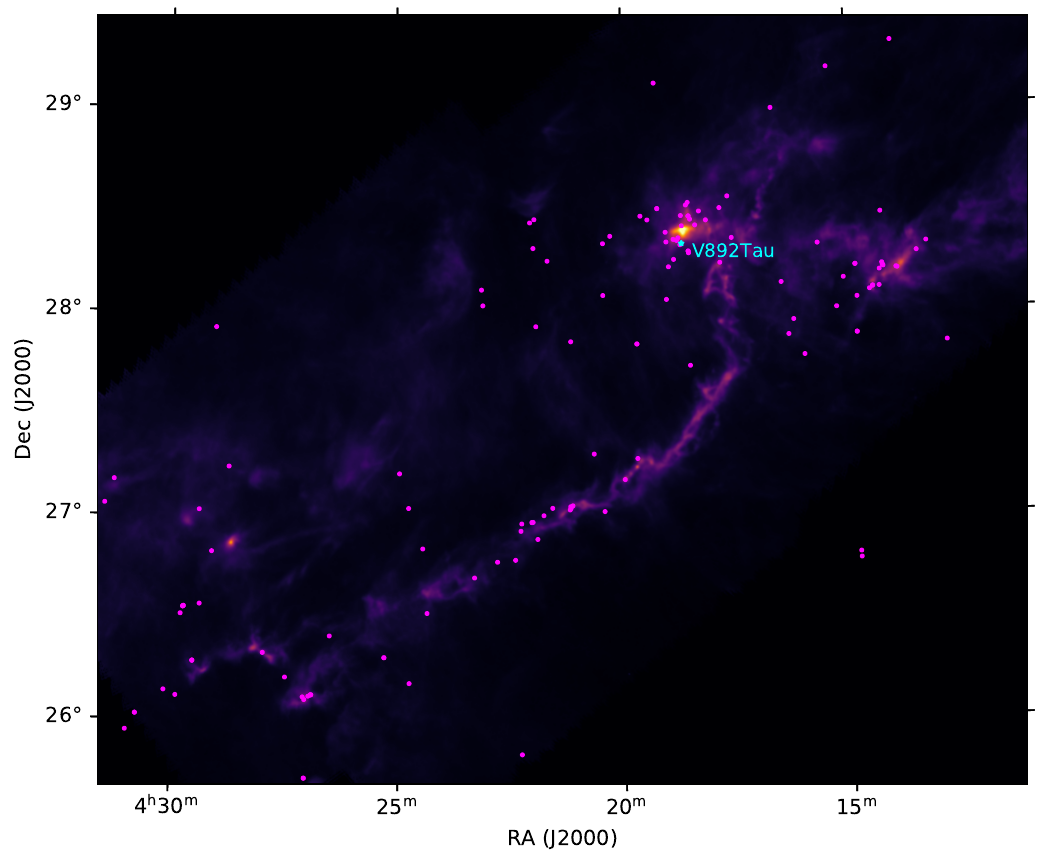} 
   \caption{V892~Tau is located in a clustered sub-region of the Taurus molecular cloud.
   \lang{The} background image from the Herschel/SPIRE survey \citep{Griffin+2010, Andre+2010} \lang{is} shown on a linear intensity scale. Magenta points show positions of pre-main-sequence stars in Taurus taken from \citet{Kenyon+2008}.}
   \label{fig:V892-locator}
\end{figure}

V892~Tau is a young triple system located in a clustered sub-region of the Taurus star-forming cloud (Figure~\ref{fig:V892-locator}). The system is composed of \mdy{two central $3$ M$_{\odot}$ stars separated by $0.06\arcsec$ surrounded by a} large circumbinary disc (CBD), \mdy{which extends up to approximately $1.5\arcsec$ from the central stars (e.g. \citealt{Monnier+2008, Long+2021})}. The \mdy{M3} star V892~Tau~NE \mdy{is projected at $4\arcsec$ \lang{north-east}
from the inner binary and completes the system \citep{EsplinLuhman2019}}. The \textit{Gaia} 
DR3 \citep{GaiaDR3} parallax measurements result in distances of $d=134.5\pm 1.5$ pc and $d=131.3\pm 5.0$ pc, respectively, for V892~Tau and V892~Tau~NE, which is consistent with a bound triple system within the error bars. 

Observations with \lang{the Atacama Large Millimeter/submillimeter Array (ALMA)} have unveiled the dust ring surrounding the inner binary of V892~Tau at a high spatial resolution \citep{Pinilla+2018}. More recently, a detailed study of the system has been conducted by \citet[hereafter \lon]{Long+2021} using that ALMA Band 6 data combined with additional 
\lang{Very Large Array (VLA)} observations: The eccentric inner binary has been resolved and its orbit tightly constrained, with two families of solutions for the longitude of the ascending node $\Omega_{in}$ and the argument of the periapsis $\omega_{in}$. The dust emission was found to have \mdy{an $0.17\arcsec$-large} inner gap, while the gas emission appears to extend to the innermost regions of the system. Both the dust and gas show hints of an inclined geometry with respect to the inner binary. The tilt of the disc with respect to the inner binary is measured at $\Delta = 8.0 \pm 4.2 \degree$ or $\Delta = 113.3 \pm 3.0 \degree$ depending of the orbital solution chosen.  The size of the gas disc is consistent with a tidal interaction with V892~Tau~NE and tentative spiral arms are distinguished in the \lang{redshifted} side of the disc. A warp is suggested by deviations from Keplerian rotation in the outer disc and could be caused by interactions between the disc and the 
\lang{inner or outer binary}. In addition, recent near-IR interferometric observations have refined the orbit of the binary and detected a narrow circumstellar disc \mdy{wide of $0.016\arcsec$} around the primary star of V892~Tau \citep{Vides+2023}. In that work, the inner CBD was also modelled and tentative evidence of a warp was invoked to explain residual patterns from geometric models. 

As \lang{a} hierarchical triple system, V892~Tau is a perfect laboratory to study the formation channels of planets from discs in multiple systems, which depends on the binary-disc interactions and dynamics. Understanding the disc dynamics in such a system will help to constrain the planet formation regions in multiple stellar systems in general, \lang{but it} would require the orbital configuration of the system to be known. Nonetheless, the orbit of the companion star V892~Tau~NE remains unconstrained at the moment. Projected at a distance of $\sim 520$au of the central binary, the orbit of V892~Tau~NE is critical to understand the dynamics of the CBD and the subsequent planet formation. 

Following that effort, here we present new ALMA observations of V892~Tau that we \lang{combined} with archival data. The observational setup and the resulting data are detailed in Section \ref{sec:observations}. We \lang{fitted} the observations with a Keplerian disc model before modelling the V892~Tau system as a triple star system with a CBD. We \lang{performed} hydrodynamical simulations of that model, testing different orbits for the outer companion, and \lang{post-processed} the simulation outputs to build synthetic observations. Methods are presented in Section \ref{sec:methods}, while an analysis of the observations, of the synthetic observations,
and their relative comparison are described in Section \ref{sec:results}. In Section \ref{sec:discussion} we give an overview of the system and discuss our results in the more general context of multiple systems. Our conclusions are summarised in Section \ref{sec:conclusion}.

\section{Observations}
\label{sec:observations}

\subsection{Data calibration, reduction, and combination}
\label{subsec : data}

The triple system V892~Tau was observed with ALMA in the context of the ALMA programme 2021.1.01137.S (PI: J. Miley) in Band 6. The observations were performed in two sessions, with the first session covered shorter baselines spanning $15-1997$ m ($12-1536 \;k\lambda$) with $42$ antennas. We call the resulting \lang{dataset} of that session Short Baselines (SB hereafter). The second session used a more extended configuration also with $42$ antennas spanning baselines from $64$ m to $8283$ m ($49-6371 \;k\lambda$) and resulted in a dataset we call Long Baselines (LB hereafter). Each 
dataset contains four
spectral windows (SPWs) with two of them set up for continuum detection centred at $217.875$ and $233.000$ GHz, with $2$ GHz of bandwidth distributed over 128 channels. The other two SPWs were centred at $230.538$ and $220.000$ GHz targeting the $^{12}$CO (2-1) and $^{13}$CO (2-1) transitions, each containing 3840 channels of $122$ kHz and $488$ kHz widths, respectively.

The raw visibility data were downloaded from the ALMA archive and subsequently calibrated using the available scripts provided by ALMA staff using the required {\sc CASA} \citep{McMullin2007} 6.2.1.7 version. The band-pass and amplitude calibrator for the observations was J$0435+2532$, and J$0438+3004$ was used as the phase calibrator. The total on-source time reached approximately $18.5$ minutes. 

The data were self-calibrated after the line-free channels were combined altogether to create continuum \lang{datasets}. We performed several phase self-calibration rounds on each continuum \lang{dataset} until the solution interval was tuned down to the integration time of $6.05$ s for the SB \lang{dataset} and of $2.02$ s for the LB \lang{dataset}. This process resulted in an increase of the peak 
signal-to-noise ratio 
(S/N) of about $6\%$ for SB and $36\%$ for LB, giving a peak S/N of $582$ and $233,$ respectively.

We downloaded the data used in \lon\ from the ALMA archive and re-calibrated it thanks to the ALMA reduction pipeline. The resulting \lang{dataset} is called the Original Long et. al \lang{dataset} (OLD hereafter). Then, we reduced the data and performed phase self-calibration rounds on the continuum data with solution intervals down to the integration time of $6.05$ s. The peak S/N increased by $9\%$ doing so to reach a value of $483$, which is a significant improvement given the already high S/N of the data.

The SB and LB visibility amplitudes were scaled with respect to the one of OLD to make sure the fluxes of all the \lang{datasets} were consistent \mdy{altogether} before combination. To do so, we used the functions \textit{estimate\_scale\_flux} and \textit{rescale\_flux} of the \textit{reduction\_utils.py} python script from the DSHARP project \footnote{\url{https://almascience.eso.org/almadata/lp/DSHARP/}}.  We then created a combined continuum model with \textit{tclean} in {\sc CASA} v6.5.3 from the three \lang{datasets}, using a Briggs weighting with a robust parameter of $0.5$ \citep{Briggs}. On the basis of that common model, the individual \lang{datasets} were self-calibrated in phase a second time. The resulting \lang{datasets} were combined to create a final continuum image with \textit{tclean} using a Briggs weighting with a robust of $-0.5$, which resulted in a RMS of $90 \mu$Jy. The resulting beam size is $0.11\arcsec \times 0.06\arcsec$ at a central frequency of $224$ GHz. \mdy{Additional details regarding the data combination process are available in the Appendix \ref{app : data combination}}

We also applied the resulting self-calibration tables to the gas observations, \mdy{from which the continuum emission was subtracted by fitting a polynomial function to line-free channels using the \textit{uvcontsub} task.} An image cube of the $^{12}$CO (2-1) line emission was then created combining the three \lang{datasets} using a Briggs weighting with a robust of $0.5$ and a channel width of $0.5$ km\,s$^{-1}$, giving a resulting beam size of $0.17\arcsec \times 0.11\arcsec$. \rmvd{The continuum emission was subtracted from the data after fitting a polynomial function to line-free channels using the \textit{uvcontsub} task.}  We summarise the details of each of the continuum and $^{12}$CO (2-1) images in Table \ref{table:comb_obs}. This paper presents the continuum observations but \lang{mainly focusses} on the analysis and discussion of the gas emission, as a detailed description of the continuum data will be addressed in a future study.

\begin{table*}
    \centering
    \caption{\lang{Observations'}
    results and image details in comparison with previous data.}
 \begin{tabular}{c c c c c c} 
OLD+SB+LB & Frequency & Beamsize & Flux \tablefootmark{1} & RMS & $R_{90\%}$\\ [0.5ex] 
 & (GHz) & $(\arcsec)$ & (mJy or Jy\,km\,s$^{-1}$ ) & (mJy\,beam$^{-1}$) & $(\arcsec)$ \\
 \hline \hline
Continuum & $224$ & $0.11 \times 0.06$ & $297.1\pm0.1$ & $0.09$ & $0.38\pm0.01$\\
$^{12}$CO (2-1) & $230.538$ & $0.17 \times 0.11$ & $15.8\pm0.1$ & $5.5$ & $1.45\pm0.02$\\
 & & & & & \\
 OLD & & & & & \\
 \hline \hline
Continuum & $224$ & $0.23 \times 0.16$ & $290.6\pm0.2$ & $0.08$ & $0.46\pm0.01$\\
$^{12}$CO (2-1) & $230.538$ & $0.23 \times 0.16$ & $14.7\pm0.4$ & $6.4$ & $1.45\pm0.02$\\
\end{tabular}
\tablefoot{The top panel shows the results achieved from the data combined in this work. The bottom panel shows the results derived in the previous work of \lon from the OLD dataset. \tablefootmark{1}{\lang{The} flux uncertainty \lang{was} measured by multiplying the noise level by the square root of the number of pixels in the integrated area.}}
\label{table:comb_obs}
\end{table*}

\subsection{Dust disc}
\label{subsec:dust_disc}

\begin{figure}
\centering
\begin{center}
    \includegraphics[width=\columnwidth, trim={0.8cm 0cm 0.7cm 0cm},clip]{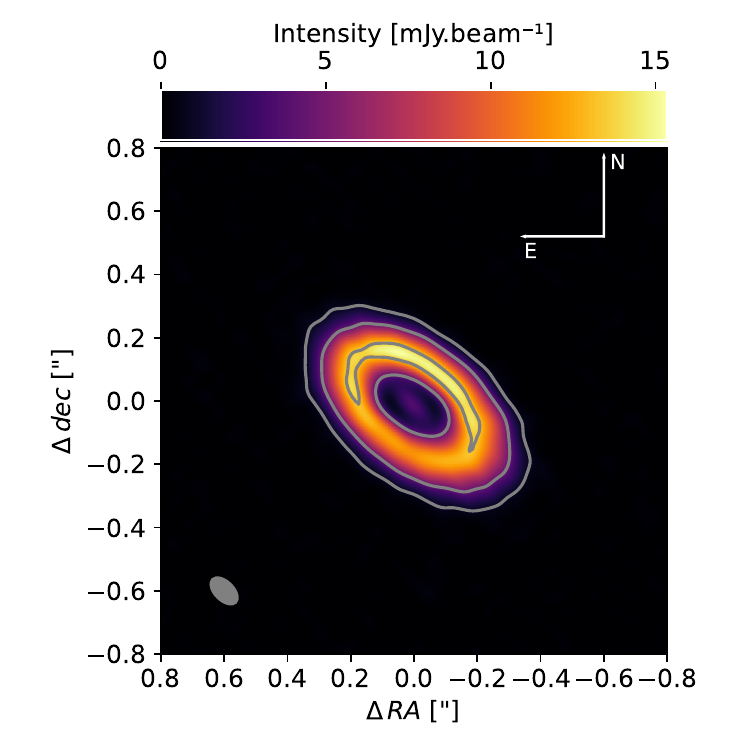}
    \includegraphics[width=\columnwidth, trim={1.4cm 0cm 2cm 1cm},clip]{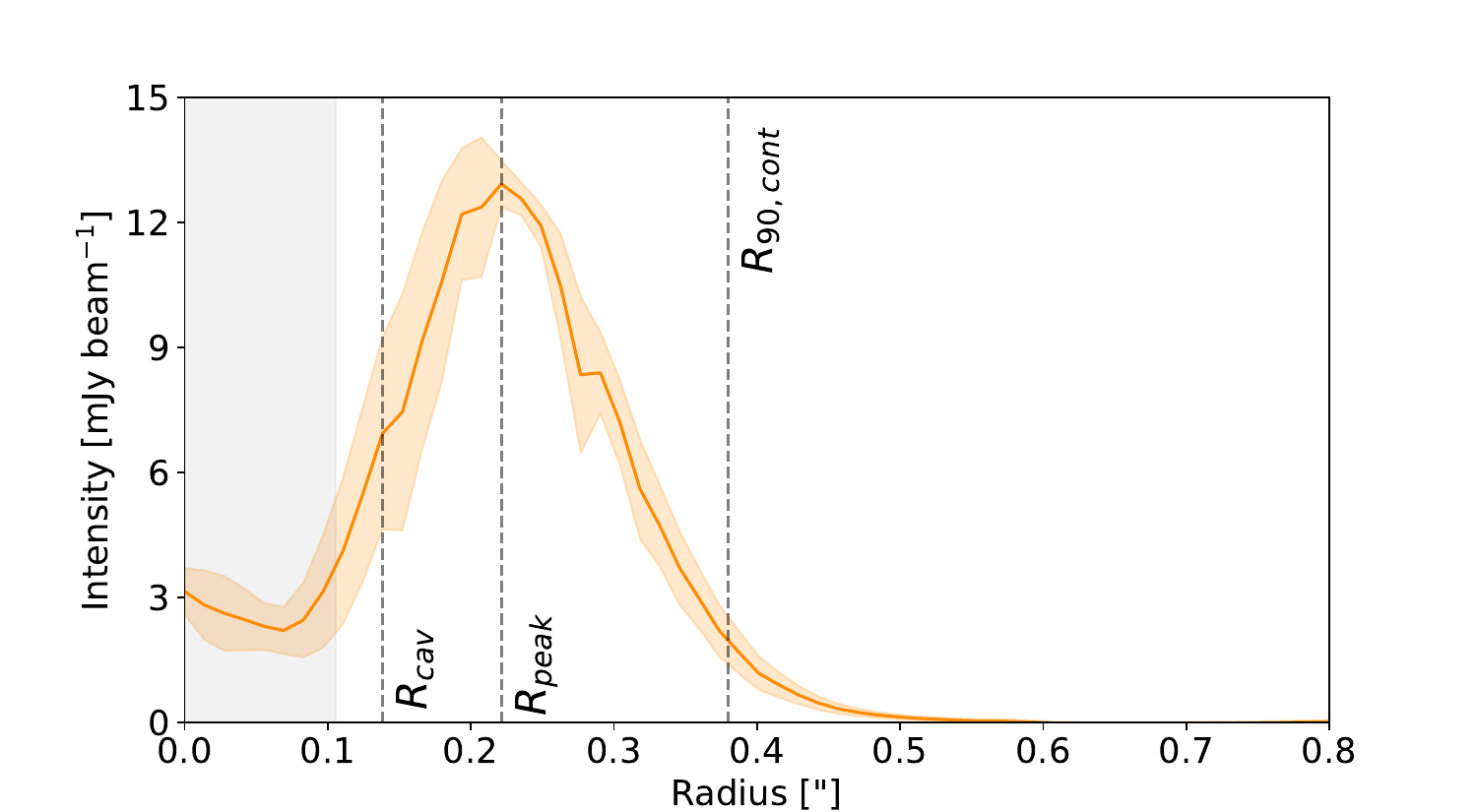}
    \caption{ALMA band 6 continuum emission \mdy{map (top) and deprojected azimuthally averaged intensity profile (bottom)} at $224$ GHz of V892~Tau. \mdy{In the top image,} contour levels are at $10\sigma, 50\sigma$, and $150\sigma$. \mdy{In the bottom plot $R_{cav}$, $R_{peak}$, and $R_{90,cont}$ correspond to the radius of the cavity, of the peak of the ring, and of the disc, respectively. The shaded orange area represents to the standard deviation of each annulus divided by the square root of the beam numbers along the annulus at each radial bin. The shaded grey area represents the major axis of the synthesised beam.}}
    \label{fig:obs_cont}
\end{center}
\end{figure}

Figure \ref{fig:obs_cont} shows the ALMA $1.3$ mm continuum image \mdy{and a deprojected brightness radial profile}. 
The observed structure emerges as a \mdy{Gaussian-like} ring located at a separation of approximately $0.22\arcsec$ from the centre of the disc, \mdy{following the position of the peak intensity radius. The inner binary has carved a cavity in the innermost parts of the system. We found a cavity radius of $R_{cav} = 0.14\pm0.02\arcsec$\lang{that} is defined as the radius at which the intensity first reaches half of its peak level with an incertitude taken as the difference with quarter peak radius.} The dusty disc has a deprojected radius of $R_{90\%} = 0.38\pm0.01\arcsec$ measured as the radius encircling $90\%$ of the total flux.

The inclination $i$ and the position angle $PA$ of the disc are respectively best-fitted to $i=54.0 \pm 1.8\degree$ and $PA=51.5\pm0.5\degree$. Thanks to an improved angular resolution, the $1.3$ mm emission is found to be more compact than in previous observations \citep{Pinilla+2018}. The cavity is spatially resolved along with an azimuthal asymmetry in the NW side of the disc. Indeed the northern side of the disc along the minor axis is $18\%$ brighter than its southern side. This trend was also found in \lon and is consistent with the reported value of $20\%$. This kind of horse-shoe asymmetry is a known consequence of binarity \citep{Ragusa+2017}. However optically thick warm dust emission from the inner rim of the disc may also produce similar patterns \citep{Ribas+2024}. This emission coming from the far side of the disc, it would indicate that the closest side to the observer is the SE side of the disc.
Like all the previous ALMA observations of V892~Tau, no emission is detected from the vicinity of V892~Tau~NE in the $1.3$ mm continuum. The sensitivity and RMS of the observations presented in this paper being similar to the previous data, this result is not surprising. 

We confirm the detection of unresolved emission in the cavity with a $30\sigma$ significance. This emission could be explained by circumstellar dusty material and could trace the circumstellar disc detected in the inner system \citep{Vides+2023}.
Future multi-wavelengths observations could allow a precise characterisation of the dust properties, while multi-epochs observations could allow the clump's dynamical behaviour \lang{to be prescribed}. Those questions are left for future investigations.

\subsection{CO emission}
\label{subsec:gas_disc}

The $^{12}$CO emission of the disc is detected at a confidence level above $3\sigma$ from channel $-6$ km\,s$^{-1}$ to channel $23$ km\,s$^{-1}$ in \lang{the kinematic local standard of rest (LSRK)} velocity. The data suffer from absorption by cloud material at the systemic velocity around $8$ km\,s$^{-1}$. The complete channel maps can be found in Appendix \ref{app : channel maps}. 
$99\%$ of the flux is contained in a radius of $1.73\arcsec$, with a maximum recoverable scale of the observational setup estimated at $2.27\arcsec$ using the $5^{th}$ percentile of the shortest baseline lengths. The gas disc has a radius of $R_{90\%} = 1.45\arcsec$ and a flux of $15.8$ Jy km s$^{-1}$, which is in good agreement with \lon.  
The \lang{north-east} side of the disc appears \lang{blueshifted} compared to the systemic velocity of $\sim 8$ km\,s$^{-1}$, which means the disc rotates in an anti-clockwise way.
The morphology of the emission in the velocity channels is similar to the observations reported in \lon: the faint lower emission surface of the disc \mdy{lacks clear visibility, indicating potential blending with the upper emission surface due to spatial resolution constraints or possible non-detection. This makes} the true orientation of the disc ambiguous and the closest side to the observer difficult to define. Irregularities at the edge of the \lang{redshifted} side are recovered as well as in the northern side of the disc. However these irregularities are not found to be part of larger structures that could have traced potential interactions with the companion star V892~Tau~NE. In the following, we model V892~Tau as a triple system and try to constrain the orbit of V892~Tau~NE based on hydrodynamical simulations. Our methods to model the $^{12}$CO emission in detail and the hydrodynamical setups used are described in the Section \ref{sec:methods} below. From this modelling, we discuss in more detail the observations in Section \ref{sec:results}.


\section{Methods}
\label{sec:methods}

\subsection{Discminer}
\label{subsec:discminer}

\mdy{In order to understand at a deeper level the kinematical information contained in the structured disc of V892~Tau, we \lang{built} a quantitative Keplerian model to be compared with the observations. To do so, we} used the python package {\sc Discminer} \citep{DiscminerI, DiscminerII} to \mdy{fit} the $^{12}$CO (2-1) line emission \mdy{and kinematics} channel by channel \mdy{by a Keplarian disc model}. {\sc Discminer} first builds a Keplerian disc model from a set of parameters described hereafter. The position offset $(x_c, y_c)$, the inclination $i$ and the position angle $PA$ define the orientation of the disc. \mdy{In this work, the PA \lang{was} defined as the angle from the northern axis to the blueshifted semi-major axis.} The velocity profile is set by the central stellar mass $M_{*}$ and the systemic velocity $v_{sys}$. The disc upper and lower surfaces are both defined by exponentially tapered power laws (Equation \ref{eq:z_prof}). The line profile width at half maximum $L_w$ and the line profile slope $L_s$ are defined as power laws as well (Equation \ref{eq:L_prof}, valid for both $L_s$ and $L_w$). The reference intensity is taken from a peak intensity power law $I_p$ in a region $\mathcal{D}(R_{out})$ that extends from $r=0$ to $r=R_{out}$, where $R_{out}$ is the outer radius of the disc (Equation \ref{eq:I_prof}).  This whole set of parameters builds the final Keplerian model of intensity $I_m$ using a Bell function kernel with a projected velocity $v_{k,los}$ along the line of sight corresponding to the channel velocity $v_{ch}$ (Equation \ref{eq:I_chan}). In the equation below, $r_0$ \mdy{is the reference radius taken as $r_0=100$ au}, \mdy{while} $z_0$, $L_0$ and $I_0$ are values of reference for the vertical height, linewidth/lineslope and intensity respectively. The critical radius $r_b$ describes the exponential cutoff of the disc surface, with $q$ describing the strength of this tapering. Finally, $p$ is the power-law index of the disc surface.

\begin{equation}
\label{eq:z_prof}
    z(r) = z_0 \left(\frac{r}{r_0}\right)^p \exp\left(\frac{-r}{r_b}\right)^q \, ,
\end{equation}
\begin{equation}
\label{eq:L_prof}
    L(r, z) = L_{0} \left(\frac{r}{r_0}\right)^{p_L} \left(\frac{|z|}{z_0}\right)^{q_L} \, ,
\end{equation}
\begin{equation}
\label{eq:I_prof}
    I_p(r, z) = I_{0} \left(\frac{r}{r_0}\right)^{p_I} \left(\frac{|z|} {z_0}\right)^{q_I}  \mathcal{D}(R_{out}) \, ,
\end{equation}
\begin{equation}
\label{eq:I_chan}
    I_m(r,z, v_{ch}) = I_p \left( 1+ \Big\rvert \frac{v_{ch} - v_{k,los}}{L_w}\Big\rvert^{2L_s}\right)^{-1} \, ,
\end{equation}

{\sc Discminer} then fits the previous model to the data in the image plane channel by channel thanks to MCMC sampling using {\sc emcee} \citep{emcee}. \rmvd{We initialised $256$ walkers, each of them going for $40000$ steps. The chains reached convergence well before the end of the sampling and the resulting auto-correlation length was $106$ steps. The best-fit parameters were taken as the median of the posterior distributions, after the first $6300$ steps were ignored and considered as a burn-in phase. The uncertainties were the $16$ and $84$ percentiles of the posterior distributions.} More details about the initial conditions \mdy{and the results} of that procedure can be found in Appendix \ref{app : Discminer}.

From the best-fit model channels, we then built moment maps from the data and model channels. First, a moment 0 map was computed by integrating the intensity in each pixel along the velocity axis. We then collapsed the cube by selecting the brightest pixel along the velocity axis for each pixel to create a peak temperature moment map (moment 8). Picking up the centroid velocity corresponding to that peak allowed \lang{for} a velocity map of the disc \lang{to be built} (moment 9). Finally, we built a moment map corresponding to the line-width around that centroid velocity. We applied a $3\sigma$ clipping in the building of data moment maps. The same mask \mdy{delimiting the detected disc area} was applied to the model moment maps. Residual maps \mdy{were built} by subtracting the masked model moment maps to the data moment maps.

\subsection{Hydrodynamical simulations}
\label{subsec:hydro_sim}

Several hints of interaction with the external star V892~Tau~NE are detected in the V892~Tau CBD (tidal truncation, tentative spiral arms, disc tilt, tentative disc warp). In order to confirm that V892~Tau~NE is bound to the V892 system and to constrain its orbit, we performed 3D hydrodynamical simulations using the {\sc Phantom} \citep{Price+2018-phantom} smoothed particle hydrodynamics (SPH, e.g. \cite{Monaghan1992}) code. 
Our model consisted in a central binary star surrounded by a CBD and an outer stellar companion orbiting further out. The model used of $10^6$ SPH particles distributed in the disc according to a surface density distribution $\Sigma \propto r^{-p}$ with $p=1$ that extends initially from $14$ au to $200$ au for a total disc mass of $0.06$ M$_{\odot}$ (\lon). The true orientation of the V892~Tau CBD is unconstrained due to an uncertainty on the inclination sign of the disc. We \lang{assumed} that inclination to be positive and of $54.6\degree$, meaning the closest side to the observer \lang{was} the SE side. The PA of the disc \lang{was} set to $53\degree$ in agreement with the observations of \lon. 
The inner border of the simulated CBD \lang{was} set to $14.2$ au, which is twice the semi-major axis of the inner binary and in line with theoretical predictions of the cavity size \citep{Miranda+2017}. The aspect ratio $H/r=0.055$ at the reference radius $R_0=100$ au and the sound speed exponent $q=0.185$ \lang{were} set to match the fitted temperature profile in \lon.

We \lang{modelled} the stars as sink particles \citep{Bate+1995} with accretion radii of $1$ au for the two components of the inner binary and of $10$ au  for the outer star. The inner binary \lang{was} initialised coplanar \mdy{in a prograde configuration} with \mdy{respect to} the disc while its other orbital parameters and its mass \lang{were} set in line with the observations (\lon), which allows us to disentangle more easily the effects triggered by the companion solely. Moreover the observed inclination of the disc to the inner binary plane is measured to be down to $4\degree$, which is close to coplanarity (\lon).
The orbit of the companion star V892~Tau~NE remains unconstrained so far. One of the goals of this work is to explore if this companion is bound to the inner binary and to constrain its eccentricity and inclination parameters $e_{out}$ and $i_{out}$ respectively. From there we \lang{modelled} the V892~Tau system as a triple system in five numerical setups, with a different orbit for the outer companion each time:
\begin{itemize}
    \item \textit{ref} (reference case) : $e_{out} = 0.2$ \& $\Delta i_{out} = 0\degree$,
    
    \item \textit{e05} (eccentric case) : $e_{out} = 0.5$ \& $\Delta i_{out} = 0\degree$,
    
    \item \textit{i30} (inclined case 1) : $e_{out} = 0.2$ \& $\Delta i_{out} = 30\degree$,
    
    \item \textit{i60} (inclined case 2) : $e_{out} = 0.2$ \& $\Delta i_{out} = 60\degree$,
    
    \item \textit{ei60} (eccentric inclined case) : $e_{out} = 0.5$ \& $\Delta i_{out} = 60\degree$,
\end{itemize}

where $\Delta i_{out}$ represents the mutual inclination between the companion and the inner binary orbital plane. It \lang{resulted} in inclination parameters with respect to the sky plane of $i_{out}=54.6\degree$ for \textit{ref} and \textit{e05}, $i_{out}=24.6\degree$ for \textit{i30} and $i_{out}=-5.4\degree$ for \textit{i60} and \textit{ei60}. The maximal eccentricity of the outer binary has been constrained to $e_{out} < 0.2$ considering a coplanar orbit and a disc extent of $R_{out}\sim200$ au (\lon). This estimation assumed a coplanar companion, but a misaligned orbit could allow for more eccentric orbits in line with the truncation of the disc. We \lang{explored} values up to $e_{out}=0.5$. We expect different eccentricity and inclination parameters to have consequences on the disc truncation and on the disc orientation.
From dynamical considerations, it is more likely that V892~Tau~NE is observed close to apoastron \citep{VanAlbada1968}. We accordingly set the argument of the periapsis to $\omega_{out}=180\degree$. Given the projected separation of the companion of $\sim520$ au \citep{GaiaDR3}, we \lang{assumed} the semi-major axis to be $a_{out}=500$ au. The longitude of the ascending node $\Omega_{out}=53\degree$ \lang{was} set in line with the observed disc PA. We \lang{assumed} the mass of V892~Tau~NE to be $0.5$ M$_\odot$, which is in agreement with the mass of similar M3 stars \citep{Luhman+2007, Flores+2022}.

\begin{figure}
\centering
\begin{center}
    \includegraphics[width=\columnwidth, trim={5.5cm 1cm 5.5cm 1cm},clip]{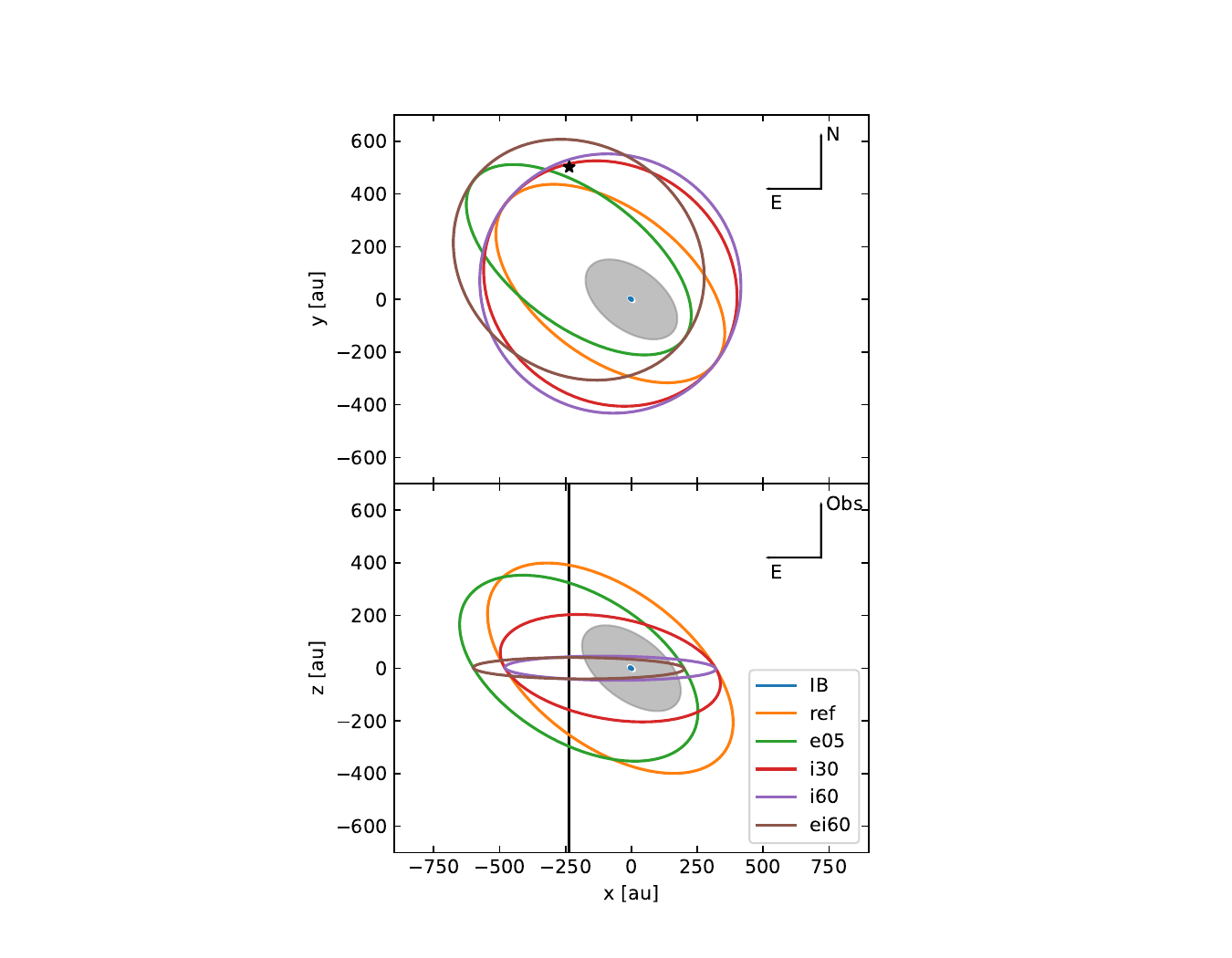}
    \caption{Projection of the initialised orbits of each tested setup in the sky-plane $xy$ and in a plane perpendicular to it, namely $xz$. The coordinates are centred on the primary. The CBD is represented by the shaded grey area. The projected position of V892~Tau~NE in the skyplane is represented by a star and by a black line in the xz plane. The orbit of the inner binary (IB) is plotted in blue and the other colours correspond to the tested orbits for V892~Tau~NE.}
    \label{fig:grid_orbits}
\end{center}
\end{figure}

Table \ref{table:setups_params} summarises the parameters defining the orbital configuration of the setups. In that table, the reference plane is the plane of the sky and the origin is the centre of mass of the inner binary. Figure \ref{fig:grid_orbits} shows the initialised orbits for each setup projected in the sky-plane and a plane perpendicular to it.
We evolved the hydrodynamical simulations for $50 $ P$_{out}$, with $P_{out}=4375$ yrs the period of the outer binary for a semi-major axis of $a_{out} = 500$ au. It corresponds to $28400$ periods of the inner binary.
At such timescales, the disc had time to relax from its initial condition and to evolve significantly.

\begin{table}[h!]
\caption{Initial parameters of the hydrodynamical simulations.} 
\centering
 \begin{tabular}{c c c} 
\multicolumn{3}{c}{Inner binary} \\
\hline 
   Parameter & Observations & Simulations \\ 
 \hline \hline

 $M_{in}$ (M$_{\odot}$) & $6.0\pm0.2$ & $6.0$ \\
 \hline
 $q_{in}$ & $\sim0.5$ & $0.5$ \\
 \hline
 $a_{in}$ (au) & $7.1\pm0.1$ & $7.1$ \\
 \hline
 $e_{in}$ & $0.27\pm0.10$ & $0.27$ \\
 \hline
 $i_{in}$ ($\degree$) & $59.3\pm2.7$ & $54.6$ \\
 \hline
 $\Omega_{in}$ ($\degree$) & $50.5^{+9.6}_{-8.8}$  & $53$ \\
 \hline
 $\omega_{in}$ ($\degree$) & $179.9^{+44.4}_{-30.3}$ & $180$ \\
 
  & & \\

\multicolumn{3}{c}{Outer binary} \\
\hline 
   Parameter & Observations & Simulations \\ 
 \hline \hline
 
 $M_{out}$ (M$_{\odot}$) & - & $6.5$\\
 \hline
 $q_{out}$ & - & $0.08$ \\
 \hline
 $a_{out}$ (au) & - & $500$\\
 \hline
 $e_{out}$ & - & \textit{ref}: $0.2$ \;\; \textit{i30}: $0.2$ \\
  &  &  \textit{e05}: $0.5$  \;\; \textit{i60}: $0.2$  \\
  &  & \textit{ei60}: $0.5$ \\
 \hline
 $i_{out}$ ($\degree$) & - & \textit{ref}:$ 54.6$ \;\; \textit{i30}: $24.6$ \\
  &  &  \textit{e05}: $54.6$  \;\; \textit{i60}: $-5.6$  \\
  &  &  \textit{ei60}: $-5.6$ \\
 \hline
 $\Omega_{out}$ ($\degree$) & - & $53$ \\
 \hline
 $\omega_{out}$ ($\degree$) & - & $180$ \\
  
    & & \\

\multicolumn{3}{c}{Circumbinary disc} \\
\hline 
   Parameter & Observations & Simulations \\ 
 
 \hline \hline
 
 $R_{in}$ (au) & - & $14.2$ \\
 \hline
 $R_{out}$ (au) & $196\pm3$ & $200$ \\
 \hline
 $R_{0}$ (au) & $100$ & $100$ \\
 \hline
 $M_d$ (M$_{\odot}$) & $\sim0.06$ & $0.06$ \\
 \hline
 $p$ & - & $1$\\
 \hline
 $q$ & $0.185$ & $0.185$ \\
 \hline
 $\Omega_d$ ($\degree$) & $53.0 \pm 0.7$ & $53$\\
 \hline
 $i_d$ ($\degree$) & $54.6 \pm 1.3$ & $54.6$ \\
 \hline
 $\frac{H}{R}\vert _{R_0}$ & $0.055$ & $0.055$ \\
 \hline
 $\alpha_{SS}$ & - & $0.005$\\ 

\end{tabular}
\tablefoot{The indices $in$, $out$ and $d$ refer to the inner binary, the outer binary and the disc respectively. '-' indicates a lack of data. The reference plane is the sky-plane. The disc parameters are taken from the centre of the inner binary. \lang{The} observational values are taken from \lon.}
\label{table:setups_params}
\end{table}

\subsection{Radiative transfer and post-processing}
\label{subsec:rt_proc}

The simulation outputs were post-processed via radiative transfer to produce intensity \lang{datacubes} of the system. Because the disc oscillates and precesses during the simulations, its inclination and position angle are evolving with time. In order to have a proper comparison with the observations, the disc was manually moved back to an average inclination of $54.6\degree$ and an average PA of $53\degree$ before the post-processing. \mdy{We report the inclination and PA values of the simulated discs before this rotation in Appendix \ref{app : manual move back}}. We used the radiative transfer code {\sc MCFOST} \citep{Pinte+2006, Pinte+2009} \lang{that} employs a Voronoi tesselation to build a grid based on the SPH particle distribution, with one cell per SPH particle. Since our SPH simulations contains gas particles only, the dust spatial distribution is assumed to follow the gas distribution.
Dust grains were assumed to be at local thermal equilibrium and subject to passive heating. They were assumed compact and following opacity laws of astrosilicates \citep{WeingartnerDraine2001} with their scattering properties computed following the Mie theory framework. We used $100$ grain sizes from $0.03\mu$m to $1000 \mu$m logarithmically spaced in scale. The global distribution was normalised by integrating over all grain sizes, assuming a typical power-law of exponent $-3.5$ \citep{Mathis1977}, and all over the grid to have a dust to gas mass ratio of $0.01$. We adopted a uniform value of $1\times10^{-4}$ for the $^{12}$CO abundance compared to H$_2$. The turbulent velocity was set to zero and the freeze-out temperature below which the grid cells do not emit, to $20$ K. 
The sink particles were considered as spherical stars that radiate isotropically as black-bodies with temperature and luminosity chosen in agreement with their mass \mdy{at an age of $3$ Myr \citep{Siess+2000} consistent with an estimated age of the system of $\sim 2$ Myr \citep{KucukAkkaya2010}}. This way each $3$ M$_{\odot}$ component of the inner binary \lang{had} a temperature of $10745$ K and a luminosity of $72$ L$_{\odot}$ while the $0.5$ M$_{\odot}$ companion \lang{had} a surface temperature of $3758$ K and a luminosity of $0.3$ L$_{\odot}$. $1.28\times10^8$ photons packets were used to compute the 3D temperature using a Monte-Carlo approach and $1.28\times10^6$ other photons packets were used to compute images with a ray-tracing method based on the temperature structure.

$^{12}$CO (2-1) channel maps were produced with a $0.075 $ km\,s$^{-1}$ resolution from $-15 $ km\,s$^{-1}$ to $15 $ km\,s$^{-1}$. The velocity axis was then shifted from the systemic velocity $v_{sys}=7.86 $ km\,s$^{-1}$. The produced \lang{datacube} was then post-processed with {\sc CASA} \citep{McMullin2007} to produce synthetic observations. 
Using the \textit{simalma} task, Measurement Sets (MS) were generated from the radiatively processed data with the right sampling of the \textit{uv} plane. The sampling in the \textit{uv} plane was taken from the antenna configurations used in the observations. Considering that the observations are made from $3$ observational \lang{datasets}, we produced synthetic data corresponding to each individual \lang{dataset}. The integration times were set to $9.25$ minutes for the synthetic SB and LB sets, and to $8$ minutes for the synthetic \lang{dataset} matching the OLD set. The precipitable water vapour index $PWV=0.6$ mm was also chosen in line with the observing conditions. 
It resulted in $3$ synthetic MS each matching an actual MS. The simulated MS were then spectrally Hanning-smoothed to match the spectral resolution of the observations. 
The $3$ synthetic \lang{datasets} had their continuum subtracted before being imaged together with the exact same \textit{tclean} parameters used in the imaging process of the observations (see Section \ref{subsec : data}). Moment 0, moment 8 (peak temperature) and moment 9 (peak velocity) maps were created from the treated synthetic cubes using the same clipping as in the observations. The central channels where the data signal is absorbed were ignored in the process.

\subsection{N-body simulations}
\label{subsec:nbody}

Additionally, we performed N-body simulations using the code {\sc rebound} \citep{rebound}. The setup reproduced the V892~Tau system with 3 stars and a CBD. Stars were modelled as to massive point particles and the disc was modelled using 10 test particles with the same inclination and position angle as the disc in the SPH setups. The test particles semi-major axis extended from $14.2$ au to $200$ au, corresponding to the borders of the disc. Their eccentricity and mass were set to zero. In agreement with the data, the test particles start the simulation with a mutual inclination of $8\degree$ with the inner binary. \lang{Five} simulations were performed with the same parameters as the SPH setups \textit{ref, e05, i30, i60}, \lang{and \textit{ei60}}, respectively. The system is then integrated with an \textit{ias15} integrator \citep{ias15} during $2000$ periods of the outer binary, corresponding to $\sim 9$ Myr.


\section{Results}
\label{sec:results}

In the following, we investigate the structure of the gaseous disc and its deviations from our Keplerian model (see Section \ref{subsec:discminer}). We focus our analysis on these deviations and to what extent the simulations are able to capture those non-Keplerian patterns.

\subsection{Channel maps}
\label{subsec:channels_result_part}

\begin{figure*}
\centering
\begin{center}
    \includegraphics[width=1.0\textwidth, trim={3cm 1.8cm 2cm 1.8cm},clip]{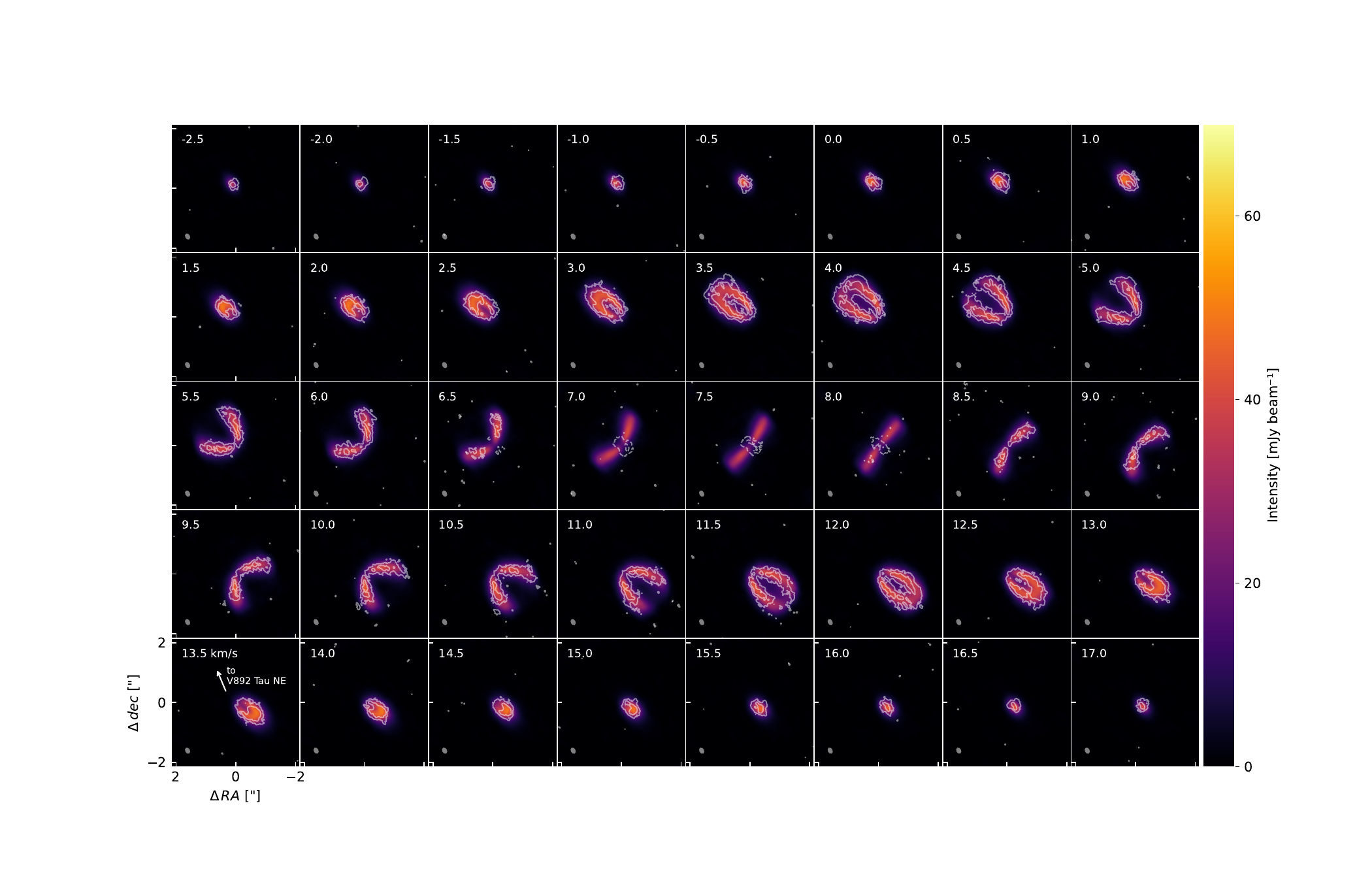}
    \caption{Channel maps of the $^{12}$CO (2-1) emission. The {\sc Discminer} best-fit model is plotted in colours and the data is represented by the white contour lines at $[3, 7] \sigma$. Negative emission at the same significance is indicated by dashed lines. The synthesised beam is plotted at the bottom left of each channel map, while the corresponding velocity is plotted at the top left. On the bottom left channel, an arrow indicates the direction to the companion star V892~Tau~NE.}
    \label{fig:channels}
\end{center}
\end{figure*}

Figure \ref{fig:channels} shows the best-fit model channels from {\sc Discminer} and their comparison to the data. The modelled emission in the channels fits well to the bulk emission of the data and the geometry of the disc is well captured by the Keplerian model. The best-fit inclination $i=54.7\degree$, position angle $PA=53.6\degree$ and dynamical mass $M_*=5.99 $ M$_{\odot}$ are in line with previous observations (\lon). The systemic velocity is found to be around the channels that are experiencing absorption from the cloud at $v_{sys}=7.86$ km\,s$^{-1}$. The disc is thin and sharply truncated at the outer edge given the best-fit lower and upper emission surfaces. The full details of the best-fit parameters and the priors used can be found in Table \ref{table:discminer_fit}.

Looking at the emission in individual velocity channels allows \lang{the deviations to be tracked} from the model at specific velocities. From $3$ km\,s$^{-1}$ to $4$ km\,s$^{-1}$, the data emission extends further than the model emission. This excess in emission is directed \lang{towards} V892~Tau~NE. The $3\sigma$ emission contour shows irregular features at the top of the channel emission up to $5.5$ km\,s$^{-1}$. From $5$ km\,s$^{-1}$ to $11.5$ km\,s$^{-1}$, the emission of the lower wing is truncated with respect to its model counterpart while the emission extent of the upper wing of each channel is well reproduced by the model. Small deviations are also observed in the \lang{south-west} part of the $12.5-14$ km\,s$^{-1}$ channels and irregular patterns are seen especially in the $12.5$ km\,s$^{-1}$ channel. These deviations could be part of a larger non-detected structure (\lon), as for instance large spiral arms commonly launched by external companion at the edge of discs. We discuss that possibility in more details in Section \ref{subsec:spirals}. Other possible explanations for these non-Keplerian features include a warped disc or a discrepancy in vertical height between the data and the model \citep{Law+2023}. 

\begin{figure}
\centering
\begin{center}
    \includegraphics[width=\columnwidth, trim={1cm 1cm 1cm 2cm},clip]{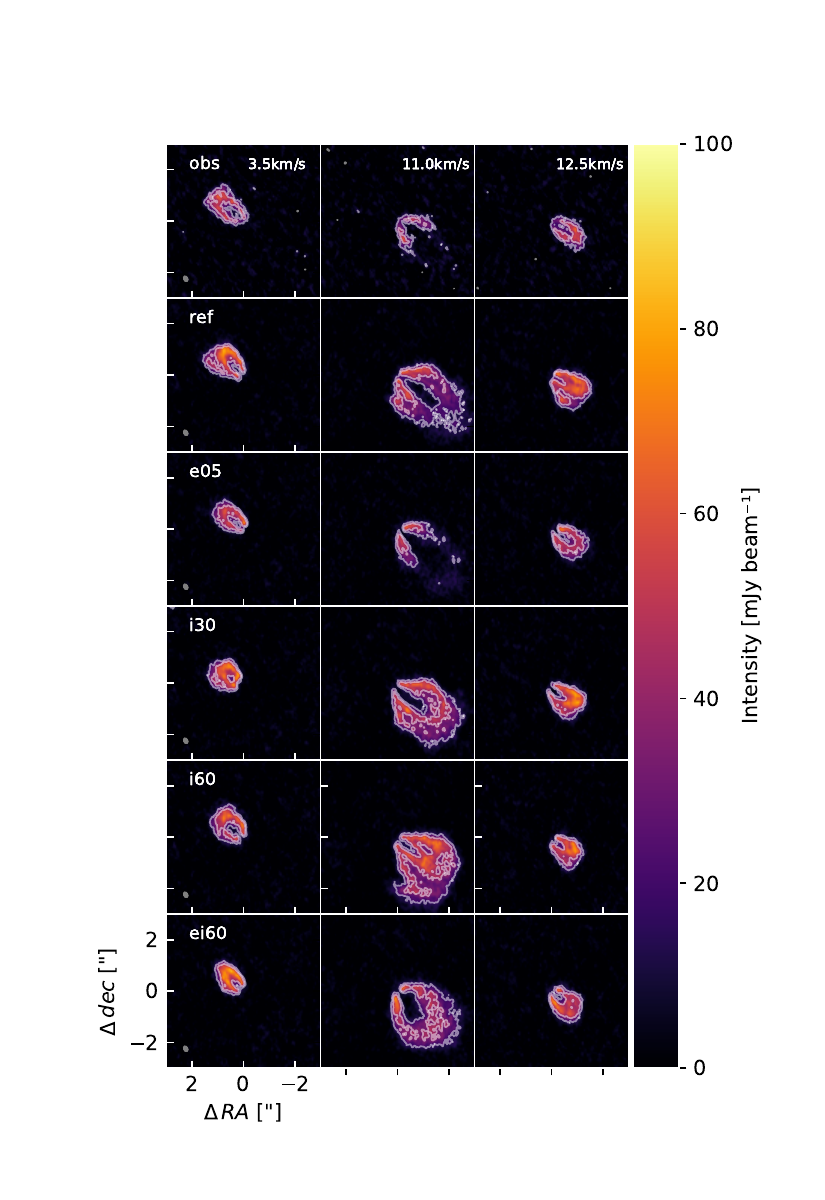}
    \caption{Comparison of channel maps at $3.5$ km\,s$^{-1}$ (left), $11$ km\,s$^{-1}$ (middle), and $12.5$ km\,s$^{-1}$ (right) of the observed emission (top row) and of the synthetic emission (following rows, in order: \textit{ref}, \textit{e05}, \textit{i30}, \textit{i60}, \lang{and} \textit{ei60}). The contours represent the $[3, 7]\sigma$ emission levels. The \lang{beam size} is represented by the grey ellipses.}
    \label{fig:channel_comp}
\end{center}
\end{figure}

Data channels at $3.5$ km\,s$^{-1}$, $11$ km\,s$^{-1}$ and $12.5$ km\,s$^{-1}$ show clear deviations from our Keplerian model. On Figure \ref{fig:channel_comp}, we look at these particular channels in the simulations and check if the deviations are reproduced or not. 
The lower surface of the disc is also visible in the simulations \textit{ref}, \textit{i30}, and \textit{i60} at $3.5$ km\,s$^{-1}$ and $12.5$ km\,s$^{-1}$, which is not the case for the other simulations \textit{e05} and \textit{ei60} and the observations. 
The emission in the $11$ km\,s$^{-1}$ synthetic channel extends further out in \textit{ref}, \textit{i30}, \textit{i60}, and \textit{ei60}. In general the signal is higher in the synthetic channels than in the data.
The closest match is made with the synthetic channels built from \textit{e05}: the emission pattern is comparable to the data in terms of size, morphology, and intensity. The \textit{e05} disc is sharply truncated and very thin due to most of the initial disc material being ejected by the companion at the beginning of the simulation. Some deviations are also recovered in the outer disc, suggesting that the kinematics observed in the data and \textit{e05} are similar.

\subsection{Disc morphology}
\label{subsec:disc_morph}

\begin{figure*}
\centering
\begin{center}
    \includegraphics[width=0.75\textwidth]{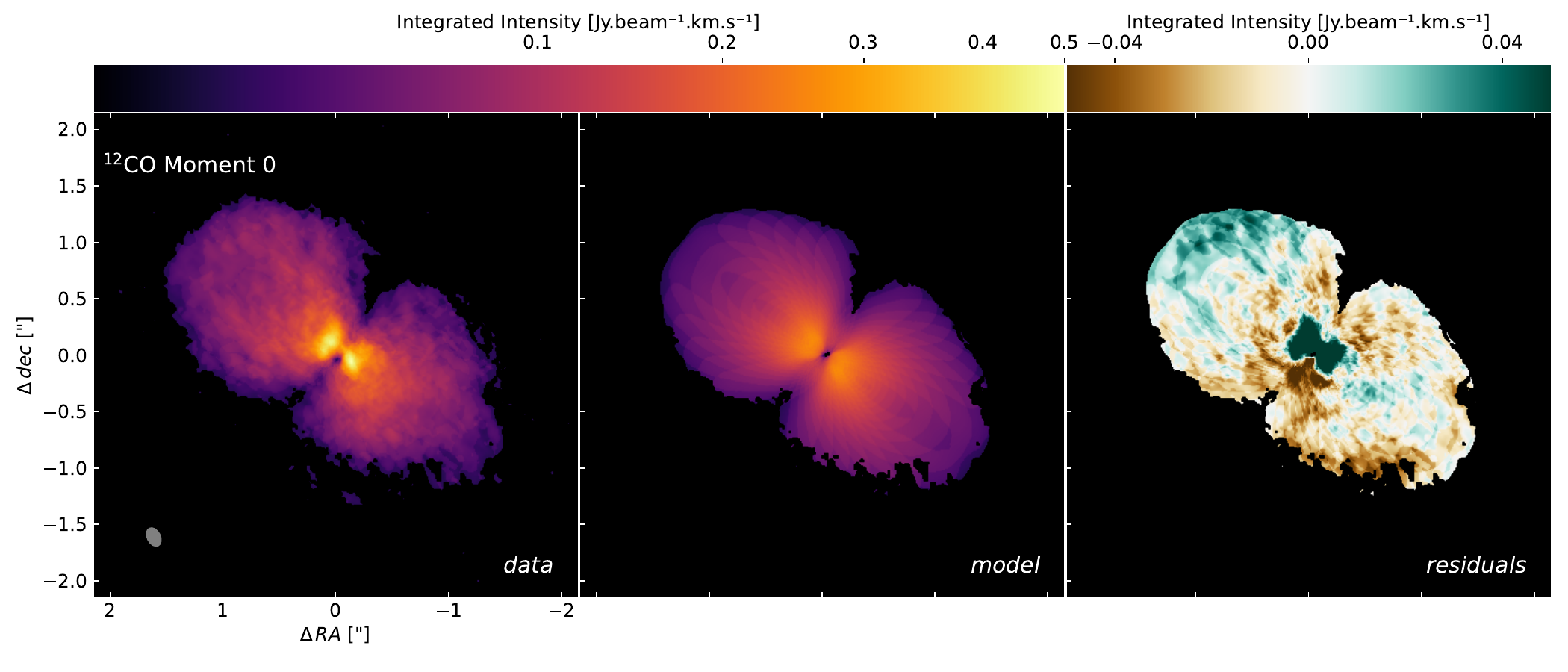}     \includegraphics[width=0.75\textwidth]{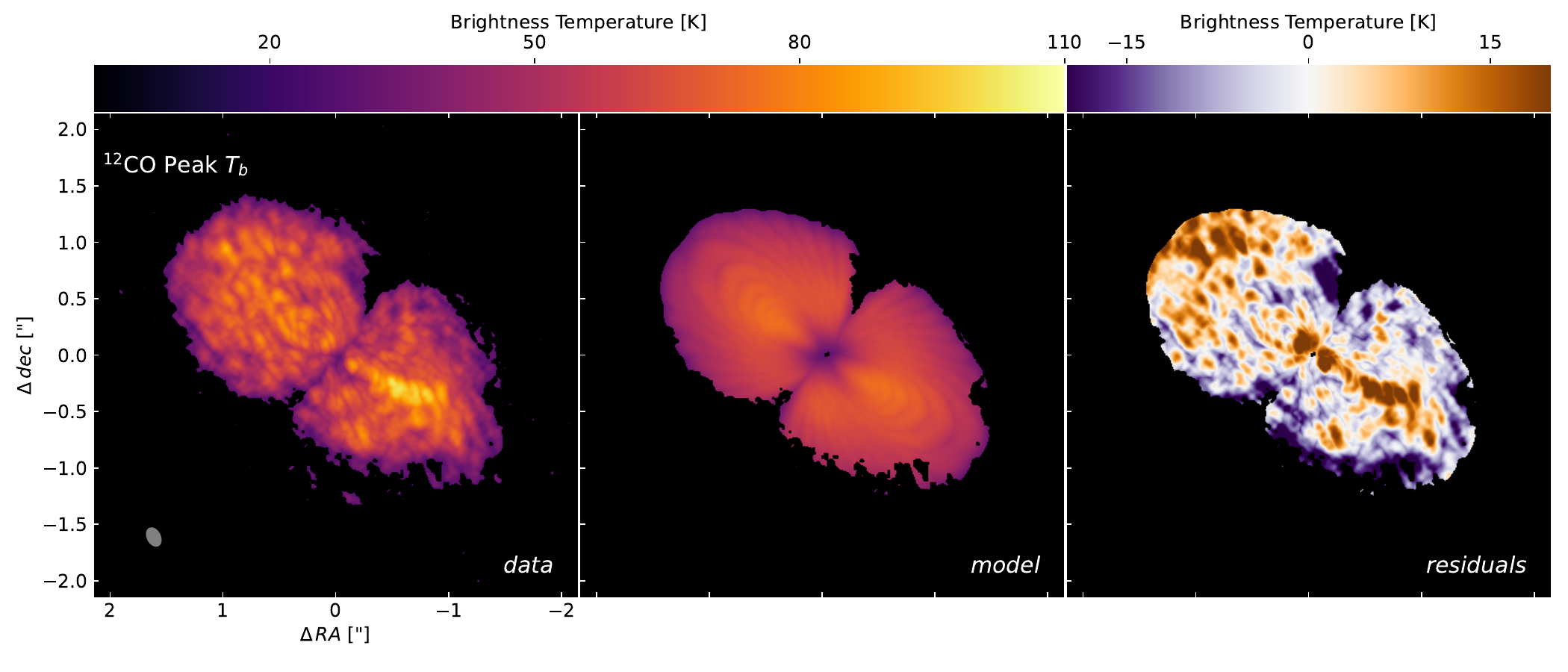}       \includegraphics[width=0.75\textwidth]{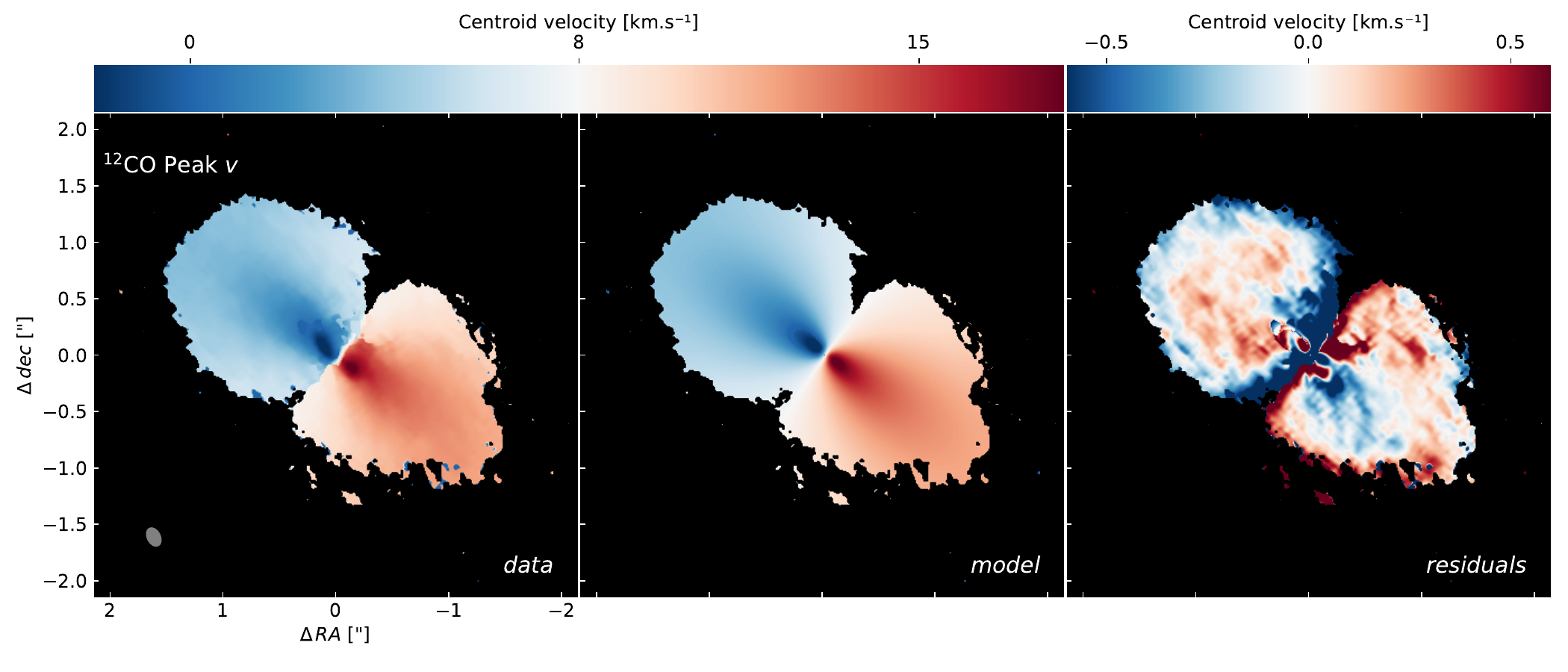}         \includegraphics[width=0.75\textwidth]{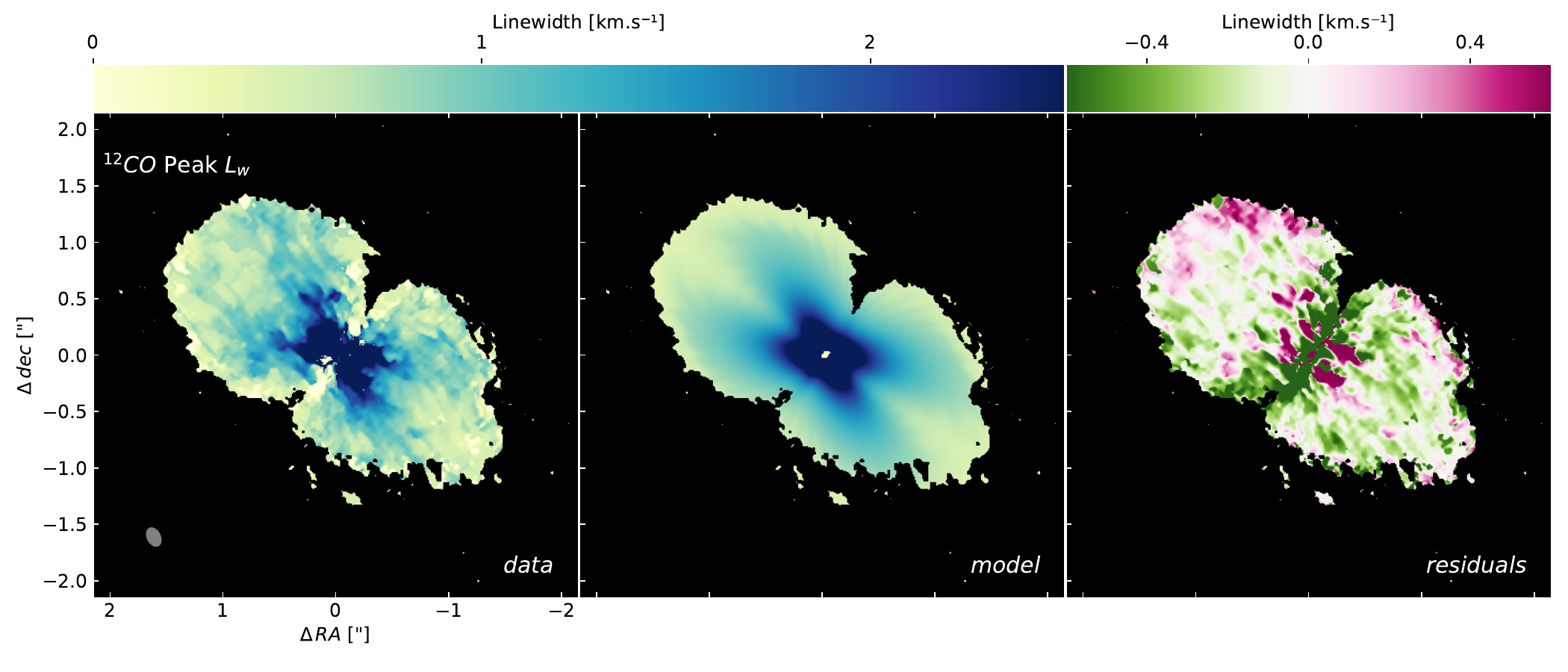}
    \caption{$^{12}$CO (2-1) moment maps of V892~Tau. The integrated intensity, the peak temperature, the peak velocity, and the line width correspond to the first, second, third, and last rows,  respectively. The first column shows moment maps computed from the data emission channels, the second column from the model and the last column shows the residuals \mdy{in the detected disc area} computed from the two previous maps. \mdy{A mask defined by the $3\sigma$ contour of the data has been applied to the model before comparison with the data.}}
    \label{fig:obs_model_panel}
\end{center}
\end{figure*}

\begin{figure}
\centering
\begin{center}
    \includegraphics[width=\columnwidth, trim={3cm 0cm 3cm 0cm},clip]{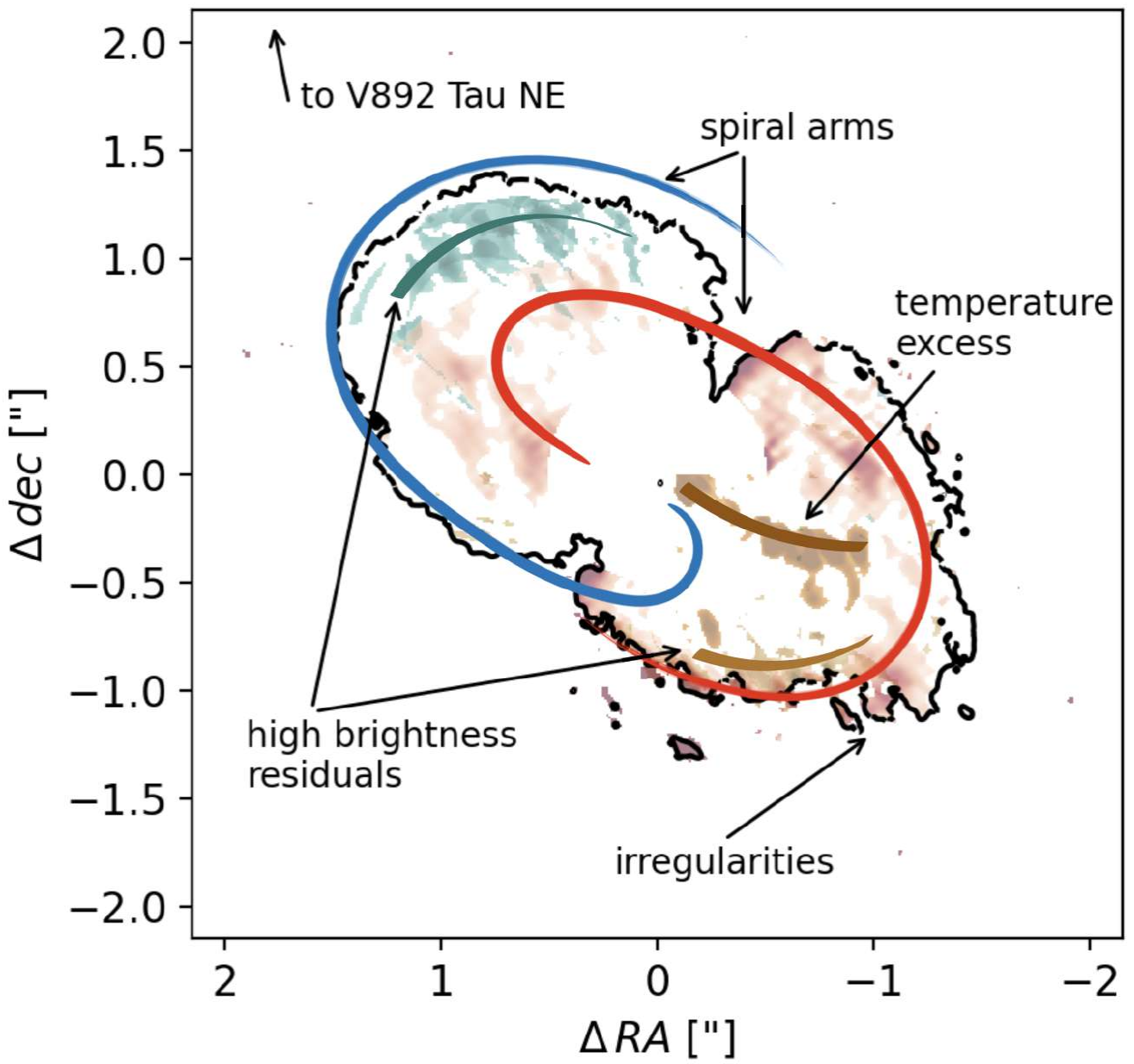}     
    \caption{
    \mdy{Sketch of the features identified in the CBD of V892~Tau from \lang{the}} residual maps shown in Figure \ref{fig:obs_model_panel}. The black line corresponds to the $3\sigma$ contour of the disc emission. The indicated high brightness residuals come from the moment 0 residuals. \mdy{The spiral arms are identified on the velocity residual map, from which only the red part is plotted and the centre masked for clarity reasons.} The temperature asymmetry coming from the peak temperature residual map is \mdy{overlaid} in dark brown. The arrow in the top left indicates the direction of V892~Tau~NE.}
    \label{fig:sketch_obs}
\end{center}
\end{figure}

The moment maps built from the best-fit modelled channels by {\sc Discminer}  are shown in Figure \ref{fig:obs_model_panel}, which includes the data moment maps and the residual maps \mdy{in the detected disc area} following the subtraction of that data and best-fit. In this section we discuss the disc morphology based on the intensity and temperature moment maps that we compare to their corresponding synthetic maps. The main features of the residual maps are highlighted on Figure \ref{fig:sketch_obs}. The velocity and linewidth moment maps will be discussed in Sections \ref{subsec:spirals} and Appendix \ref{app : incl sign}.

On the moment 0 map, the residual patterns show two main arcs at the disc edges:
one \lang{to} the north of the blueshifted side and one other \lang{to} the south of the redshifted side. The first arc represents a brightness excess while the second indicates a brightness deficit. The two arcs are quite symmetric with respect to the inner binary. Such residuals have already been identified in (\lon). The possible explanation invoked in that study was the presence of a warp in the disc and a different PA between the inner and outer disc. The arcs are also recovered in the peak temperature residual map. The latter map also shows a temperature excess oriented towards the \lang{south-west} at the centre of the redshifted side of the disc (see Fig \ref{fig:sketch_obs}) which had also been detected in (\lon). Thanks to a better angular resolution, we are able to see this asymmetry heading \lang{towards the western}
direction as it reaches the outer disc. 
Globally, the V892~Tau CBD is found to be a warmer disc than most T Tauri stars (e.g. \cite{Wolfer+2023}), heated by the two A-type stars at its heart and with a peak brightness temperatures found up to a few $110$ K.

The centre of the disc is not well reproduced by our Keplerian model according to the high residual levels in the middle of the residual maps. The modelling of the inner system of V892~Tau is complicated by the absorption of channels around the systemic velocity at the centre of the system which puts it beyond the scope of this work. 

\begin{figure*}
\centering
\begin{center}
    \includegraphics[width=1.0\textwidth,  trim={1cm 0.4cm 2cm 0cm},clip]{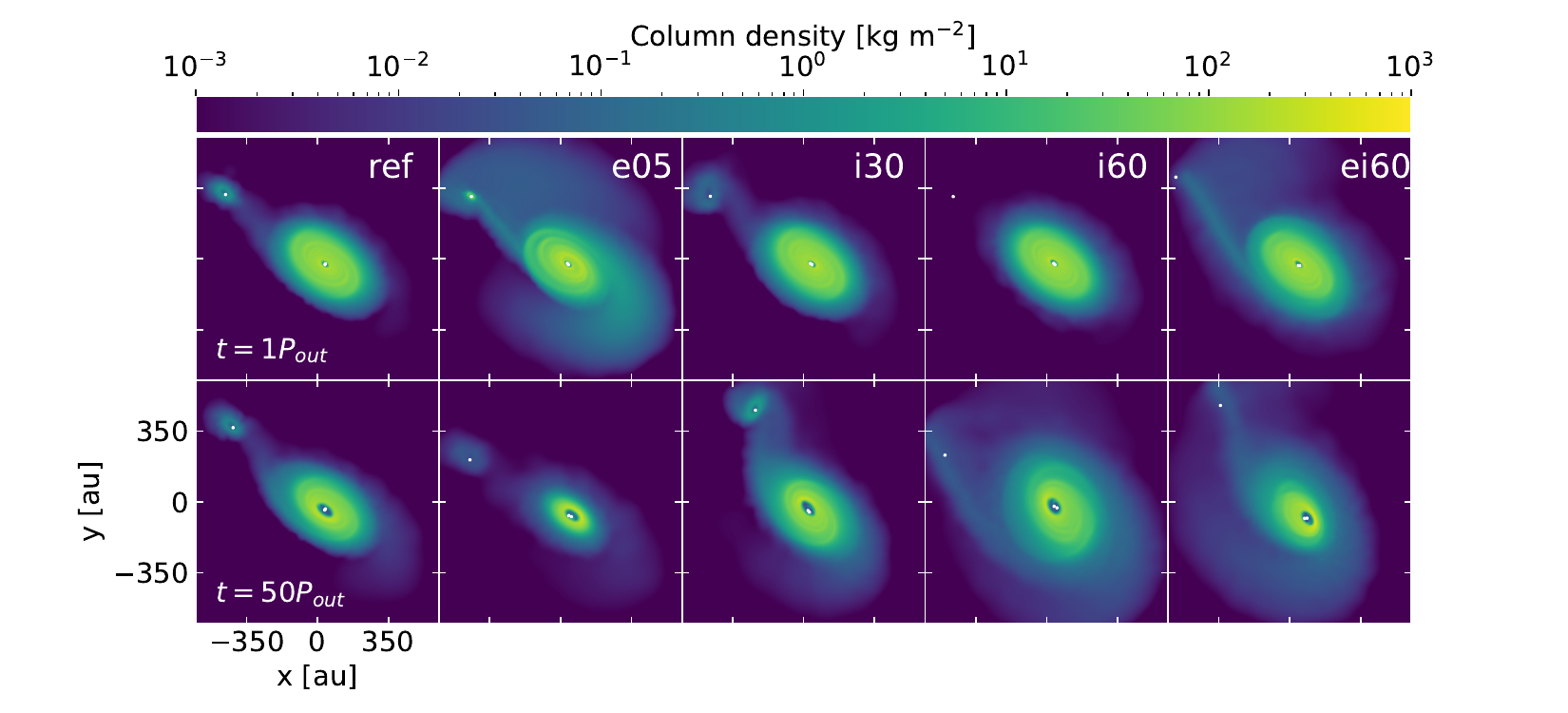}
    \caption{Rendered plots of the density spatial distribution integrated along the line of sight of the different simulated setups. \lang{The}
    \mdy{\textit{ref}, \textit{e05}, \textit{i30}, \textit{i60,} and \textit{ei60} setups are shown from left to right. The top row shows the system after one orbit of the outer companion. The bottom row displays the state of the system after approximately $50$ orbits of the outer companion, corresponding to $0.2$ Myr and $28400$ orbits of the inner binary. The white dots indicate the sink particles representing the stars.}}
    \label{fig:simgrid_rho}
\end{center}
\end{figure*}

Each tested orbit of the companion leads to a different morphology of the CBD, as shown in the snapshots of Figure \ref{fig:simgrid_rho}. While the discs in the \textit{ref} and \textit{e05} simulations are oriented close to their initial orientation, the inclination and the PA of the discs in \textit{i30}, \textit{i60}, and \textit{ei60} have significantly changed. We comment on the disc orientation in Section \ref{subsec:tilttwist} below. The disc in \textit{e05} shows a limited radial extent due the repeated passages of the companion star close to the disc, which also spread out material that forms streamers between the companion and the CBD in all the simulations. The external companion also triggers two large spiral arms at the edge of the disc while the inner binary carves a central cavity in every simulated discs. Lastly, a thin circumstellar disc is sometimes captured by the companion at the periastron but dissipates shortly after due to the large accretion radius used for the sink particle representing the companion. One can see that the \textit{i30}, \textit{i60}, and \textit{ei60} discs are mildly eccentric at the end of the simulation.

\begin{figure}
\centering
\begin{center}
    \includegraphics[width=\columnwidth, trim={1cm 1.8cm 0cm 1.8cm},clip]{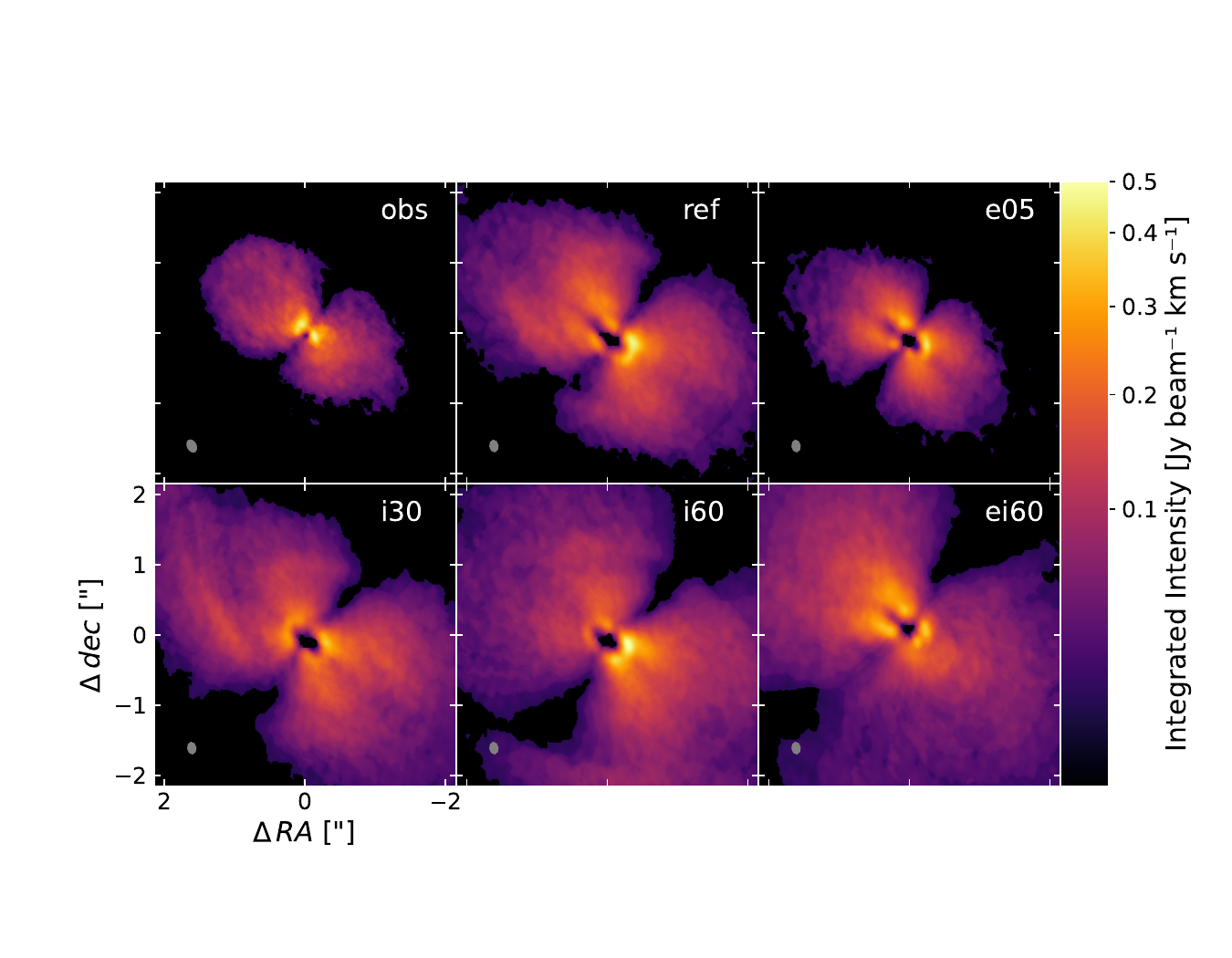}
    \caption{$^{12}$CO (2-1) integrated intensity map of the observational data (top left) compared to the synthetic maps built from the simulations \textit{ref} (top middle), \textit{e05} (top right), \textit{i30} (bottom middle), and \textit{i60} (bottom right). The discs were rotated back in the observed disc plane to allow \lang{for} a proper comparison to the observations. The \lang{beam sizes} are plotted in the bottom left corners.}
    \label{fig:simgrid_mom0_synth}
\end{center}
\end{figure}

\begin{figure}
\centering
\begin{center}
    \includegraphics[width=\columnwidth, trim={1cm 1.8cm 0cm 1.8cm},clip]{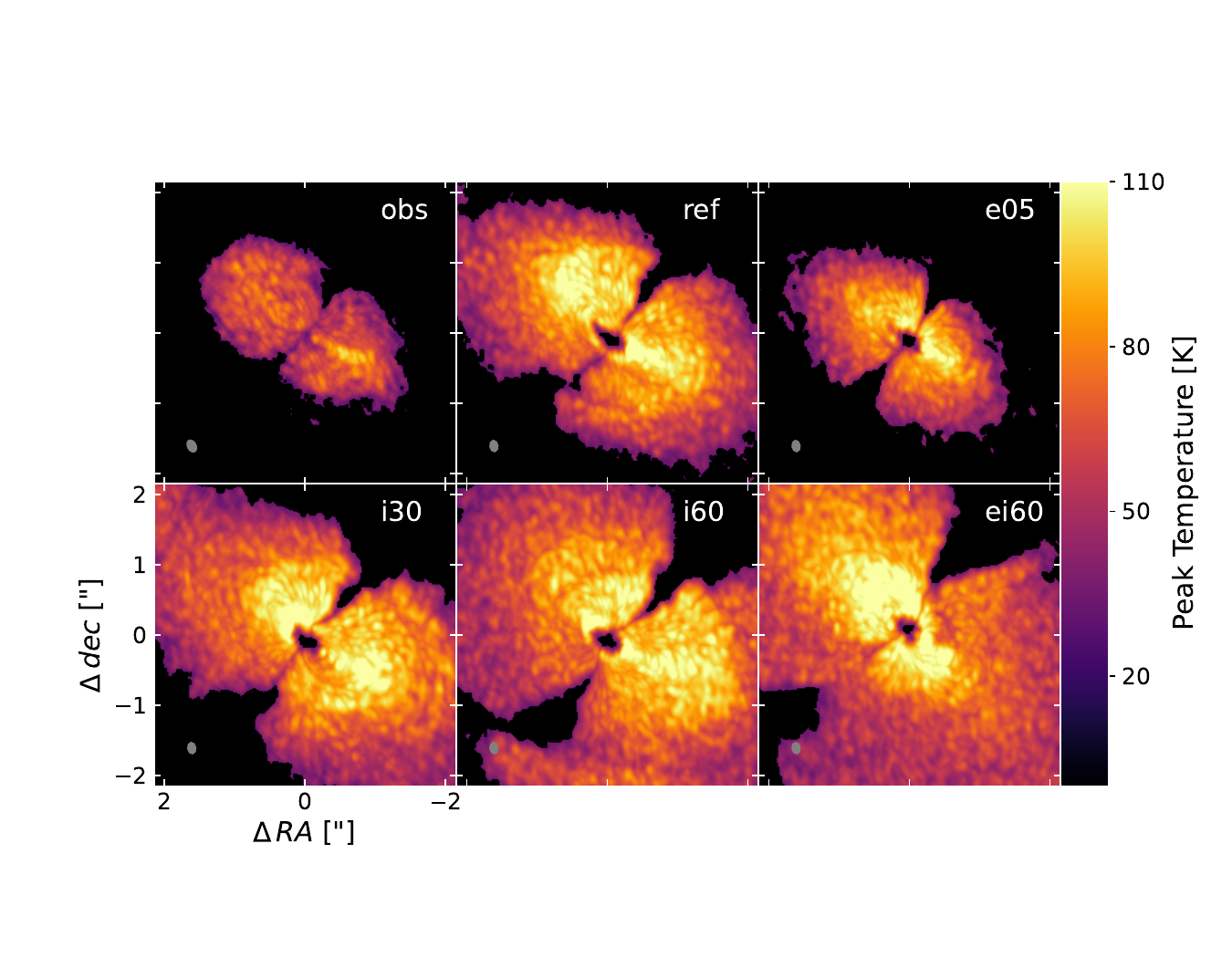}
    \caption{Same as Figure \ref{fig:simgrid_mom0_synth} \lang{but} for the $^{12}$CO (2-1) peak temperature map.}
    \label{fig:simgrid_mom8_synth}
\end{center}
\end{figure}

\begin{figure}
\centering
\begin{center}
    \includegraphics[width=\columnwidth, trim={1cm 1.8cm 0cm 1.8cm},clip]{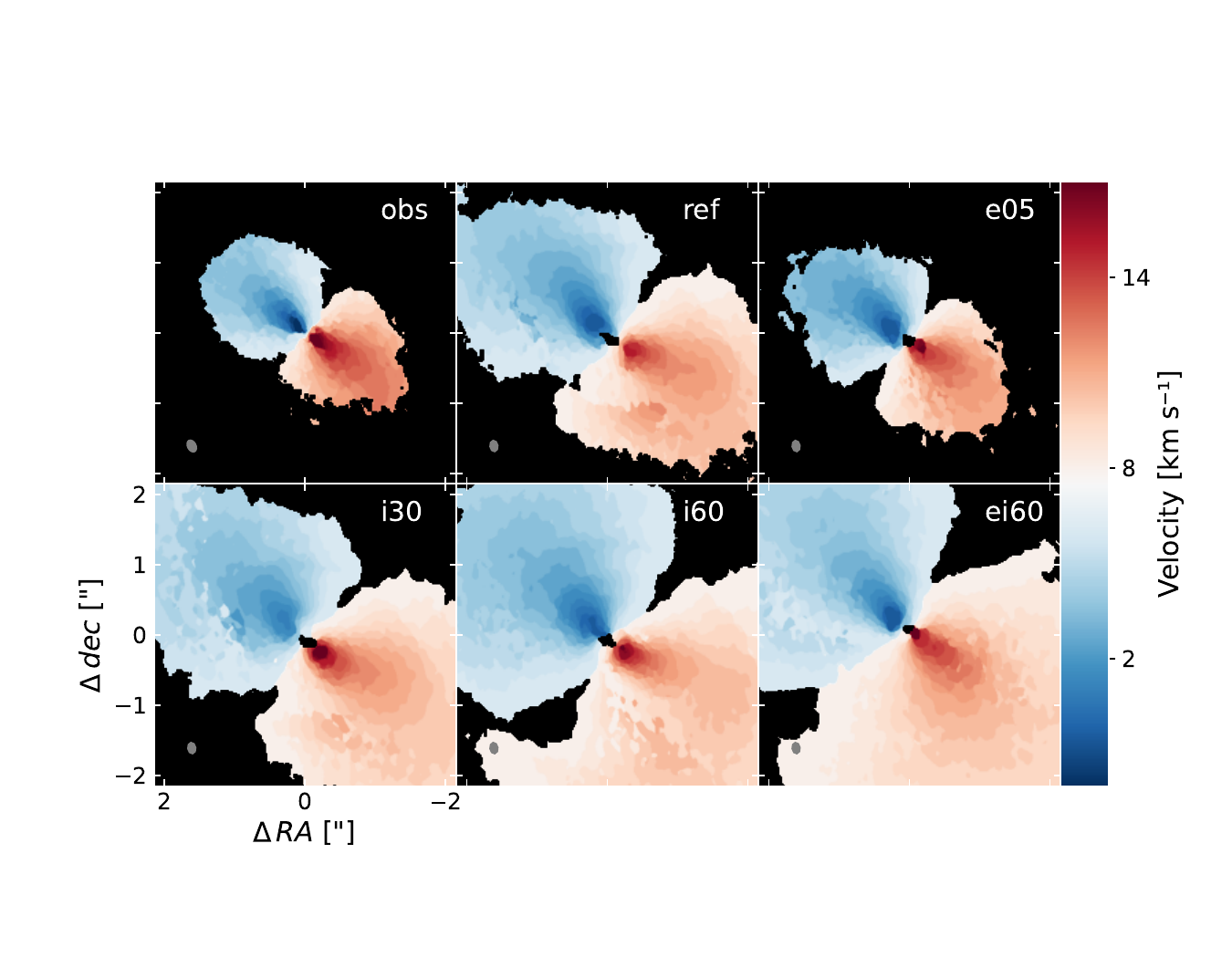}
    \caption{Same as Figure \ref{fig:simgrid_mom0_synth} \lang{but} for the $^{12}$CO (2-1) emission velocity map.}
    \label{fig:simgrid_mom9_synth}
\end{center}
\end{figure}

Figures \ref{fig:simgrid_mom0_synth}, \ref{fig:simgrid_mom8_synth} and \ref{fig:simgrid_mom9_synth} show the moment 0, moment 8 (peak temperature) and moment 9 (velocity of the brightest pixel along the spectral axis) maps respectively of the post-processed SPH simulation outputs as described in the Section \ref{subsec:rt_proc}. 
The synthetic moment 0 maps have generally a higher flux than the data by \lang{approximately} a factor $3$ but have similar levels of peak intensity except for \textit{e05} of which the flux is comparable. This is mainly due to the more extended emission in the simulations than in the observations. The CBDs of \textit{ref} and \textit{e05} exhibit a morphology comparable to the observations on the moment 0 map. The moment 0 maps of \textit{i30} and \textit{i60} show visible spirals in the outer disc and material forming streamers to the companion. Since such extended structures are not recovered in the data, it indicates that the companion triggers mild perturbations in the disc or an unbound scenario. The irregularities in the outer disc (see Fig \ref{fig:sketch_obs}) are recovered on the synthetic images of every simulation and trace faint material spread out by the companion. 
The synthetic moment 8 maps are generally brighter than the data (by a factor $3$ in brightness flux and by $50\%$ in peak value), but the average temperature in the outer disc is well reproduced. The NW side of the disc is seen warmer (by $10\%$ in average) and the warm twisted pattern in the redshifted side of the disc is recovered in all the simulations. The synthetic maps are more structured than the data: \textit{ref}, \textit{i30}, \textit{i60}, and \textit{ei60} present visible perturbations \lang{towards} the companion that correspond to streamers linking the disc and the companion. \lang{The \textit{e05} simulation} also shows subtle perturbations at the same locations, which better matches the data. 
As seen in \textit{e05}, material could stand beyond the outer disc and not being bright enough to be detected (see \mdy{bottom second snapshot from the right} on Fig \ref{fig:simgrid_rho} in comparison to the top right image on Fig \ref{fig:simgrid_mom0_synth}). In the same fashion, it is not excluded that the V892~Tau disc exhibits non-detected large scale structures which could be probed with deeper observations. Moreover the irregularities in the outer disc are similar between the data and \textit{e05} (see Fig \ref{fig:simgrid_mom0_synth} and Fig \ref{fig:simgrid_mom8_synth}) and could trace similar structures.

\rmvd{\subsection{Inclination sign of the disc}}
\label{subsec:incl_sign}

\subsection{Disc extent}
\label{subsubsec:disc extent}

\begin{figure}
\centering
\begin{center}
    \includegraphics[width=\columnwidth, trim={0cm 0cm 0.5cm 0.2cm},clip]{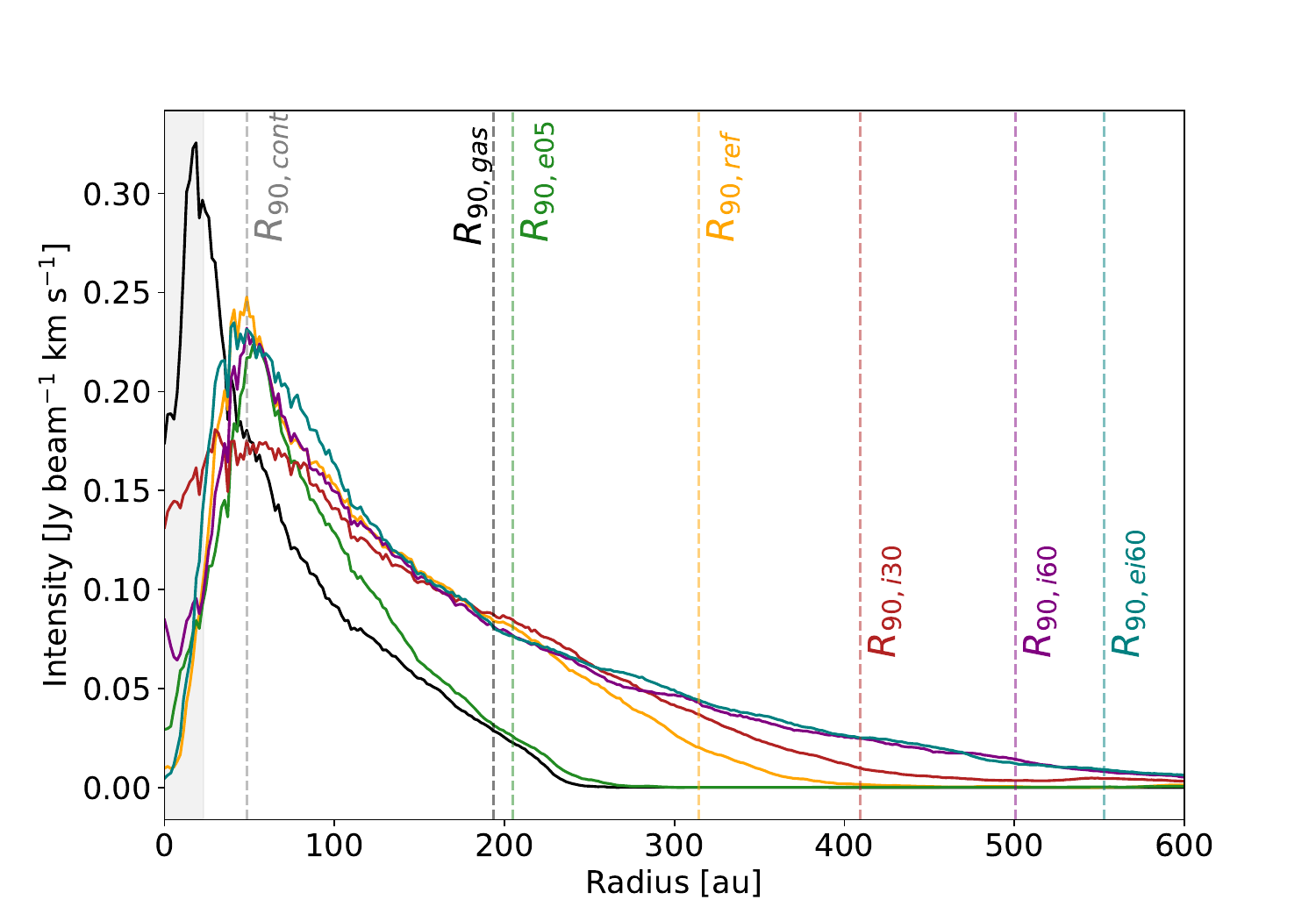}
    \caption{Azimuthally averaged radial profile of the $^{12}$CO intensity for the observations and the simulations. The \lang{vertical dashed }
    lines indicate the $R_{90\%}$ radius for the continuum, the gas emission, and the synthetic observations. The shaded \lang{grey} area \lang{corresponds} to the major axis of the synthesised beam.}
    \label{fig:radial_extent}
\end{center}
\end{figure}

The disc radial extent is measured in the $^{12}$CO observational data up to  $R_{90\%} = 1.45 \arcsec$ \lang{that} corresponds to $194$ au with the adopted distance of $d=134.5$ pc. Figure \ref{fig:radial_extent} shows the intensity profiles of the observations and of the simulated discs and the resulting $R_{90\%}$ for these discs. The intensity profile drops inside the dust ring but the imaging of a potential inner cavity is made difficult by the absorption around the systemic velocity. The disc size and the radial intensity profile measured in our observations are comparable to those in previous works (see Table \ref{table:comb_obs} and \lon).

In our hydrodynamical simulations, the inner binary quickly carves a cavity in the inner gas disc. This cavity is $4-5 a_{in}$ large , where $a_{in}$ is the semi-major axis of the inner binary. Given the eccentricity and mass ratio of the V892~Tau inner binary, the inner rim of the cavity should lie at $\sim 3.5 a_{in}$ in the case of a coplanar disc \citep{Hirsh+2020} which is in rough agreement with our simulations. The discrepancy could be explained by the large accretion radii used for the sink particles of the inner binary components. Due to the long duration of our simulations and the high computational cost of small accretion radii, modelling the inner disc precisely can not be achieved precisely with our numerical setup and we rather focus on the truncation of the disc by the external companion star.

Tidal truncation by the close outer companion swiftly takes place early in the simulations and material from the disc is spread out in the outer system. After $50$ simulated orbits of the outer companion, the spread material has been cleared in \textit{ref}, \textit{e05}, and \textit{i30}. In \textit{i60} and \textit{ei60}, the size of the bulk of the disc is set after a few orbits of the companion even if particles are still ejected from the disc at the end of the simulation. If the companion follows an eccentric orbit, its periastron is closer to the disc and the truncation is stronger, resulting in a more compact disc. When the eccentricity is lower, the disc is allowed to spread radially. In the simulations \textit{ref}, \textit{i30}, and \textit{i60}, the disc is much larger due to the truncation not being efficient enough or due to material being spread out by the companion. In the \textit{ei60} simulation, the eccentricity of the companion is higher but as is the inclination which result in a larger closest distance approach. Added to the material populating the outer system and contributing significantly to the disc emission, the estimations of the disc radius are higher in \textit{ei60}. In the end, the observations are in better agreement with the \textit{e05} simulation as both show comparable disc sizes.

\subsection{Spiral arms and kinematics}
\label{subsec:spirals}

The launch of spiral arms in a disc by an external companion is a well established result \citep{Rafikov2002} and has been observed in various systems already (e.g. HD100453 \citep{Benisty+2017}, UX Tau \citep{Menard+2020}). A binary star at the heart of a CBD can produce spiral features as well \citep{Poblete+2019}, that are most visible at the cavity edge that propagate \lang{towards} the outer disc (e.g. AB~Aur \citep{Poblete+2020}, HD~142527 \citep{Price+2018}) whereas an external companion triggers two wide open spiral arms at the Lindblad resonances, where one arm is pointing \lang{towards} the companion and the other lays in the symmetrical side of the disc. For an orbiting companion, the scale of the spirals is reduced as the disc is truncated with time (see Fig \ref{fig:simgrid_rho} and simulations in \cite{Menard+2020}).

The velocity residual map on Fig \ref{fig:obs_model_panel} shows the disc as dominated by the Keplerian rotation. \mdy{The central regions of the system exhibit pronounced residuals, which may stem from cloud absorption within the corresponding channel emission. The residual patterns suggest that two opposite spiral arms are traversing the disc. One arm delineated by red residuals extends from the inner blueshifted region to the outer redshifted side, passing through the northern section of the disc. Symmetrically the other arm is marked by blue residuals in the southern regions of the disc. We highlight these spiral patterns on the sketch shown in Figure \ref{fig:sketch_obs}. Spirals have been tentatively identified in the outer redshifted part of the disc (\lon)\lang{:} a region where the disc is irregular and where brightness residuals are high. Our observations support  the presence of a spiral arm at that location and unveil its potential symmetrical component.} 
In the light of our simulations that we describe below, we present this given geometry of trailing spirals in the disc as a tentative evidence of CBD-companion interactions. \mdy{The presence of these interactions is also supported by the rotation curve derived from the gas emission model, and that we detail in Appendix \ref{subapp : rotation curve}.}
Due to the scale and prominence of the spirals, the best match is done with the \textit{e05} setup. 
Non-Keplerian features in the $9-11.5$ km\,s$^{-1}$ channel maps could also trace a large scale spiral structure and the truncated lower wing of the channel map emission mentioned in Section \ref{subsec:channels_result_part} could be consistent with a spiral perturbation at that location.

In every simulations performed, the outer companion triggers two large spiral arms in the disc when passing by the periastron. The spirals are sustained throughout the simulation runs and are visible in the column density rendered plots of the SPH setups of Figure \ref{fig:simgrid_rho}. 
After the post-processing done, hints of spiral structures remain visible in the synthetic moment maps. In \textit{ref}, \textit{i30}, \textit{i60} and \textit{ei60} the two spiral arms are traced by the gas emission on the moment 0 map where patterns are mainly visible to the 
\lang{south} and \lang{north-west} of the CBD. The peak velocity maps also show bright emission at the same locations in these simulations.  However no clear spiral structure is seen in the moment 0 map of the \textit{e05} simulation and no clear deviation is observed in the corresponding peak velocity map of that simulation. In the small disc of the \textit{e05} simulation, the spirals disperse more rapidly and are of lower strength making their detection difficult when the companion comes close to apoastron. It suggests that faint spiral arms could be present in the V892~Tau disc.

Gravitational instability and embedded planets are also sources of spiral patterns in discs. However the triggering of the gravitational instability requires a disc massive enough for its self-gravity to dominate the gas pressure and rotation \citep{GoldreichLynden1965}. This is expected when $M_{disc} > 0.1 M_*$ \citep{KratterLodato2016}, which is not the case of the V892~Tau CBD where $M_{disc} \sim 0.01 M_*$.
Embedded planets interactions with their host disc can result in the formation of two spiral arms with one of them pointing to the planet (e.g. \cite{Dong+2015}). It has also been showed that internal massive planets could reproduce spiral features in discs \citep{Calcino+2020}. In our N-body simulations, planets in the inner system are ejected of the system by the inner binary (see Section \ref{subsec:V892TauNE}) making the scenario of a planet inside the dust cavity unlikely. Planets could survive at larger separations and trigger the observed spiral patterns, but our data does not allow us to comment on that possibility.

\subsection{Inclination and orientation of the disc}
\label{subsec:tilttwist}

The tilt $\Delta$ is defined as the angle between the disc and the inner binary orbital plane. From our fitting process, we \lang{recovered} a disc orientation (i, PA) very similar to what was derived in previous works (\lon). Thus, we \lang{assumed} the disc to have the same tilt as previously calculated. Due to an ambiguity of $180\degree$ on the inner binary position angle, the two possible tilt values are $\Delta = 8.0\pm 4.2\degree$ and $\Delta = 113.2\pm 3.0\degree$. We believe the disc to be close to coplanarity and rotating prograde with the companion star. In this way, we \lang{assumed} a tilt of $\Delta \sim 8.0\degree$ to be more likely. We further motivate this assumption in Section \ref{subsec:final_incl}.
We measure the tilt $\Delta$ and the position angle (PA or $\Omega$) of the simulated discs by averaging the tilt and PA of all the SPH particles tilt and PA having a semi-major axis between the inner border of the disc and $200$ au, which corresponds to an upper limit of the disc extent.

\begin{figure*}
\centering
\begin{center}
    \includegraphics[width=0.3\textwidth, trim={0cm 0cm 0cm 0cm},clip]{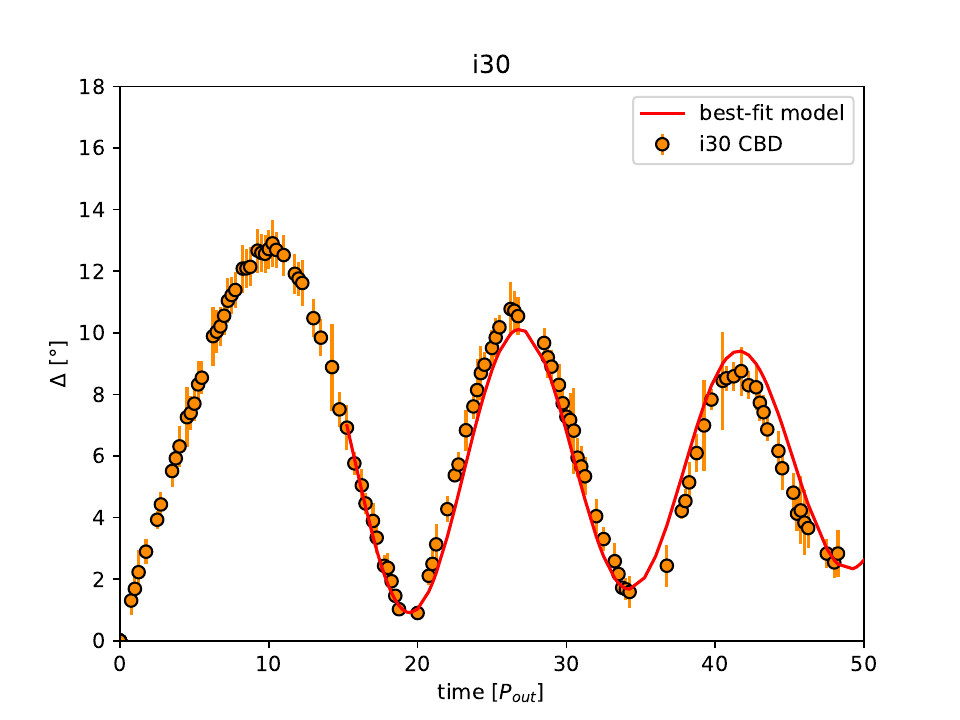}
    \includegraphics[width=0.3\textwidth, trim={0cm 0cm 0cm 0cm},clip]{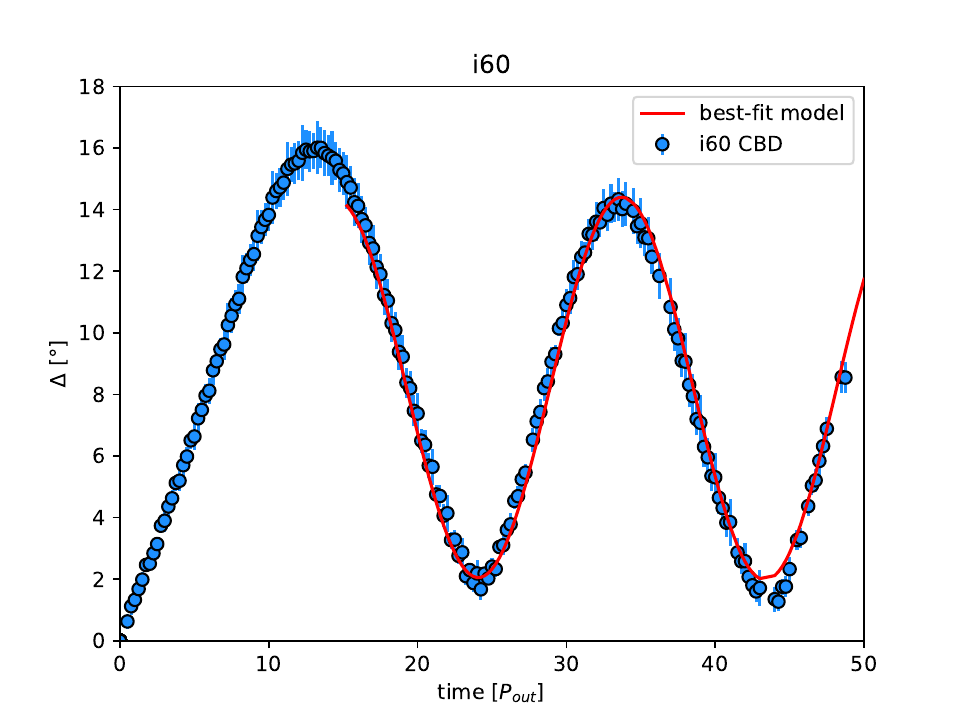}
    \includegraphics[width=0.3\textwidth, trim={0cm 0cm 0cm 0cm},clip]{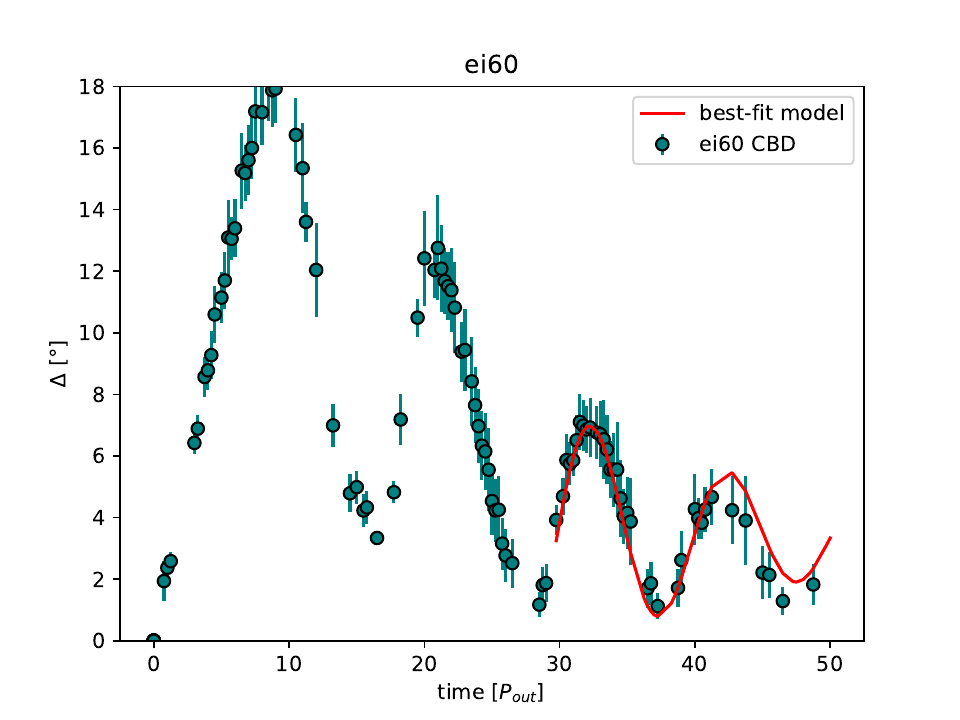}
    \includegraphics[width=0.3\textwidth, trim={0cm 0cm 0cm 0cm},clip]{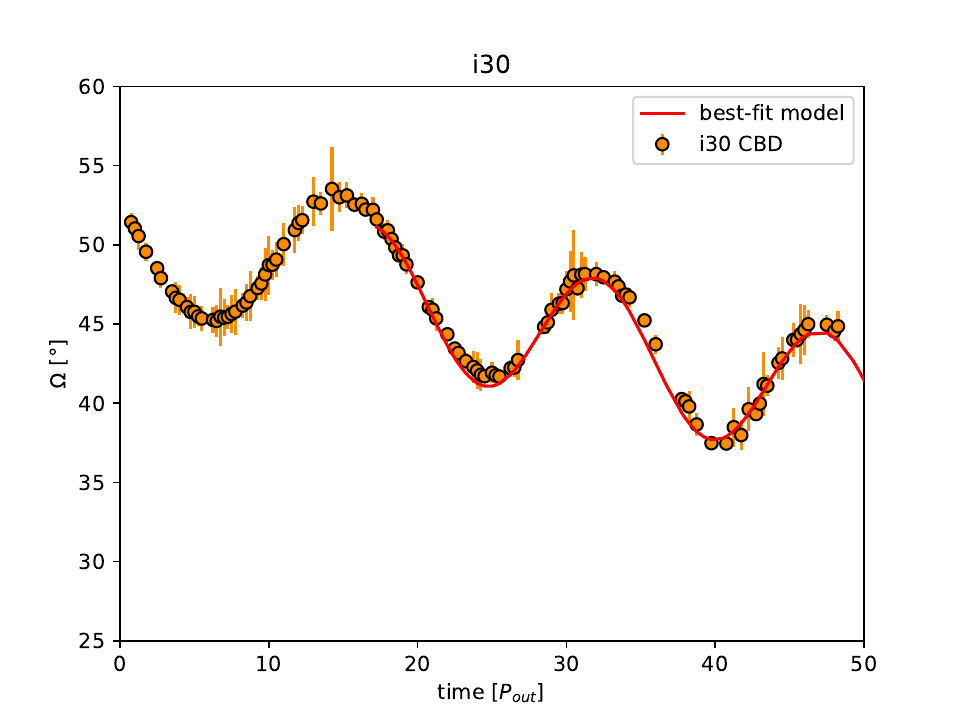}
    \includegraphics[width=0.3\textwidth, trim={0cm 0cm 0cm 0cm},clip]{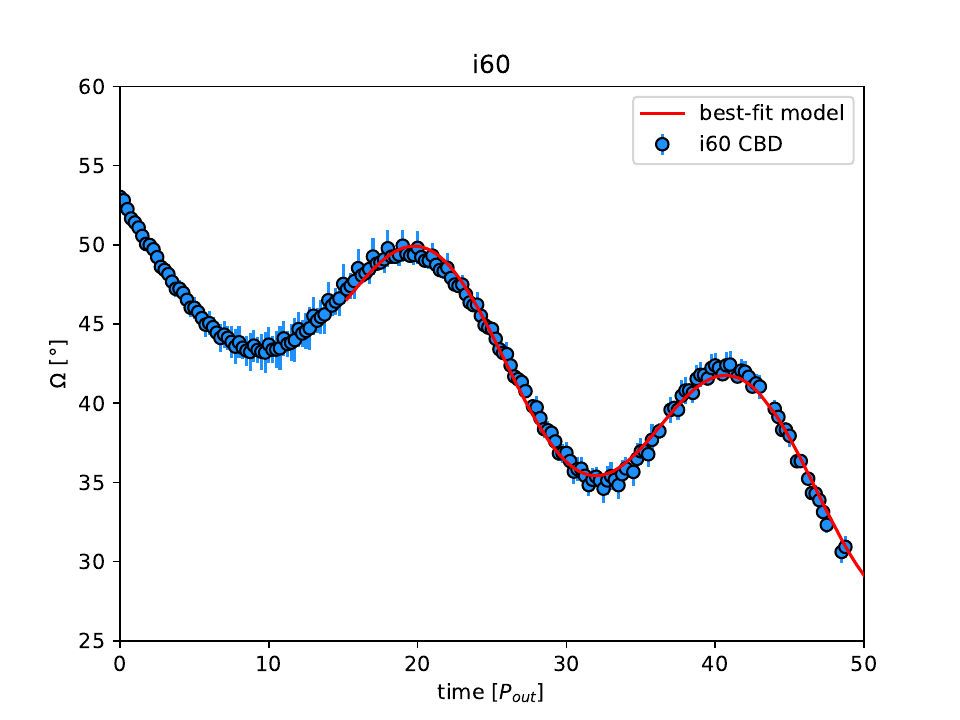}
    \includegraphics[width=0.3\textwidth, trim={0cm 0cm 0cm 0cm},clip]{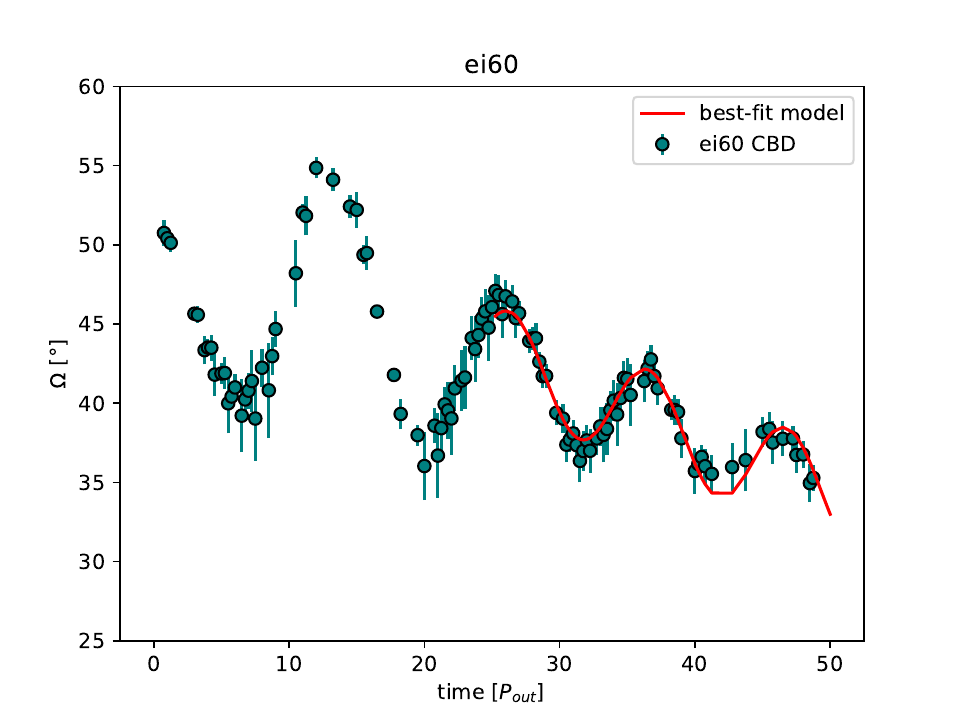}
    \caption{Tilt (top) and PA (bottom) evolution of the disc for the \textit{i30} (left), \textit{i60} (middle), and \textit{ei60} (right) setups. The data \lang{are} represented by the coloured points while the fit is shown by the red curve. The tilt evolution \lang{was} modelled by damped oscillations. The PA evolution \lang{was} modelled by oscillations on the top of a linear precession. The best-fit parameters can be found in Table \ref{table:fit params}.} 
    \label{fig:tilttwist}
\end{center}
\end{figure*}

On \lang{the} one hand, in the setups where the companion is on a coplanar orbit with the disc and the inner binary (\textit{ref} and \textit{e05}) the inclination and the PA of the disc remained the same as the initialised values during the entire simulation. The tilt $\Delta$ and PA are then of $0\degree$ and $53\degree$ respectively at the end of the simulation for these setups. 
On the other hand, in the setups where the companion is on an inclined orbit with respect to the disc and the inner binary, namely \textit{i30}, \textit{i60}, and \textit{ei60}, both the inclination and the PA of the disc oscillate periodically with time, as seen on Figure \ref{fig:tilttwist}. The tilt oscillations damp with time and tend to a non-zero value with an amplitude of the oscillations starting at about $14\degree$, $17\degree$ and $18\degree$ for \textit{i30}, \textit{i60}, and \textit{ei60} respectively. The PA of the disc decreases in average while oscillating with an amplitude of the oscillations appearing to be maintained. 

For the inclined setups \textit{i30}, \textit{i60,} and \textit{ei60,} we \lang{fitted} the inclination of the disc with a damped oscillations model (see Equation \ref{eq:tiltfit}). The PA is fitted by a linear model on the top of which are sinusoidal oscillations (see Equation \ref{eq:twistfit}). The model is adjusted to the simulation data using a non-linear least squares method. The disc relaxes from the initial conditions during the first orbits of the outer binary. We exclude at least the first $15$ orbits of the outer companion of the fitting process to neglect that effect. The best-fit results can be found in Table \ref{table:fit params}.
\begin{align}
    \label{eq:tiltfit}
    \Delta(t) = a_1 + a_2 e^{-a_3 t} \cos(2\pi a_4 t + a_5) , \\
    \label{eq:twistfit}
    \Omega(t) = b_1 + b_2 t + b_3 \cos(2\pi b_4 t + b_5) ,
\end{align}

The best-fit asymptotic tilt value is around $6\degree$ and $8\degree$ for \textit{i30} and \textit{i60} respectively. We notice that the \textit{i60} value is in excellent agreement with the value derived in \lon. Yet the \textit{i30} value is consistent as well within the errorbars. In the case of \textit{ei60}, the model indicates an asymptotic tilt value of $3.5\degree$.  
The $b_2$ parameter of the PA model can be converted into a precession timescale of $1619$ P$_{out}$, $923$ P$_{out}$ and $1003$ P$_{out}$ for \textit{i30}, \textit{i60}, and \textit{ei60}, respectively. The uncertainty of the fit for these values is around $100$ P$_{out}$. 
The $a_3$ parameter of the tilt model can be converted into a damping time \lang{that} is $85$ P$_{out}$, $226$ P$_{out}$ and $19$ P$_{out}$, which corresponds to approximately $0.3$ Myr, $0.9$ Myr and $0.08$ Myr for \textit{i30}, \textit{i60}, and \textit{ei60}, respectively. The incertitude on these values is $<20$ P$_{out}$. 
The $a_4$ and $b_4$ parameters can be converted into an oscillation period of the disc \lang{that} is fitted on the tilt profile to $14.8$ P$_{out}$ for \textit{i30}, $19.3$ P$_{out}$ for \textit{i60} and  $10.1$ P$_{out}$ for \textit{ei60}. We \lang{applied} the same process to the PA oscillations to find a period of $15.2$ P$_{out}$ for \textit{i30}, $20.9$ P$_{out}$ for \textit{i60} and $10.2$ P$_{out}$ for \textit{ei60}. The uncertainty of the fit for these values is of $<0.2 $ P$_{out}$. The oscillation period is shorter in \textit{ei60} than in \textit{i30} and \textit{i60}. We note that the oscillation period of the tilt and the PA are similar for a given simulation. We also note that our model does not account well for the \textit{ei60} disc dynamics where the oscillations are quickly damped and the disc ends up close to coplanarity with the inner binary.  We discuss in more detail those dynamical behaviours in Sections \ref{subsec:final_incl} and \ref{subsec:precession}.

The previous results indicate that an inclined companion with respect to the inner binary is able to misalign a disc initially in the inner binary plane. It proves that a misaligned geometry between the inner binary and the outer companion is needed to explain the observed tilt value of $\Delta_{obs}=8\degree$ derived in \lon, which is best reproduced by an inclination of $60\degree$ of the companion and a low eccentricity of it (\textit{i60}).
Considering the oscillations of the disc tilt, the disc could be caught in a middle of an oscillation and be measured misaligned with the inner binary. However when comparing the damping time of these oscillations to the estimated age of the system of $\sim 2$ Myr \citep{KucukAkkaya2010}, it is unlikely that the disc is still undergoing significant oscillations.

\begin{table}
    \centering
   \caption{Model functions for the tilt and the PA evolution and best-fit results for \textit{i30}, \textit{i60}\lang{,} and \textit{ei60}}
 \begin{tabular}{c | c c c} 

 & \multicolumn{3}{c}{Tilt}\\ [0.5ex] 
 & \multicolumn{3}{c}{$\Delta(t) = a_1 + a_2 e^{-a_3 t} \cos(2\pi a_4 t + a_5)$}\\ [0.5ex] 
 \hline
Setup & \textit{i30} & \textit{i60} & \textit{ei60}\\
 \hline 
$a_1$ ($\degree$) & $5.71\pm 0.10$ & $8.21\pm 0.05$ & $3.47\pm 0.11$\\

$a_2$ ($\degree$) & $6.03\pm 0.62$ & $-6.06\pm 0.22$ & $-19.59\pm 9.00$\\

$a_3$ ($10^{-2}P_{out}^{-1}$) & $1.10 \pm 0.31$ & $-0.44\pm 0.11$ & $5.32\pm 1.27$\\

$a_4$ ($10^{-2}P_{out}^{-1}$) & $6.77\pm 0.05 \;$ & $5.19\pm 0.02$ & $9.84\pm 0.15$\\

$a_5$ & $-5.18\pm 0.10$ & $4.71\pm 0.03$ & $1.96\pm 0.33$\\
 
 \multicolumn{4}{c}{ }\\

 & \multicolumn{3}{c}{Position Angle} \\ [0.5ex] 
 & \multicolumn{3}{c}{$\Omega(t) = a_1 + a_2 t + a_3 \cos(2\pi a_4 t + a_5)$} \\ [0.5ex] 
 \hline 
Setup & \textit{i30} & \textit{i60} & \textit{ei60}\\
\hline
$b_1$ ($\degree$) & $50.77\pm 0.53$ & $52.80\pm 0.20$ & $52.12\pm 0.57$\\

$b_2$ ($\degree.P_{out}^{-1}$) & $-0.22\pm 0.02$ & $-0.39\pm 0.01$ & $-0.36\pm 0.02$\\

$b_3$ ($\degree$) & $-4.24\pm 0.19$ & $5.04\pm 0.07$ & $3.10\pm 0.15$\\

$b_4$ ($10^{-2}P_{out}^{-1}$) & $6.57\pm 0.08$ & $4.78\pm 0.02$ & $9.77\pm 0.10$\\

$b_5$ & $-3.82\pm 0.18$ & $8.53\pm 0.04$ & $-9.89\pm 0.27$\\

\end{tabular}

 \begin{tabular}{c | c c c} 

\end{tabular}
\label{table:fit params}
\end{table}

\subsection{Disc warp}
\label{subsubsec:disc warp}

A warp is a non-constant inclination radial profile appearing preferentially in thinner, less viscous and larger discs \citep{PapaloizouPringle1983, Young+2023}. Disc warping is known to occur in CBDs around eccentric and inclined binaries (e.g. \cite{Nixon+2013, LodatoFacchini2013}) and in circumstellar disc perturbed by an external misaligned companion \citep{Gonzalez+2020}. In non-coplanar triple systems, these two effects occurs simultaneously. Given the previous results, a misaligned orbit of V892~Tau~NE could explain the observed disc tilt (see Section \ref{subsec:tilttwist}). If such an orbit is able to misalign the disc, it could be able in principle to trigger a warp in the disc too. In the following we investigate the presence of a warp in the CBD of V892~Tau.

In order to search for the presence of a warp in our simulated discs, the disc \lang{was} binned in semi-major axis from the inner cavity edge $r_{in}$ to $r = 200$ au and the tilt $\Delta$ \lang{was} measured in each bin taking the standard deviation of the tilt of the particles in a bin as the corresponding error. In the SPH data, the cavity semi-major axis $a_{cav}$ \lang{was} defined as $\rho(a_{cav}) = \rho_{max}/2$\mdy{ with $\rho_{max}$ defined as the maximum of the radial density profile} \citep{Artymowicz1994}. Since the disc eccentricity stays below $0.1$ during the simulation, the cavity semi-major axis can be identified as the cavity radius. At $t=50$ P$_{out}$, $a_{cav}=45 $ au for \textit{i30} and $a_{cav}=44 $ au for \textit{i60}.

\begin{figure*}
\centering
\begin{center}
    \includegraphics[width=1.0\textwidth]{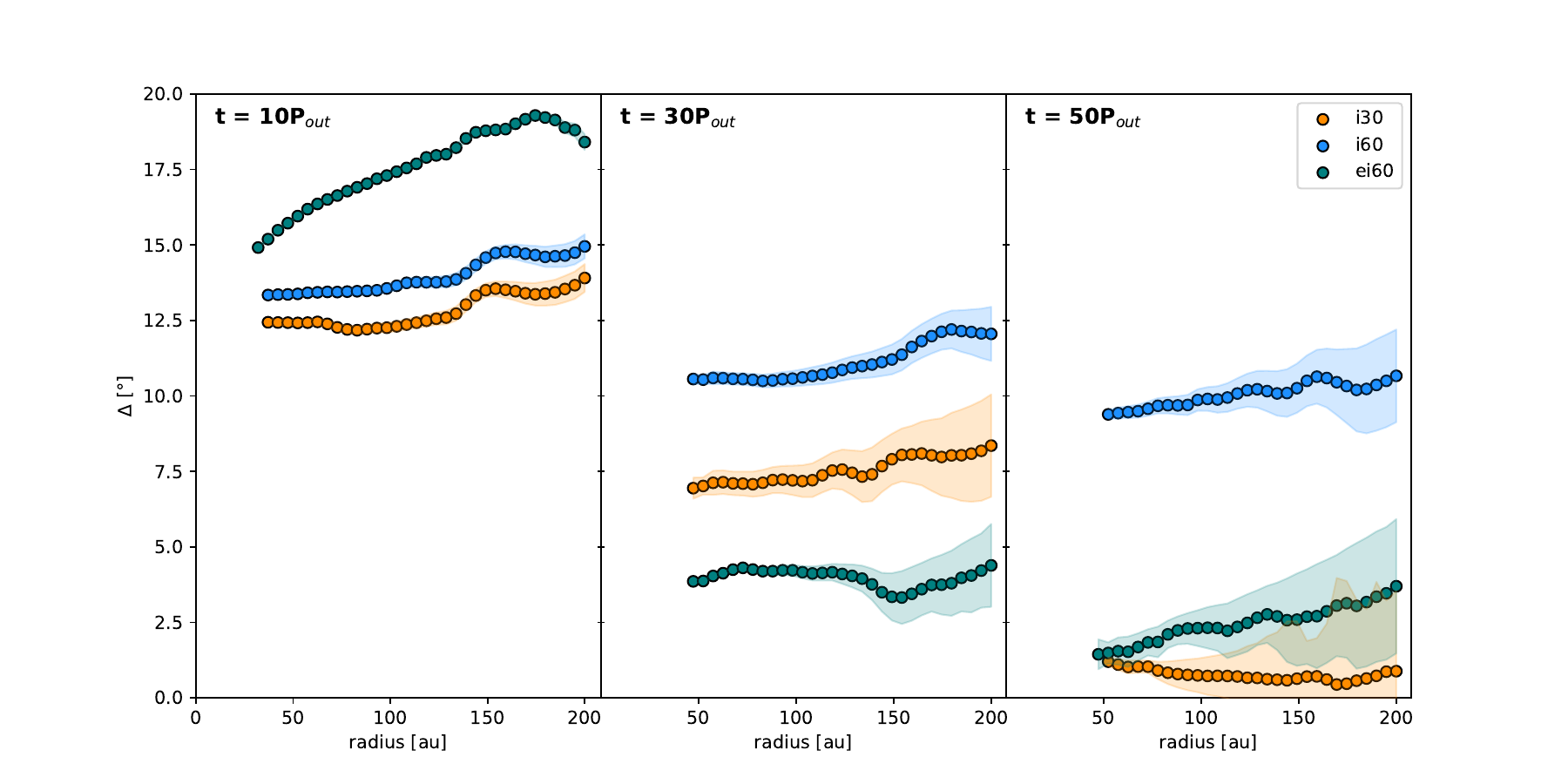}   
    \caption{Tilt radial profiles of various simulations at $t=10$ P$_{out}$ (left), $t=30$ P$_{out}$ (middle), and $t=50$ P$_{out}$ (right). The orange, blue, and teal dots correspond to \textit{i30}, \textit{i60}, and \textit{ei60}, respectively. The $1\sigma$ errors are indicated by the shaded areas.}
    \label{fig:tilt_rad}
\end{center}
\end{figure*}

We inspect and plot the radial distribution of the tilt $\Delta (r)$ of the setups \textit{i30}, \textit{i60} and \textit{ei60} at different time steps on Figure \ref{fig:tilt_rad}: $t=10$ P$_{out}$, $t=30$ P$_{out}$ and $t=50$ P$_{out}$.
The inner binary, the outer binary and the discs stayed in a coplanar configuration for the initially coplanar setups \textit{ref} and \textit{e05}.
In the setups where the companion is on an inclined orbit (namely \textit{i30}, \textit{i60}, and \textit{ei60}) a warp forms steadily during the first orbits. The tilt profile $\Delta(r)$ shows a continuously but modestly increasing trend: the inner parts of the disc are misaligned by $1\degree$ to $4\degree$  with respect to the outermost parts after a few orbits of the companion. That tilt profile is maintained during the early moments of the simulation \rmvd{for both} in all the simulations with an inclined companion. However the disc becomes planar again within the \lang{errorbars} after a time $t=30$ P$_{out}$. 
The companion bends the outer parts of the CBD at the beginning of the simulation, which creates a warp propagating in the bending wave regime \citep{PapaloizouTerquem1995} to the innermost parts of the disc. Because of the loss of material in the disc and its truncation by the companion, the disc behaves more and more rigidly with time. In this way, the strongest warping occurs at the beginning of the simulation. After a few orbits of the companion once the disc truncated enough, the tidal torque of the companion is insufficient to warp the disc: the warp is damped and the disc becomes planar again \cite{Deng+2021}.
At the end of the simulations, the disc can be considered planar within the \lang{errorbars} and the amplitude of a potential warp in the disc would be only that of a few degrees. Measuring such a low warp our synthetic data is challenging and require a more in-depth analysis that we \mdy{left for} future studies. In the meantime, we favour a planar configuration for the discs in our simulations.

A warp in the CBD has been invoked to explain deviations from simple models \citep{Long+2021, Vides+2023}. Residuals from our Keplerian model are similar to these deviations: the arcs at the disc edges on the moment 0 and peak temperature residual maps (see Figure \ref{fig:obs_model_panel} and Section \ref{subsec:disc_morph}) could be hints of a misalignment between the inner disc and the outer disc. 
According to the orbital parameters of the inner binary, the disc could be warped under the tidal torque of the binary if sufficiently misaligned (see Section \ref{subsec:final_incl}). V892~Tau~NE \lang{could also warp the disc}, if on a misaligned orbit with respect to the disc. 
We found the CBD of V892~Tau to be a thin disc with an emission surface aspect ratio of z/r$\sim$0.16 at 100 au, where most discs around young stars are found with  z/r>0.3 \mdy{in the $^{12}$CO emission \citep{Law+2021}}. The disc of V892~Tau is also moderately extended. Thus we expect warping to be difficult in the disc. 
The high temperature in the disc up to $\sim 110$ K makes it prone to damp any potential warp (e.g. \cite{Rabago+2023}). 
In the end, even if the observed residuals could be consistent with a warped disc, the combined \lang{dataset} used in this work can not robustly rule out nor confirm the presence of a warp in the disc.

\subsection{The orbit of V892~Tau~NE}
\label{subsec:V892TauNE}

\mdy{\subsubsection{The flyby scenario}}
\label{subsubsec:flyby}

\mdy{The orbit of V892~Tau~NE is currently unconstrained. This means that V892~Tau~NE could in principle be totally unbound to V892~Tau or following a flyby parabolic/hyperbolic orbit. In the unbound case, V892~Tau would be considered as a binary system. In such a configuration, a coplanar disc with respect to the inner binary and does not precess (see Appendix \ref{app : binary sims} for details). It means that the inner binary alone cannot explain the observed disc tilt, which allows us to rule out the unbound scenario. Moreover it has been recently shown that V892~Tau and V892~Tau~NE are likely related \citep{Thomas+2023}. In the case of a flyby event, efficient truncation and the formation of prominent short-lived spirals are expected \citep{Cuello+2019, Cuello+2020, Cuello+2023, Smallwood+2023}. We can study the flyby scenario in our simulations by considering the state of the disc after the first orbit of the companion. Figure \ref{fig:simgrid_rho} shows column density views of the simulated discs after the completion of the first companion orbit. In all the simulations, dense spiral arms are triggered in the disc by the passage of the companion. The disc remains close to its original size, \lang{that} is similar to the observed one, in all the simulations despite material being ejected. If the companion was to follow an inclined orbit with respect to the inner binary, Fig.~\ref{fig:tilttwist} shows that the disc would be misaligned by $1-2\deg$ at the end of the first orbit considered here as the end of the flyby event. In our simulations, the companion approaches the disc closely with a closest distance no greater than 200 au. A more distant flyby would have produced fainter spirals and lower misalignments \citep{Cuello+2019}. In any case, the subtle observed spiral patterns and the disc tilt of $\sim 8\deg$ suggest that a flyby event is unlikely to have happened in V892~Tau.}



\mdy{\subsubsection{Constraints on a bound orbit}}
\label{subsubsec:bound_orbit}

\begin{table}
    \centering
    \caption{Summary of the observed disc features and to what extent they are reproduced by the different simulations.}
 \begin{tabular}{c c c c c c}
 \hline
 Criterion & \multicolumn{5}{c}{Simulation} \\
 & \textit{ref} & \textit{e05} & \textit{i30} & \textit{i60} & \textit{ei60}\\ [0.5ex] 
 \hline
Channel maps & $\times$  & \checkmark & $\times$ & $\times$ & $\sim$ \\
Disc extent & $\times$ & \checkmark & $\times$ & $\times$ & $\times$ \\
Disc tilt & $\times$ & $\times$ &  \checkmark &  \checkmark & $\sim$\\
Spirals & $\sim$ & \checkmark & $\sim$ & $\sim$ & $\sim$ \\
\end{tabular}
\label{table:best_sim}
\end{table}

Table \ref{table:best_sim} summarises the constraints placed on the V892~Tau~NE orbit through our modelling of the V892~Tau disc. If bound, a medium eccentricity of the orbit of V892~Tau~NE is needed to explain the observed disc extent while the orbit has also to be misaligned with respect to the inner binary to explain the slight misalignment between the disc and the inner binary. \mdy{In such a configuration, the orbit of V892~Tau~NE cannot be too eccentric ($e>0.5$), as high eccentricities tend to align the disc with the inner binary (Figure \ref{fig:tilttwist}).}

From the points of the parameter space we probed with our simulations, we suggest constraints on the parameter eccentricity and mutual inclination of V892~Tau~NE of $0.2<e<0.5$ and $30\degree < \Delta i < 60\degree$ respectively.
If V892~Tau~NE was to follow an eccentric inclined orbit, we expect the CBD to oscillate and precess in the light of the discs in the setups \textit{i30}, \textit{i60}, and \textit{ei60}. According to our simulations, the CBD inclination oscillates with time and these oscillations are damped over a period of $\sim 1000$ yrs. Then the tilt of the disc reaches a non zero final value of $3\degree$ to  $8\degree$ which is in good agreement with the observed value of $\sim 8\degree$. In such a configuration, we expect the disc $PA$ to oscillate with time on top of a linear precession. We discuss the non-coplanar configuration of the system and the resulting disc dynamics in the Sections \ref{subsec:final_incl} and \ref{subsec:precession} below.

\subsection{Stability of future planets}
\label{subsec:nbody-result}

Considering the results of the previous Section \ref{subsec:V892TauNE}, we consider the dynamics of planets in a system following the \textit{ei60} orbital configuration. Although not fully encompassing all disc features, the parameter set represented by \textit{ei60} delineates the boundary of the parameter space governing an \lang{inclined eccentric} orbit of the companion. We anticipate that this particular configuration will induce the most pronounced perturbations on planets within the system, as alternative configurations would entail lower eccentricity and/or mutual inclination of the companion. Consequently, we present the following results as a worst-case scenario for the prospective planets in the system. The results for the other N-body setups can be found in Appendix \ref{app : nbody}.

\begin{figure}
\centering
\begin{center}
    \includegraphics[width=\columnwidth]{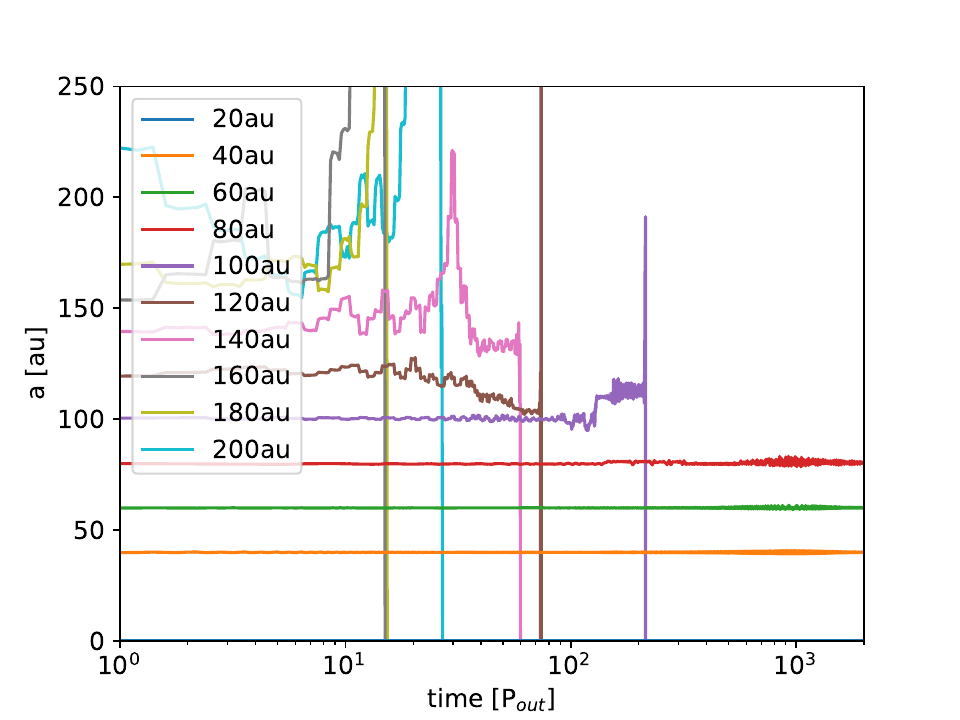}  
        \includegraphics[width=\columnwidth]{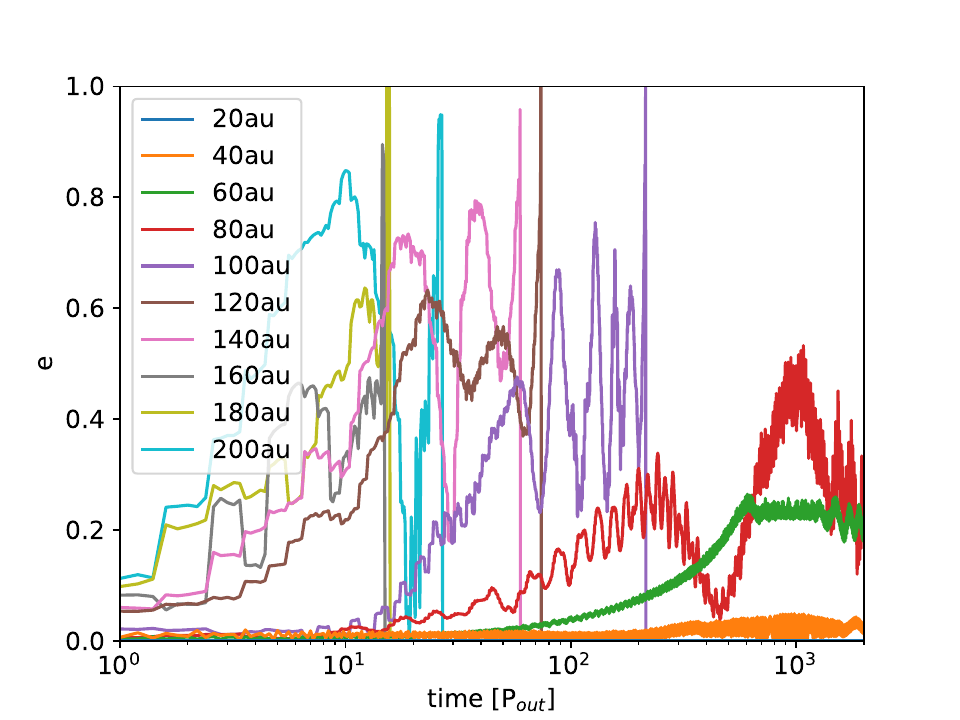}  
        \includegraphics[width=\columnwidth]{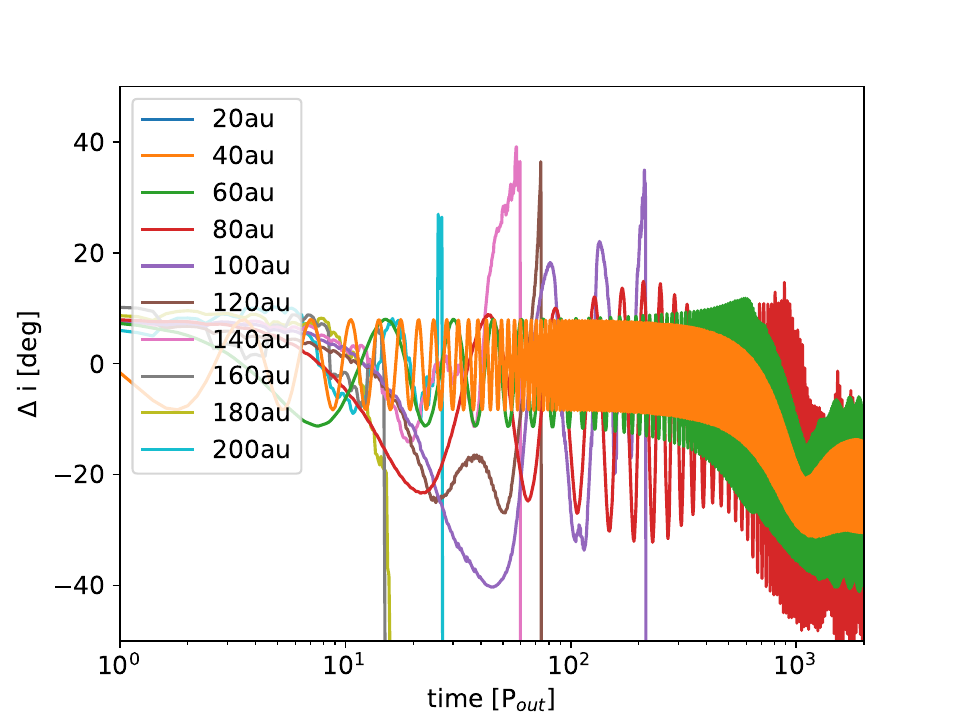}  
    \caption{Evolution of the semi-major axis (top), eccentricity (middle) and mutual inclination with the inner binary (bottom) of the test particles in the \textit{ei60} N-body setup. Each colour represents a different test particle that started at a given semi-major axis. The semi-major axis of a particle falls to $0$ once ejected from the system. The particle starting with $a=20$ au was ejected from the system before the completion of the first orbit of the outer binary.}
    \label{fig:nbody}
\end{center}
\end{figure}

The stability of circumbinary planets in a system in the \textit{ei60} configuration depends on the planets semi-major axis first. Figure \ref{fig:nbody} shows the evolution of the test particles in the N-body simulation of \textit{ei60}, which is described in Section \ref{subsec:nbody}. The test particles with a semi-major axis $a \geq 100$ au were ejected from the system during the simulation. The innermost test particle at $a = 20$ au was also gravitationally ejected from the system \mdy{by the inner binary} during the first orbit of the outer companion. Particles in between $a = 40$ au and $a = 80$ au remained stable during the simulation. \mdy{This can be explained by the competition between two dynamical effects.} \rmvd{Still, it is unclear on which timescales these particles will remain stable and to what extent secular evolution effects would affect their stability.} \mdy{First} due to the presence of the misaligned companion, the test particles undergo von Zeipel-Kozai-Lidov (vZKL hereafter) cycles \citep{vonZeipel1910, Kozai1962, Lidov1962}  with a characteristic timescale of $\sim 90$ P$_{out}$ at $100$ au \citep{Ceppi+2023}. This effect results in a rise and oscillations in eccentricity and inclination of the test particles leading to the ejection of the test particles closer \mdy{to the} external star. \mdy{Second} the inner binary induces a nodal precession of the test particles around its \mdy{angular momentum} vector \mdy{with a timescale of $\sim 98$ P$_{out}$ at $100$ au} (e.g. \cite{FaragoLaskar2010, LodatoFacchini2013}). Closer to the inner binary, the timescale of this precession is lower than the vZKL timescale, stabilising the particles in the inner half of the disc ($a \leq 100$ au) \citep{VerrierEvans2009}. 
\rmvd{At intermediate separations ($ 40 \leq a \leq 100$ au), the two effects become comparable making the fate of the test particles unclear.}
Still, the particles \mdy{found stable at the end of the simulation} \rmvd{in this region of the disc} were undergoing large oscillations in inclination (and eccentricity for the test particle at \mdy{$a=80$ au}) \lang{that} questions their future stability \mdy{on longer timescales}.
Thus \mdy{if} the current gas disc is ensuring the stability of forming planets, \rmvd{but} those planets could become unstable once the gas disc dissipated. On longer timescales, it implies the migration of circumbinary forming planets \rmvd{located in unstable regions of the disc ($<20$ au and $>80$ au)} \lang{towards} \rmvd{more} stable regions at an intermediate distance from the inner binary between $40$ au and $80$ au. We discuss these results in more details in the Section \ref{subsec:planetary_arch}.


\section{Discussion}
\label{sec:discussion}

\subsection{Final inclination of the disc}
\label{subsec:final_incl}

CBDs are expected to reach a stable coplanar or polar configuration depending on the properties of the inner binary and of the initial mutual inclination. If the initial misalignment and the eccentricity of the inner binary are low, the disc is expected to move into a coplanar configuration \citep{Facchini+2013}. On the opposite, for binaries with higher eccentricity and a high mutual inclination the CBD is expected to evolve \lang{towards} a polar configuration \citep{Aly+2015, MartinLubow2017, Cuello&Giupponne2019}.  Nonetheless CBDs have been found with a medium mutual inclination with their inner binary in a configuration that is not coplanar nor polar. For example the CBD around GG Tau A is inclined of about $25\degree$ with respected to the inner binary \citep{Kohler2011, Andrews+2014}. In SR24, the tilt of the CBD is around $35\degree$ with respect to the binary SR24N \citep{Fernandez_Lopez+2017}. These non-zero tilt values could be explained by oscillations around the final configuration of the disc before reaching the equilibrium coplanar configuration. In hierarchical triple systems as V892~Tau, the picture is modified by the evolution of the stellar orbits \lang{that} can occur on shorter timescales with respect to the disc lifetime. In particular in misaligned systems, the inclination of the orbits can change with time meaning that the intermediate tilt of CBDs could be due to the slow reaction of the disc to these variations \citep{Ceppi+2023}. 
This means that the disc of V892~Tau could eventually be seen misaligned with the inner binary even if the orbit of the companion has a different inclination. Such a misalignment would be created over the timescale of the orbit evolution that we assume to be the precession timescale of the eccentricity vector. We find that the precession timescale of the inner (outer) binary is about  $6000 $ Myr ($24 $ Myr) for V892~Tau \citep{Ceppi+2023}. Compared to a typical disc lifetime of $10$ Myr \citep{Ribas2015} and to our $0.2$ Myr-long simulation runs, the orbital configurations tested in \textit{ref} and \textit{e05} setups are not expected to become misaligned with respect to the disc.

The CBD of V892~Tau is found with a misalignment of approximately $8\degree$ or $113\degree$ with the inner binary of eccentricity $0.25$ (\lon, \citealt{Vides+2023}). The two possible solutions derive from a lack of radial velocity information on the inner binary, making its true inclination difficult to constrain.  
If considered as a CBD in a triple system with the \textit{ei60} orbital configuration, the disc of V892~Tau could go to a stable polar configuration \citep{Martin+2022, Ceppi+2023}. The large semi major axis and the low mass of the companion do not hinder the polar alignment. However, polar alignment would require the disc to be initially misaligned by $66\degree$ or more (\lon). In our simulations, the disc was initialised coplanar with the inner binary. Our simulations indicates that a misalignment of $\sim 8\degree$ is created and sustained by the outer companion. The disc oscillates around that equilibrium position. Given the proximity of our best-fit tilt value, we prefer the nearly coplanar configuration for the disc in the end. This non-zero tilt value shows that intermediate configurations are stable for CBDs, independently of secular dynamics of the stars in the system. The V892~Tau disc is in a sweet-spot of the parameter space \lang{that} would allow this kind of configuration. In order to achieve this, the semi-major axis ratio between the inner and outer binaries should be relatively large and the inclination of the companion with respect to the inner binary around a medium value of $\sim 45\degree$. Future studies to characterise CBDs inclination in triple system as a function of the companion inclination would help to better understand this problem.

\subsection{Oscillations and precession}
\label{subsec:precession}

Due to gravitational torques applied by misaligned stars, discs in multiple systems very commonly undergo precession in the sense that their PA is evolving with time. Linear precession can occur in binary systems in CBD and circumstellar discs (e.g. \cite{PapaloizouTerquem1995}). Simulations of binary systems have already observed this phenomenon (e.g. HD~100453 \citep{Gonzalez+2020, Nealon+2020}, GW~Ori \citep{Smallwood+2021}).

We find that the V892~Tau CBD is linearly precessing but that its PA also oscillates with time (see Fig \ref{fig:tilttwist}). Assuming V892~Tau and V892~Tau~NE as a binary system, the expected precession timescale of the disc is about $144 $ P$_{out}$ for \textit{i30} and $251 $ P$_{out}$ for \textit{i60} (following Eq. 6 from \citealt{Gonzalez+2020} derived from \citealt{Bate+2000} and \citealt{Terquem+1998}). This is well below the precession timescales found in Section \ref{subsec:tilttwist}. However these estimations are in line with precession timescales measured in simulations where the system is represented by only a binary. Those simulations are identical to \textit{i30} and \textit{i60} except that the inner binary is replaced by a single star of the same mass (see Appendix \ref{app : binary sims}). As seen on Figure \ref{fig:SPH_comp_binaries}, the evolution of the tilt and PA of the disc in those binary-system simulations is different from the one in the triple-system simulations. The discs in the \textit{i30} and \textit{i60} runs start to linearly precess at the same rate as in the binary-system simulations, but their precession is then slowed down by the inner binary. It indicates that the disc dynamics are set by the triplicity of the system, \lang{that} can not be simplified as a binary system.

In binary systems, a misaligned circumstellar disc is expected to align to the binary orbital plane in about a precession timescale in the presence of hydrodynamical instabilities such as the warp developed at the beginning of our simulations \citep{Bate+2000}. In the binary-system simulations, the disc starts to align on a timescale in line with that prediction. However in \textit{i30}, \textit{i60} (and \textit{ei60}) the disc oscillates and ends up misaligned with the orbital planes. The gravitational torques of both the inner and the outer binary try to align the disc in their respective planes \lang{that} results in an equilibrium position in between those planes. It shows again that when considering the disc dynamics the misaligned hierarchical triple system V892~Tau can not be reduced to a binary system.

In the \textit{i60} and \textit{ei60} simulation setups, vZKL oscillations could be triggered in the CBD \lang{that} would result in the disc eccentricity and inclination oscillating with time \citep{Martin+2014}. As the inclination oscillation pattern are similar in \textit{i30} and \textit{i60}, this strength of this effect is low in \textit{i60}. However knowing that the vZKL timescale decreases with a higher eccentricity of the outer orbit (e.g. \cite{Kiseleva+1998}), vZKL cycles could explain the dynamical behaviour. The vZKL timescale for \textit{ei60} is about $\sim 16$ P$_{out}$ at $100$ au \citep{Ceppi+2023}, which is comparable to the oscillation period in \textit{ei60} (see Figure \ref{fig:tilttwist}). Moreover the CBD of the \textit{ei60} SPH simulation shows clear eccentricity, showing the clear influence of the vZKL mechanism on this disc.

The disc is subject to the combined gravitational torques of the external companion and of the inner binary. While in the case of a binary system a linear precession is expected, in a triple system oscillations are also expected. This peculiar precession can be assessed by measuring the tilt $\Delta$ and the twist $\beta$. These oscillations, projected in the 2D sky plane, are traced by the disc inclination and PA.
Previous studies have used complex variables to examine the dynamical evolution of discs (e.g. for eccentricity modes in the context of planet-disc interactions \citep{Ragusa+2018}). A similar study of the complex variable $\Delta = |\Delta| \exp(j\beta)$, where $j$ is the complex number, could help to clarify the connection between the tilt and twist oscillations and the resulting evolution of misaligned CBDs in triple systems.

\subsection{Resulting planetary architectures}
\label{subsec:planetary_arch}

In its current configuration, the misaligned disc of V892~Tau could result in the formation of misaligned planets with respect to the inner binary. 
Though those planets would have to migrate to the inner half of the disc to survive in the system (see Section \ref{subsec:nbody-result}). It is unclear to what extent this migration would impact the orbital elements of the planet. If the planet was to tilt relatively to the disc plane, this tilt would be damped and the planet would re-align \citep{Burns1976, TanakaWard2004}. In the case of V892~Tau the resulting planets would stay close to the disc plane, inclined with respect to inner binary. The stability of inclined planets in hierarchical triple systems depends on the stellar orbital configuration. In triple systems with low semi-major axis ratio ($<20$) and eccentricity of the outer orbit, the outer bound of the stability zone is limited. If the mutual inclination between the companion star and the inner binary is not zero, the planets stability region extends vertically allowing for high inclination \citep{Busetti+2018}. In that configuration, circular planets that form coplanar with the inner binary end up with their $i$ and $\Omega$ parameters oscillating on timescales several order of magnitude longer than the orbital period \citep{Busetti+2018}. The inclination between the planet and the inner binary also sets the stability regime. Planets are generally found stable close to coplanarity with the inner binary, regardless of the inclination of the outer star. In the \textit{ei60} configuration of V892~Tau, the stability region extends up to $\sim 100 $ au from the inner binary and up to misalignments of the planet of $\sim 30\degree$ \citep{Gianuzzi+2024}. Putting that result together with the results described in Section \ref{subsec:nbody-result}, we expect future planets in the V892~Tau system to form at intermediate distance from the centre of the system between $40$ au to $80$ au with mild inclination with respect to the inner binary.

The dust disc of V892~Tau is one of the most massive class II disc (\lon), potentially presenting an abundant pebble reservoir. The horseshoe asymmetry seen in the continuum could be explained by an over-density created by the inner binary, which could enhance planet formation at that particular location. Misaligned and warped discs around eccentric binaries are also subject to dust concentrations triggered by the difference in precession timescale between dust and gas, which results in dust traps and favours planet formation \citep{Aly+2021, Aly+2023}. More advanced dust+gas hydrodynamical simulations would allow \lang{for} the presence of warp instabilities in the V892~Tau CBD \lang{to be assessed and this would help to understand} planet formation in triple systems in general.

\mdy{\subsection{Circumbinary discs interactions with an external companion}}
\label{subsec:CBDs in triple}

In the system of V892~Tau, the inner binary dominates the mass budget and the low mass companion V892~Tau~NE tidally interacts with the CBD. It induces  spiral patterns within the disc, a truncation of the outer disc, a slight misalignment with the inner binary but no sustained warped geometry. In the following we explore how, given observations of a CBD, similar structures can be explained by ongoing interactions with an hypothetical companion. This companion may either follow a bound orbit around the inner binary or have been involved in a flyby event. In the absence of identifiable candidates in the vicinity of the CBD, \textit{Gaia} astrometry remains a viable method to search for potential past external flyby occurrences.

Repeated or not, the passage of a star close to the CBD would dramatically reduce the size of the CBD by tidal truncation (e.g. \citealt{Artymowicz1994, MirandaLai2015}), providing a potential explanation of small CBDs. 
The efficiency of truncation increases with a higher mass ratio between the inner binary and the outer companion, as well as with greater eccentricity or a smaller semi-major axis of the companion. 
During a close encounter, the companion may also accrete material from the CBD, leading to the formation of a new disc and associated streamers (e.g. VLA 1623-2417 \citealt{Mercimek+2023}, FU Ori-like systems \citealt{Borchert+2022}).

CBD-binary misalignments are common and expected in binary systems and can find their origin in interactions with surrounding stars \citep{Czekala+2019, Elsender+2023}. Indeed an external perturber would exert a gravitational torque on the CBD, if misaligned with respect to each other, aligning the CBD with the orbital plane of the perturber \citep{PapaloizouTerquem1995}. Should the misalignment between the CBD and the inner binary reach a critical angle, the CBD may transition towards a polar configuration relatively to the (eccentric) inner binary \citep{MartinLubow2017, Martin+2022, Ceppi+2023}. The disc can also become warped in the process or break at a given radius depending on the inner binary parameters (e.g. \citealt{Nixon+2013, LodatoFacchini2013}). All these effects are enhanced by large companion to inner binary mass ratios and high misalignment degrees to the perturber, or in the case of a retrograde flyby \citep{Cuello+2019}.

The observation of two \lang{diametrically} 
\lang{opposed} spirals in the CBDs, as in AS 205 \citep{Kurtovic+2018}, suggests the occurrence of a flyby event. Indeed, any close encounter would invariably induce spiral density waves within the disc through tidal effects \citep{Rafikov2002}. Prograde encounters and close flybys tend to induce the formation of prominent spiral arms, whereas bound companions typically result in less conspicuous spiral arms (e.g. \citealt{Cuello+2019, Menard+2020}). 

However if none of the aforementioned structures are detected in a CBD, it remains plausible that the disc has undergone relaxation over time due to prior interactions with neighbouring stars.
Lastly, some multiple systems show a lack of CBD (e.g. HT Lup; \citealt{Kurtovic+2018}, Sz65\&Sz66; \citealt{Miley+2024}). This absence could be attributed to the considerable separation between the binary components \lang{precluding} the formation of a CBD \citep{Harris+2012, Elsender+2023}. Moreover the hypothetical existence of such a disc suggests rapid dispersal, possibly stemming from intense interactions with a neighbouring star. These interactions could range from dramatic events like disc-penetrating flybys \citep{Cuello+2023} to the influence of additional radiation enhancing photoevaporation \citep{Dai+2018}.
In any case, a detailed systematic study of CBD in triple systems exploring the wide range of available orbital configurations will help to properly understand the previously cited effects.

\subsection{Limitations and caveats}
\label{subsec:limits}

The main limitation of our work is the low number of orbital configurations tested. Considering the total lack of orbital information on V892~Tau~NE, we proposed orbital arrangements that could trigger the observed structures in the CBD. Admittedly our goal was not to thoroughly sample the available parameter space for the orbit of V892~Tau~NE, which would have been too computationally demanding with SPH simulations. Instead we chose to model V892~Tau as a triple system with a CBD in order to put constraints on the companion's orbit. Without any astrometric measurements available and considering the well-characterised inner binary orbit, disc forward modelling constitutes the most reliable method to infer the long-period orbit of V892~Tau NE.

We ensured an optimal quality of the data by combining previous \lang{datasets} with new observations. Our disc model (surface density exponent, aspect ratio, viscosity, etc.) and the values used in the radiative calculations (dust to gas mass ratio, CO abundance, freeze-out temperature, etc...) are all typical values found in the literature for class II protoplanetary discs. We used gas only simulations for computational reasons. Therefore the dynamics of large dust grains (marginally coupled to gas) is not fully consistent. At any rate, in this work we focus on an in-depth study of the gas morphology and kinematics, which is marginally affected by our assumption regarding the gas-dust coupling.

\rmvd{As stated previously, the orbit of V892~Tau~NE is currently unconstrained. This means that V892~Tau~NE could in principle be unrelated to V892~Tau or could be following a flyby orbit. Our study favours a bound orbit for V892~Tau~NE, as we can rule out the unrelated case because the inner binary only is not able to reproduce the tilt of the disc (see Appendix \ref{app : binary sims}). Moreover, it has been recently shown that V892~Tau and V892~Tau~NE are likely related \citep{Thomas+2023}. Moreover, disc-grazing flybys cause efficient disc truncation and trigger prominent and short-lived spirals \citep{Cuello+2019, Cuello+2020, Cuello+2023, Smallwood+2023}. Given the disc extent and the low strength of the observed spiral-shaped features in V892~Tau, the flyby scenario seems unlikely.}

Differences between the data and our synthetic maps could come from our numerical setup. Due to a lack of numerical resolution, the inner system is not well reproduced in our simulations. Absorption of the parent cloud at the centre of the system also limits its precise modelling. For example the circumprimary disc can not be captured due to an accretion radius of the sink particles being much larger than its expected physical size. This could explain the flux difference between our images and the data, as this narrow disc could capture incoming radiation of the primary. Also the large accretion radii of the sinks translate into an inner cavity wider than the one observed. 

Last, the expense of the presented hydrodynamical simulations prohibits us from evolving the system until its equilibrium state. However our simulations reached a state close to this equilibrium, as the size of the bulk of the disc does not significantly change in the last $20$ orbits of the companion. Therefore our simulations are long enough to reasonably infer the long-term disc behaviour. For longer evolutionary times, we expect the companion to clear even more material in the outer regions, which should lead to smaller disc sizes --- potentially in better agreement with the observations, especially for \textit{ei60}.


\section{Conclusion}
\label{sec:conclusion}

In this paper, we presented new ALMA Band 6 observations of V892~Tau at a high spatial and spectral resolution. We confirmed the presence of structures in the disc, \lang{which} we interpreted as hints of \lang{an} interaction with the companion star V892~Tau~NE. Using 3D SPH simulations, we modelled V892~Tau as a triple system with a CBD and produced synthetic ALMA maps comparable to the observations. We investigated the available orbits for V892~Tau~NE on the basis of their ability to reproduce the data. Once the orbital configuration \lang{was} constrained, we discussed the resulting disc dynamics and planetary architectures.

Our main findings are summarised as follows:
\begin{enumerate}
    \item Our observations confirm the hints of \lang{an} ongoing interaction between the circumbinary disc around V892~Tau and V892~Tau~NE. The disc extent is consistent with tidal truncation models. Even if dominated by Keplerian rotation, the CBD harbours non-Keplerian structures that are seen in the channel maps and moment maps, such as spiral features. \rmvd{Based on our modelling, we propose that the closest side to the observer is the SE disc side.} 
    \mdy{Finally, the data neither conclusively support nor refute the existence of a warp in the disc.}
    \item An eccentric inclined orbit of V892~Tau NE reproduces features compatible with the disc observations. A non-zero mutual inclination of the companion with respect to the inner binary \lang{best explains} the slight tilt of the disc. \mdy{Given our limited parameter space sampling,} this tilt \mdy{seems} \lang{to be reproduced best} by a mutual inclination of $60\degree$. The eccentricity of the companion sets the disc extent, which \mdy{seems} \lang{to be reproduced best} by eccentricity values around $0.5$. 
    \item The misalignment between the inner binary and V892~Tau~NE triggers disc damped oscillations in inclination and a non-linear precession. At equilibrium, the disc mid-plane is close to being coplanar with the inner binary, but on a slightly misaligned plane with respect to the orbital planes of the stars. Remarkably, the observed dynamical state of the disc cannot be explained using binary system models alone.
    \item Given the likely orbital configuration of V892~Tau and assuming planets will form in the disc, these circumbinary planets would become unstable in the very inner parts and in the outer parts of the disc --- as soon as the gas dissipates. Any surviving planet should have \lang{an} intermediate semi-major axis between $40$ au and $80$ au and is expected to undergo significant oscillations in eccentricity and inclination due to the interactions with the external companion.
\end{enumerate}

Based on our investigation, V892~Tau constitutes the archetype of a hierarchical triple system with a CBD, standing out as a unique \lang{test bed} for planet formation in multiple stellar systems. However, to thoroughly infer the available formation channels of circumbinary planets in such systems, detailed studies on dust dynamics and growth are required. In addition, future near-IR observations would be key to further characterise the disc content and morphology, which would add complementary and independent constraints on the disc structure. By combining this information to gas-dust hydrodynamical simulations coupled to dust growth algorithms, we could in principle identify the possible regions of planet formation in V892~Tau. This would pave the way for a global understanding of planet formation and dynamics in the context of multiple systems.


\begin{acknowledgements}
    This project has received funding from the European Research Council (ERC) under the European Union Horizon Europe programme (grant agreement No. 101042275, project Stellar-MADE). The Geryon2 cluster housed at the Centro de Astro-Ingenier\'ia UC was used for the calculations performed in this paper. The BASAL PFB-06 CATA, Anillo ACT-86, FONDEQUIP AIC-57, and QUIMAL 130008 provided funding for several improvements to the Geryon/Geryon2 cluster. \mdy{Some of the computations presented in this paper were performed using the GRICAD infrastructure (\url{https://gricad.univ-grenoble-alpes.fr}), which is supported by Grenoble research communities.} A.R. has been supported by the UK Science and Technology research Council (STFC) via the consolidated grant ST/W000997/1 and by the European Union’s Horizon 2020 research and innovation programme under the Marie Sklodowska-Curie grant agreement No. 823823 (RISE DUSTBUSTERS project). R.N. acknowledges support from UKRI/EPSRC through a Stephen Hawking Fellowship (EP/T017287/1). \mdy{Support for AI was provided by NASA through the NASA Hubble Fellowship grant No. HST-HF2-51532.001-A awarded by the Space Telescope Science Institute, which is operated by the Association of Universities for Research in Astronomy, Inc., for NASA, under contract NAS5-26555.}
    This paper makes use of the following ALMA data: ADS/JAO.ALMA\#2021.1.01137.S., ADS/JAO.ALMA\#2013.1.00498.S. ALMA is a partnership of ESO (representing its member states), NSF (USA) and NINS (Japan), together with NRC (Canada), MOST and ASIAA (Taiwan), and KASI (Republic of Korea), in cooperation with the Republic of Chile. The Joint ALMA Observatory is operated by ESO, AUI/NRAO and NAOJ.
    The data underlying this article will be shared on reasonable request to the corresponding author. The code {\sc Phantom} used in this work is publicly available at \url{https://github.com/danieljprice/phantom}. \mdy{The {\sc Discminer} code used in this study is open source and can be found at \url{https://github.com/andizq/discminer}}. This work makes use of {\sc Splash} \citep{splash}, {\sc Numpy} \citep{numpy} and {\sc Matplotlib} \citep{Matplotlib}. We also acknowledge the use of {\sc Sarracen} \citep{Sarracen}. AA would like to extend his sincere appreciation to Jean-François Gonzalez and Enrico Ragusa for their fruitful discussions throughout the course of this research. We would like to express our profound gratitude to the anonymous referee for their insightful comments and suggestions which have greatly contributed to the improvement of this work.
\end{acknowledgements}

\bibliographystyle{aa} 
\bibliography{biblio}

\begin{thebibliography}{106}
\expandafter\ifx\csname natexlab\endcsname\relax\def\natexlab#1{#1}\fi

\bibitem[{{Aly} {et~al.}(2015){Aly}, {Dehnen}, {Nixon}, \& {King}}]{Aly+2015}
{Aly}, H., {Dehnen}, W., {Nixon}, C., \& {King}, A. 2015, \mnras, 449, 65

\bibitem[{{Aly} {et~al.}(2021){Aly}, {Gonzalez}, {Nealon}, {Longarini}, {Lodato}, \& {Price}}]{Aly+2021}
{Aly}, H., {Gonzalez}, J.-F., {Nealon}, R., {et~al.} 2021, \mnras, 508, 2743

\bibitem[{{Aly} {et~al.}(2023){Aly}, {Nealon}, \& {Gonzalez}}]{Aly+2023}
{Aly}, H., {Nealon}, R., \& {Gonzalez}, J.-F. 2023, \mnras [\eprint[arXiv]{2311.06182}]

\bibitem[{{Andr{\'e}} {et~al.}(2010){Andr{\'e}}, {Men'shchikov}, {Bontemps}, {K{\"o}nyves}, {Motte}, {Schneider}, {Didelon}, {Minier}, {Saraceno}, {Ward-Thompson}, {di Francesco}, {White}, {Molinari}, {Testi}, {Abergel}, {Griffin}, {Henning}, {Royer}, {Mer{\'\i}n}, {Vavrek}, {Attard}, {Arzoumanian}, {Wilson}, {Ade}, {Aussel}, {Baluteau}, {Benedettini}, {Bernard}, {Blommaert}, {Cambr{\'e}sy}, {Cox}, {di Giorgio}, {Hargrave}, {Hennemann}, {Huang}, {Kirk}, {Krause}, {Launhardt}, {Leeks}, {Le Pennec}, {Li}, {Martin}, {Maury}, {Olofsson}, {Omont}, {Peretto}, {Pezzuto}, {Prusti}, {Roussel}, {Russeil}, {Sauvage}, {Sibthorpe}, {Sicilia-Aguilar}, {Spinoglio}, {Waelkens}, {Woodcraft}, \& {Zavagno}}]{Andre+2010}
{Andr{\'e}}, P., {Men'shchikov}, A., {Bontemps}, S., {et~al.} 2010, \aap, 518, L102

\bibitem[{{Andrews} {et~al.}(2014){Andrews}, {Chandler}, {Isella}, {Birnstiel}, {Rosenfeld}, {Wilner}, {P{\'e}rez}, {Ricci}, {Carpenter}, {Calvet}, {Corder}, {Deller}, {Dullemond}, {Greaves}, {Harris}, {Henning}, {Kwon}, {Lazio}, {Linz}, {Mundy}, {Sargent}, {Storm}, \& {Testi}}]{Andrews+2014}
{Andrews}, S.~M., {Chandler}, C.~J., {Isella}, A., {et~al.} 2014, \apj, 787, 148

\bibitem[{{Artymowicz} \& {Lubow}(1994)}]{Artymowicz1994}
{Artymowicz}, P. \& {Lubow}, S.~H. 1994, \apj, 421, 651

\bibitem[{{Bate}(2018)}]{Bate2018}
{Bate}, M.~R. 2018, \mnras, 475, 5618

\bibitem[{{Bate} {et~al.}(2000){Bate}, {Bonnell}, {Clarke}, {Lubow}, {Ogilvie}, {Pringle}, \& {Tout}}]{Bate+2000}
{Bate}, M.~R., {Bonnell}, I.~A., {Clarke}, C.~J., {et~al.} 2000, \mnras, 317, 773

\bibitem[{{Bate} {et~al.}(1995){Bate}, {Bonnell}, \& {Price}}]{Bate+1995}
{Bate}, M.~R., {Bonnell}, I.~A., \& {Price}, N.~M. 1995, \mnras, 277, 362

\bibitem[{{Benisty} {et~al.}(2017){Benisty}, {Stolker}, {Pohl}, {de Boer}, {Lesur}, {Dominik}, {Dullemond}, {Langlois}, {Min}, {Wagner}, {Henning}, {Juhasz}, {Pinilla}, {Facchini}, {Apai}, {van Boekel}, {Garufi}, {Ginski}, {M{\'e}nard}, {Pinte}, {Quanz}, {Zurlo}, {Boccaletti}, {Bonnefoy}, {Beuzit}, {Chauvin}, {Cudel}, {Desidera}, {Feldt}, {Fontanive}, {Gratton}, {Kasper}, {Lagrange}, {LeCoroller}, {Mouillet}, {Mesa}, {Sissa}, {Vigan}, {Antichi}, {Buey}, {Fusco}, {Gisler}, {Llored}, {Magnard}, {Moeller-Nilsson}, {Pragt}, {Roelfsema}, {Sauvage}, \& {Wildi}}]{Benisty+2017}
{Benisty}, M., {Stolker}, T., {Pohl}, A., {et~al.} 2017, \aap, 597, A42

\bibitem[{{Borchert} {et~al.}(2022){Borchert}, {Price}, {Pinte}, \& {Cuello}}]{Borchert+2022}
{Borchert}, E. M.~A., {Price}, D.~J., {Pinte}, C., \& {Cuello}, N. 2022, \mnras, 517, 4436

\bibitem[{{Briggs}(1995)}]{Briggs}
{Briggs}, D.~S. 1995, in American Astronomical Society Meeting Abstracts, Vol. 187, American Astronomical Society Meeting Abstracts, 112.02

\bibitem[{{Burns}(1976)}]{Burns1976}
{Burns}, J.~A. 1976, American Journal of Physics, 44, 944

\bibitem[{{Busetti} {et~al.}(2018){Busetti}, {Beust}, \& {Harley}}]{Busetti+2018}
{Busetti}, F., {Beust}, H., \& {Harley}, C. 2018, \aap, 619, A91

\bibitem[{{Calcino} {et~al.}(2020){Calcino}, {Christiaens}, {Price}, {Pinte}, {Davis}, {van der Marel}, \& {Cuello}}]{Calcino+2020}
{Calcino}, J., {Christiaens}, V., {Price}, D.~J., {et~al.} 2020, \mnras, 498, 639

\bibitem[{{Ceppi} {et~al.}(2023){Ceppi}, {Longarini}, {Lodato}, {Cuello}, \& {Lubow}}]{Ceppi+2023}
{Ceppi}, S., {Longarini}, C., {Lodato}, G., {Cuello}, N., \& {Lubow}, S.~H. 2023, \mnras, 520, 5817

\bibitem[{{Cuello} {et~al.}(2019){Cuello}, {Dipierro}, {Mentiplay}, {Price}, {Pinte}, {Cuadra}, {Laibe}, {M{\'e}nard}, {Poblete}, \& {Montesinos}}]{Cuello+2019}
{Cuello}, N., {Dipierro}, G., {Mentiplay}, D., {et~al.} 2019, \mnras, 483, 4114

\bibitem[{{Cuello} \& {Giuppone}(2019)}]{Cuello&Giupponne2019}
{Cuello}, N. \& {Giuppone}, C.~A. 2019, \aap, 628, A119

\bibitem[{{Cuello} {et~al.}(2020){Cuello}, {Louvet}, {Mentiplay}, {Pinte}, {Price}, {Winter}, {Nealon}, {M{\'e}nard}, {Lodato}, {Dipierro}, {Christiaens}, {Montesinos}, {Cuadra}, {Laibe}, {Cieza}, {Dong}, \& {Alexander}}]{Cuello+2020}
{Cuello}, N., {Louvet}, F., {Mentiplay}, D., {et~al.} 2020, \mnras, 491, 504

\bibitem[{{Cuello} {et~al.}(2023){Cuello}, {M{\'e}nard}, \& {Price}}]{Cuello+2023}
{Cuello}, N., {M{\'e}nard}, F., \& {Price}, D.~J. 2023, European Physical Journal Plus, 138, 11

\bibitem[{{Czekala} {et~al.}(2019){Czekala}, {Chiang}, {Andrews}, {Jensen}, {Torres}, {Wilner}, {Stassun}, \& {Macintosh}}]{Czekala+2019}
{Czekala}, I., {Chiang}, E., {Andrews}, S.~M., {et~al.} 2019, \apj, 883, 22

\bibitem[{{Dai} {et~al.}(2018){Dai}, {Liu}, {Wu}, {Xie}, {Yang}, {Zhang}, \& {Zhou}}]{Dai+2018}
{Dai}, Y.-Z., {Liu}, H.-G., {Wu}, W.-B., {et~al.} 2018, \mnras, 480, 4080

\bibitem[{{Deng} {et~al.}(2021){Deng}, {Ogilvie}, \& {Mayer}}]{Deng+2021}
{Deng}, H., {Ogilvie}, G.~I., \& {Mayer}, L. 2021, \mnras, 500, 4248

\bibitem[{{Dong} {et~al.}(2015){Dong}, {Zhu}, {Rafikov}, \& {Stone}}]{Dong+2015}
{Dong}, R., {Zhu}, Z., {Rafikov}, R.~R., \& {Stone}, J.~M. 2015, \apjl, 809, L5

\bibitem[{{Duch{\^e}ne} \& {Kraus}(2013)}]{DucheneKraus2013}
{Duch{\^e}ne}, G. \& {Kraus}, A. 2013, \araa, 51, 269

\bibitem[{{Elsender} {et~al.}(2023){Elsender}, {Bate}, {Lakeland}, {Jensen}, \& {Lubow}}]{Elsender+2023}
{Elsender}, D., {Bate}, M.~R., {Lakeland}, B.~S., {Jensen}, E. L.~N., \& {Lubow}, S.~H. 2023, \mnras, 523, 4353

\bibitem[{{Esplin} \& {Luhman}(2019)}]{EsplinLuhman2019}
{Esplin}, T.~L. \& {Luhman}, K.~L. 2019, \aj, 158, 54

\bibitem[{{Facchini} {et~al.}(2013){Facchini}, {Lodato}, \& {Price}}]{Facchini+2013}
{Facchini}, S., {Lodato}, G., \& {Price}, D.~J. 2013, \mnras, 433, 2142

\bibitem[{{Farago} \& {Laskar}(2010)}]{FaragoLaskar2010}
{Farago}, F. \& {Laskar}, J. 2010, \mnras, 401, 1189

\bibitem[{{Fern{\'a}ndez-L{\'o}pez} {et~al.}(2017){Fern{\'a}ndez-L{\'o}pez}, {Zapata}, \& {Gabbasov}}]{Fernandez_Lopez+2017}
{Fern{\'a}ndez-L{\'o}pez}, M., {Zapata}, L.~A., \& {Gabbasov}, R. 2017, \apj, 845, 10

\bibitem[{{Flores} {et~al.}(2022){Flores}, {Connelley}, {Reipurth}, \& {Duch{\^e}ne}}]{Flores+2022}
{Flores}, C., {Connelley}, M.~S., {Reipurth}, B., \& {Duch{\^e}ne}, G. 2022, \apj, 925, 21

\bibitem[{{Foreman-Mackey} {et~al.}(2013){Foreman-Mackey}, {Hogg}, {Lang}, \& {Goodman}}]{emcee}
{Foreman-Mackey}, D., {Hogg}, D.~W., {Lang}, D., \& {Goodman}, J. 2013, \pasp, 125, 306

\bibitem[{{Gaia Collaboration} {et~al.}(2023){Gaia Collaboration}, {Vallenari}, {Brown}, {Prusti}, {de Bruijne}, {Arenou}, {Babusiaux}, {Biermann}, {Creevey}, {Ducourant}, {Evans}, {Eyer}, {Guerra}, {Hutton}, {Jordi}, {Klioner}, {Lammers}, {Lindegren}, {Luri}, {Mignard}, {Panem}, {Pourbaix}, {Randich}, {Sartoretti}, {Soubiran}, {Tanga}, {Walton}, {Bailer-Jones}, {Bastian}, {Drimmel}, {Jansen}, {Katz}, {Lattanzi}, {van Leeuwen}, {Bakker}, {Cacciari}, {Casta{\~n}eda}, {De Angeli}, {Fabricius}, {Fouesneau}, {Fr{\'e}mat}, {Galluccio}, {Guerrier}, {Heiter}, {Masana}, {Messineo}, {Mowlavi}, {Nicolas}, {Nienartowicz}, {Pailler}, {Panuzzo}, {Riclet}, {Roux}, {Seabroke}, {Sordo}, {Th{\'e}venin}, {Gracia-Abril}, {Portell}, {Teyssier}, {Altmann}, {Andrae}, {Audard}, {Bellas-Velidis}, {Benson}, {Berthier}, {Blomme}, {Burgess}, {Busonero}, {Busso}, {C{\'a}novas}, {Carry}, {Cellino}, {Cheek}, {Clementini}, {Damerdji}, {Davidson}, {de Teodoro}, {Nu{\~n}ez Campos}, {Delchambre}, {Dell'Oro}, {Esquej},
  {Fern{\'a}ndez-Hern{\'a}ndez}, {Fraile}, {Garabato}, {Garc{\'\i}a-Lario}, {Gosset}, {Haigron}, {Halbwachs}, {Hambly}, {Harrison}, {Hern{\'a}ndez}, {Hestroffer}, {Hodgkin}, {Holl}, {Jan{\ss}en}, {Jevardat de Fombelle}, {Jordan}, {Krone-Martins}, {Lanzafame}, {L{\"o}ffler}, {Marchal}, {Marrese}, {Moitinho}, {Muinonen}, {Osborne}, {Pancino}, {Pauwels}, {Recio-Blanco}, {Reyl{\'e}}, {Riello}, {Rimoldini}, {Roegiers}, {Rybizki}, {Sarro}, {Siopis}, {Smith}, {Sozzetti}, {Utrilla}, {van Leeuwen}, {Abbas}, {{\'A}brah{\'a}m}, {Abreu Aramburu}, {Aerts}, {Aguado}, {Ajaj}, {Aldea-Montero}, {Altavilla}, {{\'A}lvarez}, {Alves}, {Anders}, {Anderson}, {Anglada Varela}, {Antoja}, {Baines}, {Baker}, {Balaguer-N{\'u}{\~n}ez}, {Balbinot}, {Balog}, {Barache}, {Barbato}, {Barros}, {Barstow}, {Bartolom{\'e}}, {Bassilana}, {Bauchet}, {Becciani}, {Bellazzini}, {Berihuete}, {Bernet}, {Bertone}, {Bianchi}, {Binnenfeld}, {Blanco-Cuaresma}, {Blazere}, {Boch}, {Bombrun}, {Bossini}, {Bouquillon}, {Bragaglia}, {Bramante}, {Breedt},
  {Bressan}, {Brouillet}, {Brugaletta}, {Bucciarelli}, {Burlacu}, {Butkevich}, {Buzzi}, {Caffau}, {Cancelliere}, {Cantat-Gaudin}, {Carballo}, {Carlucci}, {Carnerero}, {Carrasco}, {Casamiquela}, {Castellani}, {Castro-Ginard}, {Chaoul}, {Charlot}, {Chemin}, {Chiaramida}, {Chiavassa}, {Chornay}, {Comoretto}, {Contursi}, {Cooper}, {Cornez}, {Cowell}, {Crifo}, {Cropper}, {Crosta}, {Crowley}, {Dafonte}, {Dapergolas}, {David}, {David}, {de Laverny}, {De Luise}, {De March}, {De Ridder}, {de Souza}, {de Torres}, {del Peloso}, {del Pozo}, {Delbo}, {Delgado}, {Delisle}, {Demouchy}, {Dharmawardena}, {Di Matteo}, {Diakite}, {Diener}, {Distefano}, {Dolding}, {Edvardsson}, {Enke}, {Fabre}, {Fabrizio}, {Faigler}, {Fedorets}, {Fernique}, {Fienga}, {Figueras}, {Fournier}, {Fouron}, {Fragkoudi}, {Gai}, {Garcia-Gutierrez}, {Garcia-Reinaldos}, {Garc{\'\i}a-Torres}, {Garofalo}, {Gavel}, {Gavras}, {Gerlach}, {Geyer}, {Giacobbe}, {Gilmore}, {Girona}, {Giuffrida}, {Gomel}, {Gomez}, {Gonz{\'a}lez-N{\'u}{\~n}ez},
  {Gonz{\'a}lez-Santamar{\'\i}a}, {Gonz{\'a}lez-Vidal}, {Granvik}, {Guillout}, {Guiraud}, {Guti{\'e}rrez-S{\'a}nchez}, {Guy}, {Hatzidimitriou}, {Hauser}, {Haywood}, {Helmer}, {Helmi}, {Sarmiento}, {Hidalgo}, {Hilger}, {H{\l}adczuk}, {Hobbs}, {Holland}, {Huckle}, {Jardine}, {Jasniewicz}, {Jean-Antoine Piccolo}, {Jim{\'e}nez-Arranz}, {Jorissen}, {Juaristi Campillo}, {Julbe}, {Karbevska}, {Kervella}, {Khanna}, {Kontizas}, {Kordopatis}, {Korn}, {K{\'o}sp{\'a}l}, {Kostrzewa-Rutkowska}, {Kruszy{\'n}ska}, {Kun}, {Laizeau}, {Lambert}, {Lanza}, {Lasne}, {Le Campion}, {Lebreton}, {Lebzelter}, {Leccia}, {Leclerc}, {Lecoeur-Taibi}, {Liao}, {Licata}, {Lindstr{\o}m}, {Lister}, {Livanou}, {Lobel}, {Lorca}, {Loup}, {Madrero Pardo}, {Magdaleno Romeo}, {Managau}, {Mann}, {Manteiga}, {Marchant}, {Marconi}, {Marcos}, {Marcos Santos}, {Mar{\'\i}n Pina}, {Marinoni}, {Marocco}, {Marshall}, {Martin Polo}, {Mart{\'\i}n-Fleitas}, {Marton}, {Mary}, {Masip}, {Massari}, {Mastrobuono-Battisti}, {Mazeh}, {McMillan}, {Messina}, {Michalik},
  {Millar}, {Mints}, {Molina}, {Molinaro}, {Moln{\'a}r}, {Monari}, {Mongui{\'o}}, {Montegriffo}, {Montero}, {Mor}, {Mora}, {Morbidelli}, {Morel}, {Morris}, {Muraveva}, {Murphy}, {Musella}, {Nagy}, {Noval}, {Oca{\~n}a}, {Ogden}, {Ordenovic}, {Osinde}, {Pagani}, {Pagano}, {Palaversa}, {Palicio}, {Pallas-Quintela}, {Panahi}, {Payne-Wardenaar}, {Pe{\~n}alosa Esteller}, {Penttil{\"a}}, {Pichon}, {Piersimoni}, {Pineau}, {Plachy}, {Plum}, {Poggio}, {Pr{\v{s}}a}, {Pulone}, {Racero}, {Ragaini}, {Rainer}, {Raiteri}, {Rambaux}, {Ramos}, {Ramos-Lerate}, {Re Fiorentin}, {Regibo}, {Richards}, {Rios Diaz}, {Ripepi}, {Riva}, {Rix}, {Rixon}, {Robichon}, {Robin}, {Robin}, {Roelens}, {Rogues}, {Rohrbasser}, {Romero-G{\'o}mez}, {Rowell}, {Royer}, {Ruz Mieres}, {Rybicki}, {Sadowski}, {S{\'a}ez N{\'u}{\~n}ez}, {Sagrist{\`a} Sell{\'e}s}, {Sahlmann}, {Salguero}, {Samaras}, {Sanchez Gimenez}, {Sanna}, {Santove{\~n}a}, {Sarasso}, {Schultheis}, {Sciacca}, {Segol}, {Segovia}, {S{\'e}gransan}, {Semeux}, {Shahaf}, {Siddiqui}, {Siebert},
  {Siltala}, {Silvelo}, {Slezak}, {Slezak}, {Smart}, {Snaith}, {Solano}, {Solitro}, {Souami}, {Souchay}, {Spagna}, {Spina}, {Spoto}, {Steele}, {Steidelm{\"u}ller}, {Stephenson}, {S{\"u}veges}, {Surdej}, {Szabados}, {Szegedi-Elek}, {Taris}, {Taylor}, {Teixeira}, {Tolomei}, {Tonello}, {Torra}, {Torra}, {Torralba Elipe}, {Trabucchi}, {Tsounis}, {Turon}, {Ulla}, {Unger}, {Vaillant}, {van Dillen}, {van Reeven}, {Vanel}, {Vecchiato}, {Viala}, {Vicente}, {Voutsinas}, {Weiler}, {Wevers}, {Wyrzykowski}, {Yoldas}, {Yvard}, {Zhao}, {Zorec}, {Zucker}, \& {Zwitter}}]{GaiaDR3}
{Gaia Collaboration}, {Vallenari}, A., {Brown}, A.~G.~A., {et~al.} 2023, \aap, 674, A1

\bibitem[{{Gianuzzi} {et~al.}(2024){Gianuzzi}, {Cuello}, {Giuppone}, \& {Sucerqueia}}]{Gianuzzi+2024}
{Gianuzzi}, E., {Cuello}, N., {Giuppone}, C., \& {Sucerqueia}, M. 2024, \aap, in preparation

\bibitem[{{Goldreich} \& {Lynden-Bell}(1965)}]{GoldreichLynden1965}
{Goldreich}, P. \& {Lynden-Bell}, D. 1965, \mnras, 130, 97

\bibitem[{{Gonzalez} {et~al.}(2020){Gonzalez}, {van der Plas}, {Pinte}, {Cuello}, {Nealon}, {M{\'e}nard}, {Revol}, {Rodet}, {Langlois}, \& {Maire}}]{Gonzalez+2020}
{Gonzalez}, J.-F., {van der Plas}, G., {Pinte}, C., {et~al.} 2020, \mnras, 499, 3837

\bibitem[{{Griffin} {et~al.}(2010){Griffin}, {Abergel}, {Abreu}, {Ade}, {Andr{\'e}}, {Augueres}, {Babbedge}, {Bae}, {Baillie}, {Baluteau}, {Barlow}, {Bendo}, {Benielli}, {Bock}, {Bonhomme}, {Brisbin}, {Brockley-Blatt}, {Caldwell}, {Cara}, {Castro-Rodriguez}, {Cerulli}, {Chanial}, {Chen}, {Clark}, {Clements}, {Clerc}, {Coker}, {Communal}, {Conversi}, {Cox}, {Crumb}, {Cunningham}, {Daly}, {Davis}, {de Antoni}, {Delderfield}, {Devin}, {di Giorgio}, {Didschuns}, {Dohlen}, {Donati}, {Dowell}, {Dowell}, {Duband}, {Dumaye}, {Emery}, {Ferlet}, {Ferrand}, {Fontignie}, {Fox}, {Franceschini}, {Frerking}, {Fulton}, {Garcia}, {Gastaud}, {Gear}, {Glenn}, {Goizel}, {Griffin}, {Grundy}, {Guest}, {Guillemet}, {Hargrave}, {Harwit}, {Hastings}, {Hatziminaoglou}, {Herman}, {Hinde}, {Hristov}, {Huang}, {Imhof}, {Isaak}, {Israelsson}, {Ivison}, {Jennings}, {Kiernan}, {King}, {Lange}, {Latter}, {Laurent}, {Laurent}, {Leeks}, {Lellouch}, {Levenson}, {Li}, {Li}, {Lilienthal}, {Lim}, {Liu}, {Lu}, {Madden}, {Mainetti}, {Marliani},
  {McKay}, {Mercier}, {Molinari}, {Morris}, {Moseley}, {Mulder}, {Mur}, {Naylor}, {Nguyen}, {O'Halloran}, {Oliver}, {Olofsson}, {Olofsson}, {Orfei}, {Page}, {Pain}, {Panuzzo}, {Papageorgiou}, {Parks}, {Parr-Burman}, {Pearce}, {Pearson}, {P{\'e}rez-Fournon}, {Pinsard}, {Pisano}, {Podosek}, {Pohlen}, {Polehampton}, {Pouliquen}, {Rigopoulou}, {Rizzo}, {Roseboom}, {Roussel}, {Rowan-Robinson}, {Rownd}, {Saraceno}, {Sauvage}, {Savage}, {Savini}, {Sawyer}, {Scharmberg}, {Schmitt}, {Schneider}, {Schulz}, {Schwartz}, {Shafer}, {Shupe}, {Sibthorpe}, {Sidher}, {Smith}, {Smith}, {Smith}, {Spencer}, {Stobie}, {Sudiwala}, {Sukhatme}, {Surace}, {Stevens}, {Swinyard}, {Trichas}, {Tourette}, {Triou}, {Tseng}, {Tucker}, {Turner}, {Vaccari}, {Valtchanov}, {Vigroux}, {Virique}, {Voellmer}, {Walker}, {Ward}, {Waskett}, {Weilert}, {Wesson}, {White}, {Whitehouse}, {Wilson}, {Winter}, {Woodcraft}, {Wright}, {Xu}, {Zavagno}, {Zemcov}, {Zhang}, \& {Zonca}}]{Griffin+2010}
{Griffin}, M.~J., {Abergel}, A., {Abreu}, A., {et~al.} 2010, \aap, 518, L3

\bibitem[{{Harris} \& {Tricco}(2023)}]{Sarracen}
{Harris}, A. \& {Tricco}, T. 2023, The Journal of Open Source Software, 8, 5263

\bibitem[{{Harris} {et~al.}(2020){Harris}, {Millman}, {van der Walt}, {Gommers}, {Virtanen}, {Cournapeau}, {Wieser}, {Taylor}, {Berg}, {Smith}, {Kern}, {Picus}, {Hoyer}, {van Kerkwijk}, {Brett}, {Haldane}, {del R{\'\i}o}, {Wiebe}, {Peterson}, {G{\'e}rard-Marchant}, {Sheppard}, {Reddy}, {Weckesser}, {Abbasi}, {Gohlke}, \& {Oliphant}}]{numpy}
{Harris}, C.~R., {Millman}, K.~J., {van der Walt}, S.~J., {et~al.} 2020, \nat, 585, 357

\bibitem[{{Harris} {et~al.}(2012){Harris}, {Andrews}, {Wilner}, \& {Kraus}}]{Harris+2012}
{Harris}, R.~J., {Andrews}, S.~M., {Wilner}, D.~J., \& {Kraus}, A.~L. 2012, \apj, 751, 115

\bibitem[{{Hirsh} {et~al.}(2020){Hirsh}, {Price}, {Gonzalez}, {Ubeira-Gabellini}, \& {Ragusa}}]{Hirsh+2020}
{Hirsh}, K., {Price}, D.~J., {Gonzalez}, J.-F., {Ubeira-Gabellini}, M.~G., \& {Ragusa}, E. 2020, \mnras, 498, 2936

\bibitem[{{Hunter}(2007)}]{Matplotlib}
{Hunter}, J.~D. 2007, Computing in Science and Engineering, 9, 90

\bibitem[{{Izquierdo} {et~al.}(2021){Izquierdo}, {Testi}, {Facchini}, {Rosotti}, \& {van Dishoeck}}]{DiscminerI}
{Izquierdo}, A.~F., {Testi}, L., {Facchini}, S., {Rosotti}, G.~P., \& {van Dishoeck}, E.~F. 2021, \aap, 650, A179

\bibitem[{{Izquierdo} {et~al.}(2023){Izquierdo}, {Testi}, {Facchini}, {Rosotti}, {van Dishoeck}, {W{\"o}lfer}, \& {Paneque-Carre{\~n}o}}]{DiscminerII}
{Izquierdo}, A.~F., {Testi}, L., {Facchini}, S., {et~al.} 2023, \aap, 674, A113

\bibitem[{{Kenyon} {et~al.}(2008){Kenyon}, {G{\'o}mez}, \& {Whitney}}]{Kenyon+2008}
{Kenyon}, S.~J., {G{\'o}mez}, M., \& {Whitney}, B.~A. 2008, in Handbook of Star Forming Regions, Volume I, ed. B.~{Reipurth}, Vol.~4, 405

\bibitem[{{Kiseleva} {et~al.}(1998){Kiseleva}, {Eggleton}, \& {Mikkola}}]{Kiseleva+1998}
{Kiseleva}, L.~G., {Eggleton}, P.~P., \& {Mikkola}, S. 1998, \mnras, 300, 292

\bibitem[{{K{\"o}hler}(2011)}]{Kohler2011}
{K{\"o}hler}, R. 2011, \aap, 530, A126

\bibitem[{{Kozai}(1962)}]{Kozai1962}
{Kozai}, Y. 1962, \aj, 67, 591

\bibitem[{{Kratter} \& {Lodato}(2016)}]{KratterLodato2016}
{Kratter}, K. \& {Lodato}, G. 2016, \araa, 54, 271

\bibitem[{{K{\"u}{\c{c}}{\"u}k} \& {Akkaya}(2010)}]{KucukAkkaya2010}
{K{\"u}{\c{c}}{\"u}k}, I. \& {Akkaya}, I. 2010, \rmxaa, 46, 109

\bibitem[{{Kurtovic} {et~al.}(2018){Kurtovic}, {P{\'e}rez}, {Benisty}, {Zhu}, {Zhang}, {Huang}, {Andrews}, {Dullemond}, {Isella}, {Bai}, {Carpenter}, {Guzm{\'a}n}, {Ricci}, \& {Wilner}}]{Kurtovic+2018}
{Kurtovic}, N.~T., {P{\'e}rez}, L.~M., {Benisty}, M., {et~al.} 2018, \apjl, 869, L44

\bibitem[{{Law} {et~al.}(2021){Law}, {Teague}, {Loomis}, {Bae}, {{\"O}berg}, {Czekala}, {Andrews}, {Aikawa}, {Alarc{\'o}n}, {Bergin}, {Bergner}, {Booth}, {Bosman}, {Calahan}, {Cataldi}, {Cleeves}, {Furuya}, {Guzm{\'a}n}, {Huang}, {Ilee}, {Le Gal}, {Liu}, {Long}, {M{\'e}nard}, {Nomura}, {P{\'e}rez}, {Qi}, {Schwarz}, {Soto}, {Tsukagoshi}, {Yamato}, {van't Hoff}, {Walsh}, {Wilner}, \& {Zhang}}]{Law+2021}
{Law}, C.~J., {Teague}, R., {Loomis}, R.~A., {et~al.} 2021, \apjs, 257, 4

\bibitem[{{Law} {et~al.}(2023){Law}, {Teague}, {{\"O}berg}, {Rich}, {Andrews}, {Bae}, {Benisty}, {Facchini}, {Flaherty}, {Isella}, {Jin}, {Hashimoto}, {Huang}, {Loomis}, {Long}, {Mu{\~n}oz-Romero}, {Paneque-Carre{\~n}o}, {P{\'e}rez}, {Qi}, {Schwarz}, {Stadler}, {Tsukagoshi}, {Wilner}, \& {van der Plas}}]{Law+2023}
{Law}, C.~J., {Teague}, R., {{\"O}berg}, K.~I., {et~al.} 2023, \apj, 948, 60

\bibitem[{{Lidov}(1962)}]{Lidov1962}
{Lidov}, M.~L. 1962, \planss, 9, 719

\bibitem[{{Lodato} \& {Facchini}(2013)}]{LodatoFacchini2013}
{Lodato}, G. \& {Facchini}, S. 2013, \mnras, 433, 2157

\bibitem[{{Long} {et~al.}(2021){Long}, {Andrews}, {Vega}, {Wilner}, {Chandler}, {Ragusa}, {Teague}, {P{\'e}rez}, {Calvet}, {Carpenter}, {Henning}, {Kwon}, {Linz}, \& {Ricci}}]{Long+2021}
{Long}, F., {Andrews}, S.~M., {Vega}, J., {et~al.} 2021, \apj, 915, 131

\bibitem[{{Luhman}(2007)}]{Luhman+2007}
{Luhman}, K.~L. 2007, \apjs, 173, 104

\bibitem[{{Martin} {et~al.}(2022){Martin}, {Lepp}, {Lubow}, {Kenworthy}, {Kennedy}, \& {Vallet}}]{Martin+2022}
{Martin}, R.~G., {Lepp}, S., {Lubow}, S.~H., {et~al.} 2022, \apjl, 927, L26

\bibitem[{{Martin} \& {Lubow}(2017)}]{MartinLubow2017}
{Martin}, R.~G. \& {Lubow}, S.~H. 2017, \apjl, 835, L28

\bibitem[{{Martin} {et~al.}(2014){Martin}, {Nixon}, {Lubow}, {Armitage}, {Price}, {Do{\u{g}}an}, \& {King}}]{Martin+2014}
{Martin}, R.~G., {Nixon}, C., {Lubow}, S.~H., {et~al.} 2014, \apjl, 792, L33

\bibitem[{{Mathis} {et~al.}(1977){Mathis}, {Rumpl}, \& {Nordsieck}}]{Mathis1977}
{Mathis}, J.~S., {Rumpl}, W., \& {Nordsieck}, K.~H. 1977, \apj, 217, 425

\bibitem[{{McMullin} {et~al.}(2007){McMullin}, {Waters}, {Schiebel}, {Young}, \& {Golap}}]{McMullin2007}
{McMullin}, J.~P., {Waters}, B., {Schiebel}, D., {Young}, W., \& {Golap}, K. 2007, in Astronomical Society of the Pacific Conference Series, Vol. 376, Astronomical Data Analysis Software and Systems XVI, ed. R.~A. {Shaw}, F.~{Hill}, \& D.~J. {Bell}, 127

\bibitem[{{M{\'e}nard} {et~al.}(2020){M{\'e}nard}, {Cuello}, {Ginski}, {van der Plas}, {Villenave}, {Gonzalez}, {Pinte}, {Benisty}, {Boccaletti}, {Price}, {Boehler}, {Chripko}, {de Boer}, {Dominik}, {Garufi}, {Gratton}, {Hagelberg}, {Henning}, {Langlois}, {Maire}, {Pinilla}, {Ruane}, {Schmid}, {van Holstein}, {Vigan}, {Zurlo}, {Hubin}, {Pavlov}, {Rochat}, {Sauvage}, \& {Stadler}}]{Menard+2020}
{M{\'e}nard}, F., {Cuello}, N., {Ginski}, C., {et~al.} 2020, \aap, 639, L1

\bibitem[{{Mercimek} {et~al.}(2023){Mercimek}, {Podio}, {Codella}, {Chahine}, {L{\'o}pez-Sepulcre}, {Ohashi}, {Loinard}, {Johnstone}, {Menard}, {Cuello}, {Caselli}, {Zamponi}, {Aikawa}, {Bianchi}, {Busquet}, {Pineda}, {Bouvier}, {De Simone}, {Zhang}, {Sakai}, {Chandler}, {Ceccarelli}, {Alves}, {Dur{\'a}n}, {Fedele}, {Murillo}, {Jim{\'e}nez-Serra}, \& {Yamamoto}}]{Mercimek+2023}
{Mercimek}, S., {Podio}, L., {Codella}, C., {et~al.} 2023, \mnras, 522, 2384

\bibitem[{{Miley} {et~al.}(2024){Miley}, {Carpenter}, {Booth}, {Jennings}, {Haworth}, {Vioque}, {Andrews}, {Wilner}, {Benisty}, {Huang}, {Perez}, {Guzman}, {Ricci}, \& {Isella}}]{Miley+2024}
{Miley}, J.~M., {Carpenter}, J., {Booth}, R., {et~al.} 2024, \aap, 682, A55

\bibitem[{{Miranda} \& {Lai}(2015)}]{MirandaLai2015}
{Miranda}, R. \& {Lai}, D. 2015, \mnras, 452, 2396

\bibitem[{{Miranda} {et~al.}(2017){Miranda}, {Mu{\~n}oz}, \& {Lai}}]{Miranda+2017}
{Miranda}, R., {Mu{\~n}oz}, D.~J., \& {Lai}, D. 2017, \mnras, 466, 1170

\bibitem[{{Monaghan}(1992)}]{Monaghan1992}
{Monaghan}, J.~J. 1992, \araa, 30, 543

\bibitem[{{Monnier} {et~al.}(2008){Monnier}, {Tannirkulam}, {Tuthill}, {Ireland}, {Cohen}, {Danchi}, \& {Baron}}]{Monnier+2008}
{Monnier}, J.~D., {Tannirkulam}, A., {Tuthill}, P.~G., {et~al.} 2008, \apjl, 681, L97

\bibitem[{{Nealon} {et~al.}(2020){Nealon}, {Cuello}, {Gonzalez}, {van der Plas}, {Pinte}, {Alexander}, {M{\'e}nard}, \& {Price}}]{Nealon+2020}
{Nealon}, R., {Cuello}, N., {Gonzalez}, J.-F., {et~al.} 2020, \mnras, 499, 3857

\bibitem[{{Nixon} {et~al.}(2013){Nixon}, {King}, \& {Price}}]{Nixon+2013}
{Nixon}, C., {King}, A., \& {Price}, D. 2013, \mnras, 434, 1946

\bibitem[{{Nowak} {et~al.}(2024){Nowak}, {Rowther}, {Lacour}, {Meru}, {Nealon}, \& {Price}}]{Nowak+2024}
{Nowak}, M., {Rowther}, S., {Lacour}, S., {et~al.} 2024, \aap, 683, A6

\bibitem[{{Offner} {et~al.}(2022){Offner}, {Moe}, {Kratter}, {Sadavoy}, {Jensen}, \& {Tobin}}]{Offner+2022}
{Offner}, S. S.~R., {Moe}, M., {Kratter}, K.~M., {et~al.} 2022, arXiv e-prints, arXiv:2203.10066

\bibitem[{{Papaloizou} \& {Pringle}(1983)}]{PapaloizouPringle1983}
{Papaloizou}, J.~C.~B. \& {Pringle}, J.~E. 1983, \mnras, 202, 1181

\bibitem[{{Papaloizou} \& {Terquem}(1995)}]{PapaloizouTerquem1995}
{Papaloizou}, J. C.~B. \& {Terquem}, C. 1995, \mnras, 274, 987

\bibitem[{{Pinilla} {et~al.}(2018){Pinilla}, {Tazzari}, {Pascucci}, {Youdin}, {Garufi}, {Manara}, {Testi}, {van der Plas}, {Barenfeld}, {Canovas}, {Cox}, {Hendler}, {P{\'e}rez}, \& {van der Marel}}]{Pinilla+2018}
{Pinilla}, P., {Tazzari}, M., {Pascucci}, I., {et~al.} 2018, \apj, 859, 32

\bibitem[{{Pinte} {et~al.}(2009){Pinte}, {Harries}, {Min}, {Watson}, {Dullemond}, {Woitke}, {M{\'e}nard}, \& {Dur{\'a}n-Rojas}}]{Pinte+2009}
{Pinte}, C., {Harries}, T.~J., {Min}, M., {et~al.} 2009, \aap, 498, 967

\bibitem[{{Pinte} {et~al.}(2006){Pinte}, {M{\'e}nard}, {Duch{\^e}ne}, \& {Bastien}}]{Pinte+2006}
{Pinte}, C., {M{\'e}nard}, F., {Duch{\^e}ne}, G., \& {Bastien}, P. 2006, \aap, 459, 797

\bibitem[{{Pinte} {et~al.}(2018){Pinte}, {Price}, {M{\'e}nard}, {Duch{\^e}ne}, {Dent}, {Hill}, {de Gregorio-Monsalvo}, {Hales}, \& {Mentiplay}}]{Pinte+2018}
{Pinte}, C., {Price}, D.~J., {M{\'e}nard}, F., {et~al.} 2018, \apjl, 860, L13

\bibitem[{{Poblete} {et~al.}(2020){Poblete}, {Calcino}, {Cuello}, {Mac{\'\i}as}, {Ribas}, {Price}, {Cuadra}, \& {Pinte}}]{Poblete+2020}
{Poblete}, P.~P., {Calcino}, J., {Cuello}, N., {et~al.} 2020, \mnras, 496, 2362

\bibitem[{{Poblete} {et~al.}(2019){Poblete}, {Cuello}, \& {Cuadra}}]{Poblete+2019}
{Poblete}, P.~P., {Cuello}, N., \& {Cuadra}, J. 2019, \mnras, 489, 2204

\bibitem[{{Price}(2007)}]{splash}
{Price}, D.~J. 2007, \pasa, 24, 159

\bibitem[{{Price} {et~al.}(2018{\natexlab{a}}){Price}, {Cuello}, {Pinte}, {Mentiplay}, {Casassus}, {Christiaens}, {Kennedy}, {Cuadra}, {Sebastian Perez}, {Marino}, {Armitage}, {Zurlo}, {Juhasz}, {Ragusa}, {Laibe}, \& {Lodato}}]{Price+2018}
{Price}, D.~J., {Cuello}, N., {Pinte}, C., {et~al.} 2018{\natexlab{a}}, \mnras, 477, 1270

\bibitem[{{Price} {et~al.}(2018{\natexlab{b}}){Price}, {Wurster}, {Tricco}, {Nixon}, {Toupin}, {Pettitt}, {Chan}, {Mentiplay}, {Laibe}, {Glover}, {Dobbs}, {Nealon}, {Liptai}, {Worpel}, {Bonnerot}, {Dipierro}, {Ballabio}, {Ragusa}, {Federrath}, {Iaconi}, {Reichardt}, {Forgan}, {Hutchison}, {Constantino}, {Ayliffe}, {Hirsh}, \& {Lodato}}]{Price+2018-phantom}
{Price}, D.~J., {Wurster}, J., {Tricco}, T.~S., {et~al.} 2018{\natexlab{b}}, \pasa, 35, e031

\bibitem[{{Rabago} {et~al.}(2023){Rabago}, {Zhu}, {Lubow}, \& {Martin}}]{Rabago+2023}
{Rabago}, I., {Zhu}, Z., {Lubow}, S., \& {Martin}, R.~G. 2023, arXiv e-prints, arXiv:2310.00459

\bibitem[{{Rafikov}(2002)}]{Rafikov2002}
{Rafikov}, R.~R. 2002, \apj, 569, 997

\bibitem[{{Ragusa} {et~al.}(2017){Ragusa}, {Dipierro}, {Lodato}, {Laibe}, \& {Price}}]{Ragusa+2017}
{Ragusa}, E., {Dipierro}, G., {Lodato}, G., {Laibe}, G., \& {Price}, D.~J. 2017, \mnras, 464, 1449

\bibitem[{{Ragusa} {et~al.}(2018){Ragusa}, {Rosotti}, {Teyssandier}, {Booth}, {Clarke}, \& {Lodato}}]{Ragusa+2018}
{Ragusa}, E., {Rosotti}, G., {Teyssandier}, J., {et~al.} 2018, \mnras, 474, 4460

\bibitem[{{Rein} \& {Liu}(2012)}]{rebound}
{Rein}, H. \& {Liu}, S.~F. 2012, \aap, 537, A128

\bibitem[{{Rein} \& {Spiegel}(2015)}]{ias15}
{Rein}, H. \& {Spiegel}, D.~S. 2015, \mnras, 446, 1424

\bibitem[{{Reipurth} {et~al.}(2014){Reipurth}, {Clarke}, {Boss}, {Goodwin}, {Rodr{\'\i}guez}, {Stassun}, {Tokovinin}, \& {Zinnecker}}]{Reipurth+2014}
{Reipurth}, B., {Clarke}, C.~J., {Boss}, A.~P., {et~al.} 2014, in Protostars and Planets VI, ed. H.~{Beuther}, R.~S. {Klessen}, C.~P. {Dullemond}, \& T.~{Henning}, 267--290

\bibitem[{{Ribas} {et~al.}(2015){Ribas}, {Bouy}, \& {Mer{\'\i}n}}]{Ribas2015}
{Ribas}, {\'A}., {Bouy}, H., \& {Mer{\'\i}n}, B. 2015, \aap, 576, A52

\bibitem[{{Ribas} {et~al.}(2024){Ribas}, {Clarke}, \& {Zagaria}}]{Ribas+2024}
{Ribas}, {\'A}., {Clarke}, C.~J., \& {Zagaria}, F. 2024, \mnras, 532, 1752

\bibitem[{{Siess} {et~al.}(2000){Siess}, {Dufour}, \& {Forestini}}]{Siess+2000}
{Siess}, L., {Dufour}, E., \& {Forestini}, M. 2000, \aap, 358, 593

\bibitem[{{Smallwood} {et~al.}(2021){Smallwood}, {Nealon}, {Chen}, {Martin}, {Bi}, {Dong}, \& {Pinte}}]{Smallwood+2021}
{Smallwood}, J.~L., {Nealon}, R., {Chen}, C., {et~al.} 2021, \mnras, 508, 392

\bibitem[{{Smallwood} {et~al.}(2023){Smallwood}, {Yang}, {Zhu}, {Martin}, {Dong}, {Cuello}, \& {Isella}}]{Smallwood+2023}
{Smallwood}, J.~L., {Yang}, C.-C., {Zhu}, Z., {et~al.} 2023, \mnras, 521, 3500

\bibitem[{{Tanaka} \& {Ward}(2004)}]{TanakaWard2004}
{Tanaka}, H. \& {Ward}, W.~R. 2004, \apj, 602, 388

\bibitem[{{Terquem}(1998)}]{Terquem+1998}
{Terquem}, C. E.~J.~M.~L.~J. 1998, \apj, 509, 819

\bibitem[{{Thomas} {et~al.}(2023){Thomas}, {Rodgers}, {van der Bliek}, {Doppmann}, {Bouvier}, {Salvo}, {Beuzit}, \& {Rigaut}}]{Thomas+2023}
{Thomas}, S.~J., {Rodgers}, B., {van der Bliek}, N.~S., {et~al.} 2023, \aj, 165, 135

\bibitem[{{van Albada}(1968)}]{VanAlbada1968}
{van Albada}, T.~S. 1968, \bain, 20, 47

\bibitem[{{Verrier} \& {Evans}(2009)}]{VerrierEvans2009}
{Verrier}, P.~E. \& {Evans}, N.~W. 2009, \mnras, 394, 1721

\bibitem[{{Vides} {et~al.}(2023){Vides}, {Sallum}, {Eisner}, {Skemer}, \& {Murray-Clay}}]{Vides+2023}
{Vides}, C., {Sallum}, S., {Eisner}, J., {Skemer}, A., \& {Murray-Clay}, R. 2023, arXiv e-prints, arXiv:2310.02241

\bibitem[{{von Zeipel}(1910)}]{vonZeipel1910}
{von Zeipel}, H. 1910, Astronomische Nachrichten, 183, 345

\bibitem[{{Weingartner} \& {Draine}(2001)}]{WeingartnerDraine2001}
{Weingartner}, J.~C. \& {Draine}, B.~T. 2001, \apj, 563, 842

\bibitem[{{W{\"o}lfer} {et~al.}(2023){W{\"o}lfer}, {Facchini}, {van der Marel}, {van Dishoeck}, {Benisty}, {Bohn}, {Francis}, {Izquierdo}, \& {Teague}}]{Wolfer+2023}
{W{\"o}lfer}, L., {Facchini}, S., {van der Marel}, N., {et~al.} 2023, \aap, 670, A154

\bibitem[{{Young} {et~al.}(2023){Young}, {Stevenson}, {Nixon}, \& {Rice}}]{Young+2023}
{Young}, A.~K., {Stevenson}, S., {Nixon}, C.~J., \& {Rice}, K. 2023, \mnras, 525, 2616

\end{thebibliography}

\clearpage

\begin{appendix}

\section{Data combination}
\label{app : data combination}

The data on which this paper is based come from a combination of data from two observational programs. A first \lang{dataset} (OLD) has been acquired in the context of the ALMA programm $2013.1.00498.S$ in 2013, and \lang{analysed} in details in \lon. A second one has been acquired in end-2021 with longer baselines (LB) being covered during the ALMA programm $2021.1.01137.S$. A third \lang{dataset} has been acquired mid-2022 in the context of the same ALMA programm and covering shorter baselines (SB).

The data reduction and calibration are described in Section \ref{subsec : data} and we add details about the self-calibration process of the data in the following.
The three raw \lang{datasets} were downloaded from the ALMA archive and calibrated using the available calibration scripts. At first, we reduced and self-calibrated each \lang{dataset} independently. The line channels were flagged and the SPW averaged together to create continuum MS. Using the \textit{gaincal} and \textit{applycal}, self-calibration rounds in phase were conducted on the continuum data. The first calibration tables were created for combined SPWs, combined \lang{polarisation} and a infinite solution interval to maximise the S/N. In the following rounds, the SPWs, \lang{polarisations} were progressively decombined and the solution interval progressively turned down to the integration time. Between each rounds, the RMS was measured on images produced with \textit{tclean}.

We ensured that the RMS was decreasing between the rounds without any decrease of the total flux and peak intensity. 

Second, we created a common model for the three resulting continuum \lang{datasets} using the \textit{tclean} task. Before doing so, the observations were centred together with the \textit{phaseshift} task using the centre of the cavity as a common reference. The different MS were also rescaled to a common flux reference using the DSHARP utilities \footnote{\url{https://almascience.eso.org/almadata/lp/DSHARP/}}. From this model, additional self-calibration rounds in phase were performed on the individual MS. It ensured that the three \lang{datasets} were properly centred together. The calibration were stopped before the RMS stopped decreasing between the rounds. The \textit{gaincal} parameters were turned down to a solution interval of two times the integration time, combined SPWs and independent \lang{polarisations}. The final continuum image was created using \textit{tclean} from the three resulting MS of the previous process.

Third, $^{12}$CO channels were separated from the original MS. The continuum level was subtracted from each MS using the task \textit{uvcontsub}. The resulting gas MS were scaled to a common flux reference using the same correction factors that the continuum MS. The calibration tables of the previous self-calibration rounds were applied to the corresponding gas MS. Finally, the $^{12}$CO datacube was created using the \textit{tclean} task, that ensures a spectral smoothing of the MS to a common spectral resolution of $0.5$ km\,s$^{-1}$.

\begin{figure}[h]
\centering
    \includegraphics[width=\columnwidth, trim={2cm 5cm 2cm 5cm},clip]{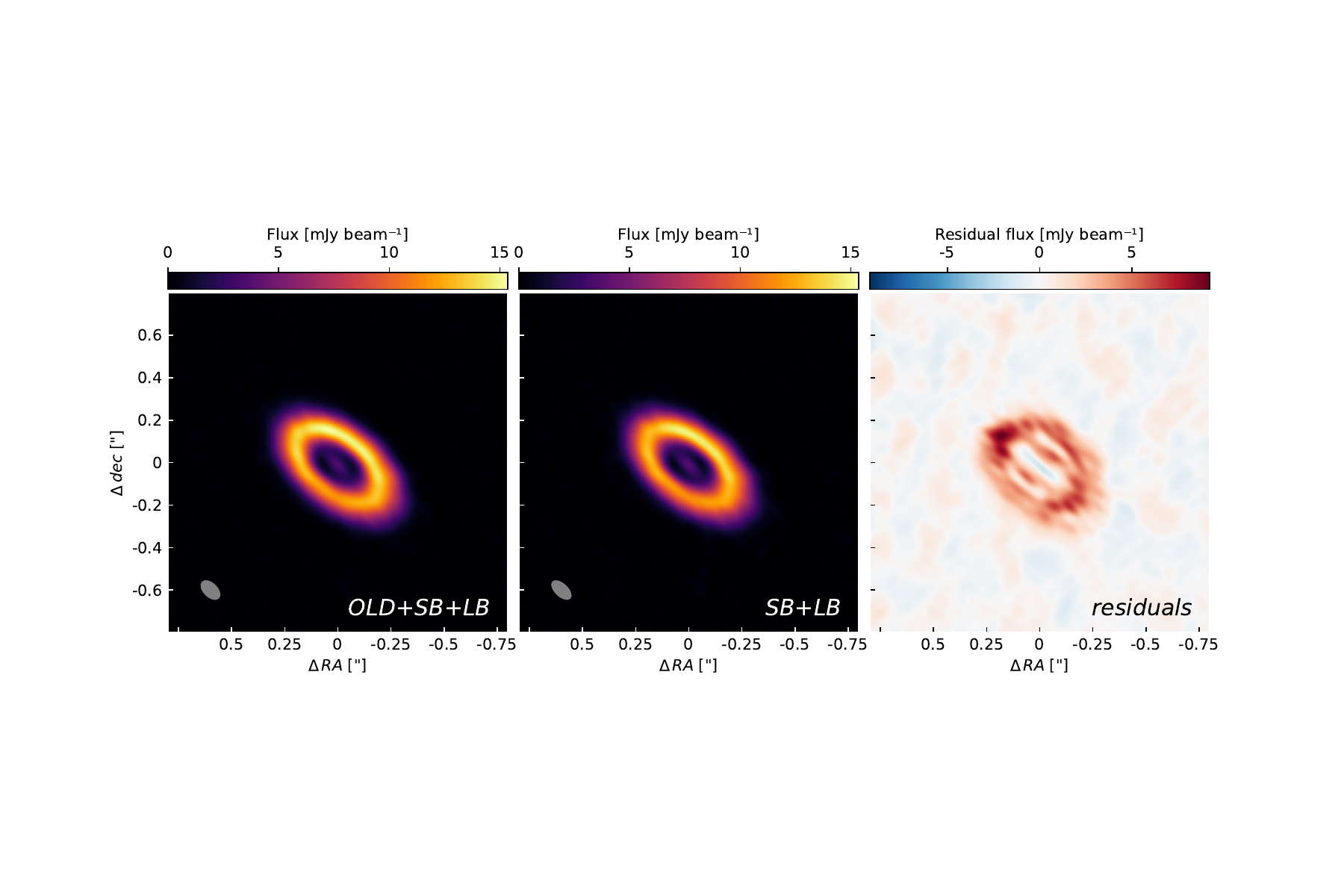}
     \includegraphics[width=\columnwidth, trim={2cm 5cm 2cm 5cm},clip]{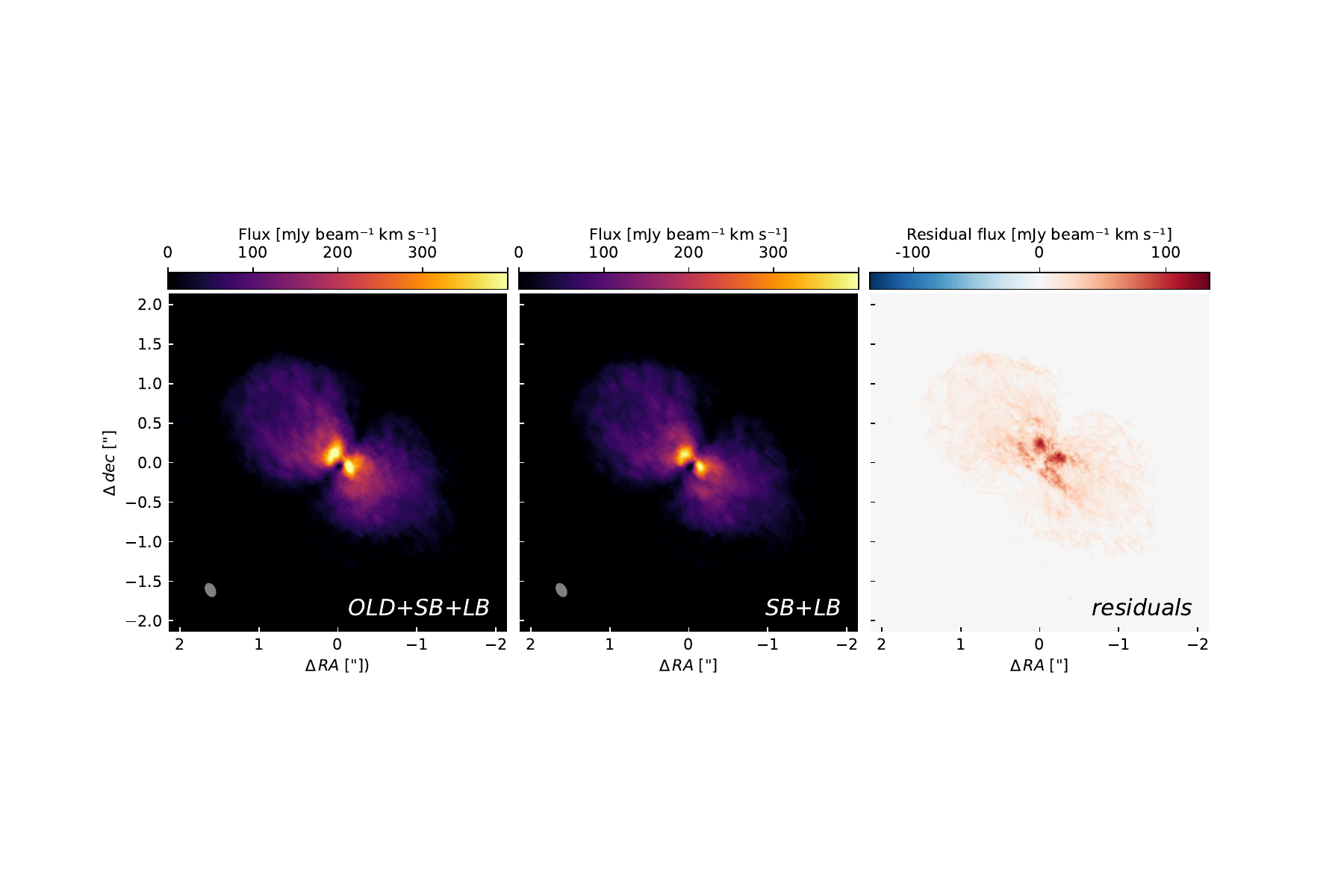}
    \caption{Comparison of the ALMA band 6 continuum (top) and $^{12}$CO (2-1) moment 0 (bottom) images resulting from the different data-sets combination. The left image in each panel has been created from SB, LB and the data-set used in \lon. The middle image in each panel results from the combination of SB and LB only. The images on the right are residual maps, which are uniformly positive, highlighting the benefits of the data combination.}
    \label{fig:reduc_comp}
\end{figure}

Figure \ref{fig:reduc_comp} shows how the combination of the \lang{datasets} affects the final continuum image and $^{12}$CO moment 0 images. The combination of the \lang{dataset} used in \lon\ with our newer data enhances the final S/N compared to our data (SB +LB) alone. This \lang{S/N increase allows the continuum to be imaged} with a more robust parameter controlling the Briggs weighting when applying the \textit{tclean} task, and thus for more angular resolution than in previous studies.

\FloatBarrier

\clearpage
\section{Channel maps}
\label{app : channel maps}
\FloatBarrier
\
\begin{figure*}[b]
 \centering
     \includegraphics[width=\textwidth, trim={3cm 1.8cm 2cm 1.8cm}, clip]{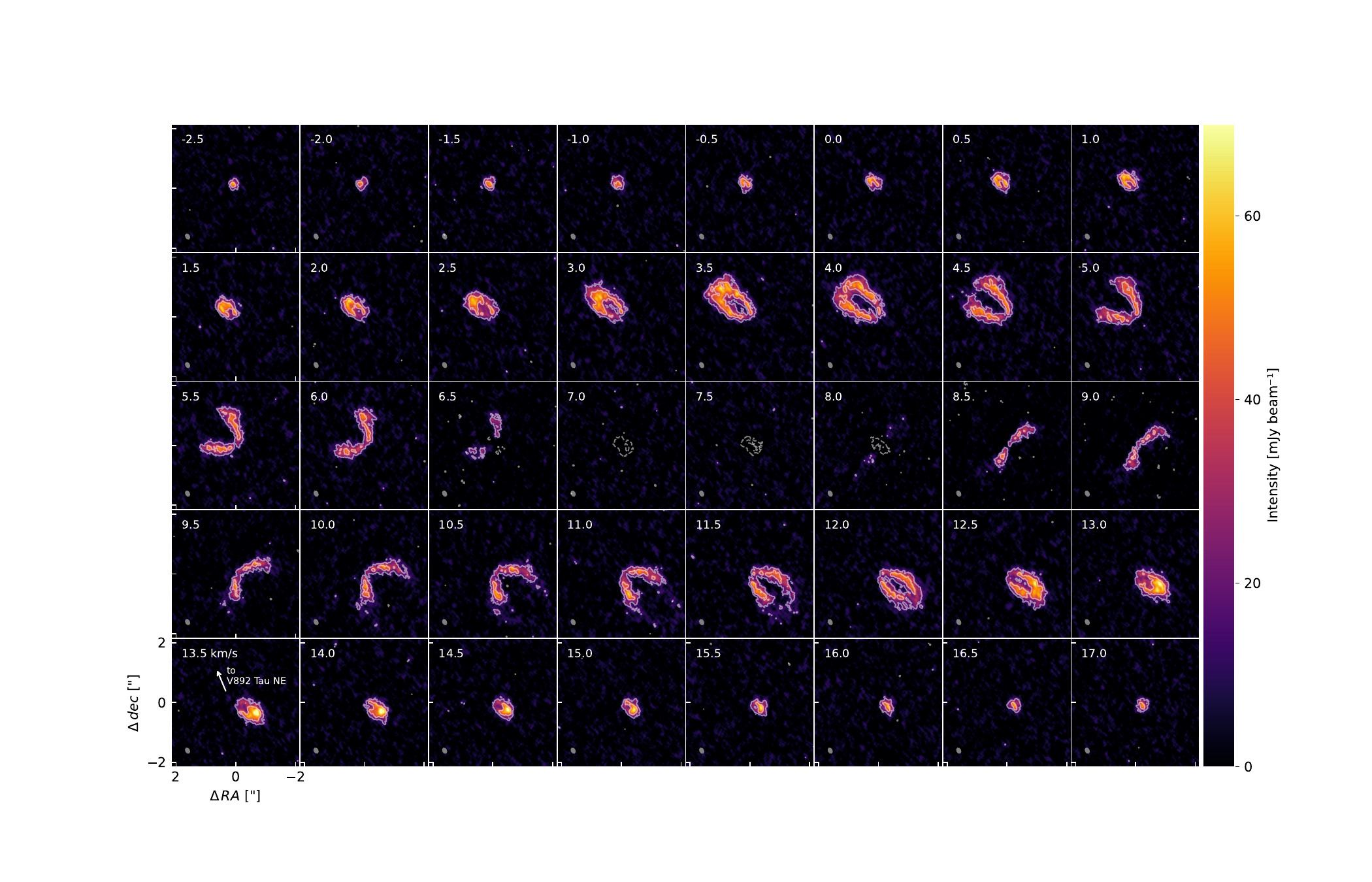}
     \caption{$^{12}$CO channel maps of V892~Tau. The contours shows the $[3,7]\sigma$ emission levels. The dashed line represents corresponding negative emission levels. The velocity of the emission channels is indicated at the top left. The synthesised beam is plotted at the bottom left of each channel.}
     \label{fig:channel_maps}
\end{figure*}

\clearpage
\section{Gas emission fitting with Discminer}
\label{app : Discminer}

\mdy{\subsection{Initial conditions and results}}
\label{subapp : model}

The initial values of the MCMC sampling were chosen in agreement with the values derived in \lon when possible, namely for the orientation and kinematic parameters. The initial values of the parameters defining the emission surfaces were set to initialise a thin disc, in agreement with the observations. The rest of the initial values were put in the middle of the prior range or at typical values. \mdy{The reference radius $r_0$ was taken as $100$ au.}
Priors were defined to limit the sampling of the parameter space and the computational cost. The range of priors for all the parameters was set to a limit the disc to a reasonable extent.
Following the need of a precise model, we initialised $256$ walkers with a total of $40000$ steps. \mdy{The chains reached convergence well before the
end of the sampling and the resulting auto-correlation length was
$106$ steps.} The first $6300$ steps are considered as a burn-in phase where the walkers explore the parameter space. After that, the walkers reached convergence and remained concentrated around the final value.
The best-fit parameters are taken as the median of the posterior distributions. The errorbars are the $16$ and $84$ percentiles of the posterior distributions. Table \ref{table:discminer_fit} \lang{summarises} the results of the fit.

\begin{table}
    \centering
    \caption{Parameters of the best-fit model found by {\sc Discminer} to the $^{12}$CO (2-1) emission.}
 \begin{tabular}{c c c} 
Parameter & \mdy{Search range} & Best-fit \\ [0.5ex] 
 \hline \hline
$i$ & $[20;80]^\circ$ & $54.7 ^{+0.07\circ}_{-0.40} $\\
$PA$ & $[20;80]^\circ$ & $53.6 ^{+0.07\circ}_{-0.11}$\\
$x_c$ & \mdy{$[-50;50]$ au} & $4.26 \pm ^{+0.16}_{-0.82}$ au\\
$y_c$ & \mdy{$[-50;50]$ au} & $0.52 \pm ^{+0.22}_{-0.17}$ au\\
 & & \\
$M_{*}$ & $[5;7]$ M$_{\odot}$ & $5.99 ^{+0.03}_{-0.01}$ M$_{\odot}$\\
$v_{sys}$ & $[7;9]$ km\,s$^{-1}$ & $7.86 ^{+0.01}_{-0.01}$ km\,s$^{-1}$\\
 & & \\
$z_{0,low}$ & $[0;30]$ au & $13.7 ^{+0.55}_{-1.34}$ au\\
$p_{low}$  & $[0;3]$ & $1.35 ^{+0.02}_{-0.06}$\\
$q_{low}$  & $[0;3]$ & $1.31 ^{+0.02}_{-0.07}$\\
$R_{b,low}$  & $[100;400]$ au & $217 ^{+2}_{-2}$ au\\
 & & \\
$z_{0,up}$ & $[0;30]$ au & $14.7 ^{+0.59}_{-1.34}$ au\\
$p_{up}$  & $[0;3]$ & $1.21 ^{+0.01}_{-0.03}$\\
$q_{up}$  & $[0;3]$ & $0.97 ^{+0.01}_{-0.02}$\\
$R_{b,up}$  & $[100;400]$ au & $221 ^{+4}_{-4}$ au\\
 & & \\
$I_{0}$ & \mdy{$[0;1000]$ Jy pixel$^{-1}$} & $3.80 ^{+0.52}_{-0.28}$ Jy \mdy{pixel}$^{-1}$\\
$p_{I}$  & \mdy{$[-10;10]$} & $-2.10 ^{+0.05}_{-0.03}$\\
$q_{I}$  & \mdy{$[0;5]$} & $2.16 ^{+0.02}_{-0.02}$\\
$R_{out}$  & $[150;400]$ au & $223 ^{+4}_{-1}$ au\\
 & & \\
$L_{w,0}$ & \mdy{$[0.005;5]$ km\,s$^{-1}$} & $0.19 ^{+0.01}_{-0.01}$ km\,s$^{-1}$\\
$p_{L_w}$  & \mdy{$[-5;5]$} & $-0.41 ^{+0.05}_{-0.03}$\\
$q_{L_w}$  & \mdy{$[-5;5]$} & $-0.53 ^{+0.03}_{-0.03}$\\
 & & \\
$L_{s,0}$ & \mdy{$[0.005;20]$} & $1.26 ^{+0.03}_{-0.02}$\\
$p_{L_s}$  & \mdy{$[-5;5]$} & $-0.01 ^{+0.04}_{-0.02}$\\
$q_{L_s}$  & $0.00$ & Not fitted\\
\end{tabular}
\tablefoot{\mdy{The reference radius is taken as $r_0=100$ au and the pixel scale is $0.01\arcsec$.}}
\label{table:discminer_fit}
\end{table}

\FloatBarrier

\rmvd{\subsection{Flipped model}
\label{subapp : flipped model}

With the same method as described in Sections \ref{subsec:discminer} and \ref{app : Discminer}, we ran an MCMC sampling to fit the disc observations with a model with a negative inclination sign. This makes the closest side to the observer being the NW side of the disc. The sampling was run with the same other initial conditions and boundaries for the parameters as the original run.

The resulting fitted values were similar to the original model apart from the inclination sign. It lead to a disc model quite identical to the original one in the outer disc. Differences can be seen between the two models in the central parts of the disc, for which our data are not reliable due to cloud absorption. For reasons detailed in Section \ref{app : incl_sign}, the fitted linewidth residuals could trace the nearside of the disc. However, the two models resulted in the same patterns of residuals, as shown by the Figure \ref{fig:lw_res} below.
This could be explained by the disc being significantly flat and its lower emission surface not being detected, which prohibits us to conclude on the true orientation of the disc. }

\begin{figure*}
\centering
\begin{center}
    \includegraphics[width=1.0\textwidth, trim={3cm 1.8cm 2cm 1.8cm}, clip]{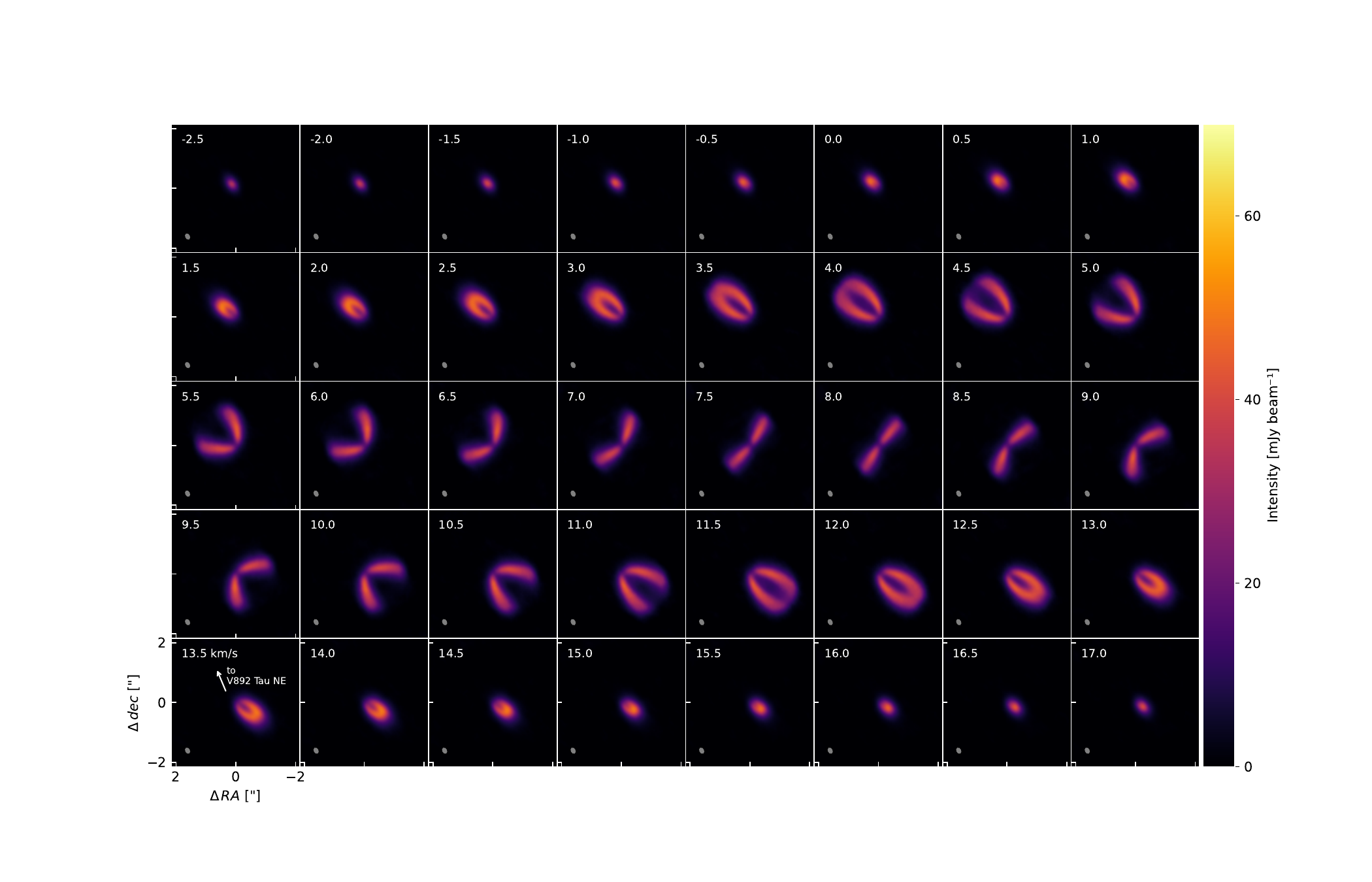}
    \caption{Same as Figure \ref{fig:channel_maps} for {\sc Discminer} best-fit channel maps.}
    \label{fig:discminer_channel_maps}
\end{center}
\end{figure*}

\FloatBarrier

\mdy{\subsection{Rotation curve}}
\label{subapp : rotation curve}

\begin{figure}
\centering
\begin{center}
    \includegraphics[width=1\columnwidth, trim={0cm 3cm 0cm 3cm},clip]{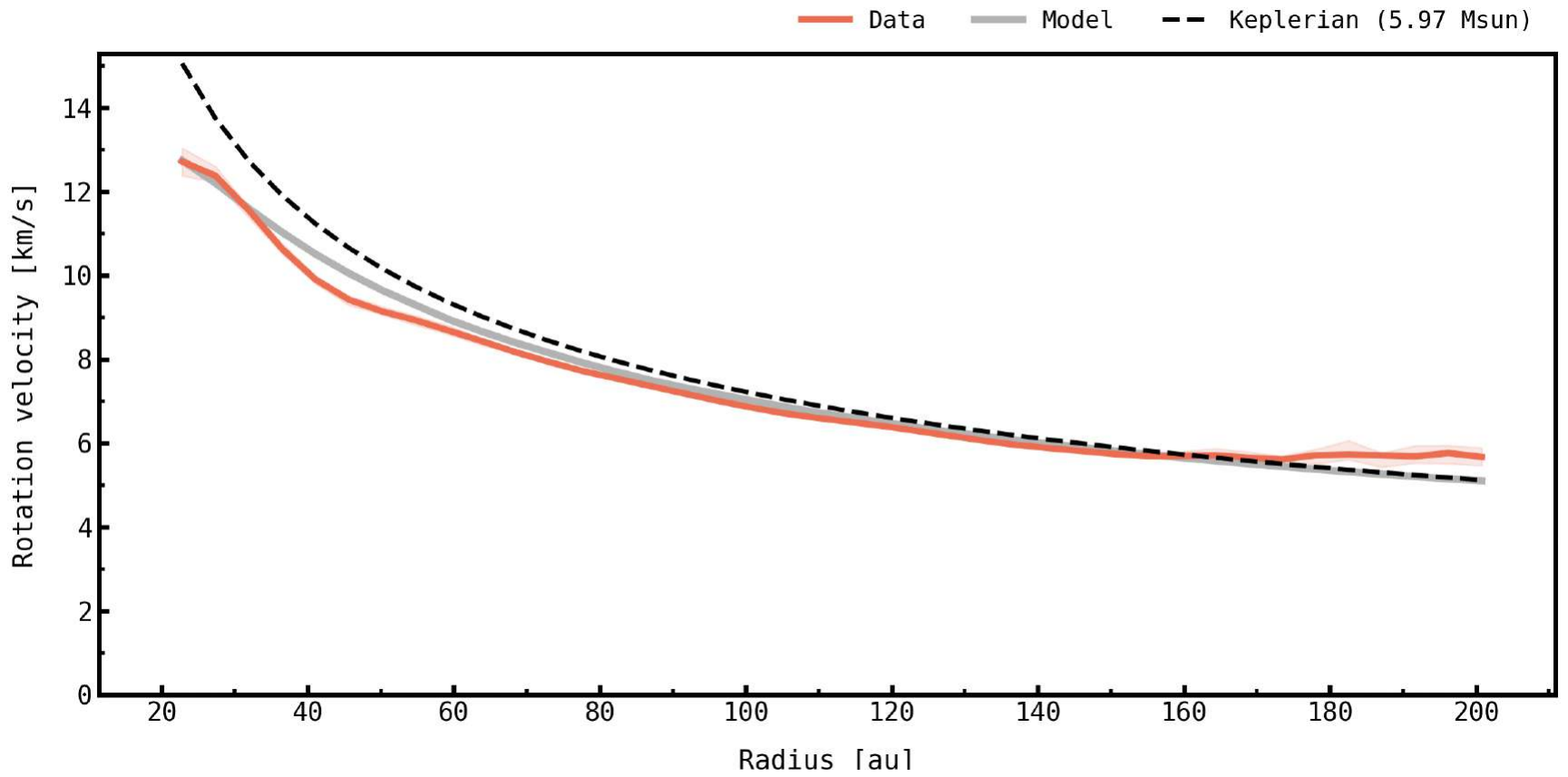}
    \includegraphics[width=1\columnwidth, trim={0cm 3cm 0cm 3cm},clip]{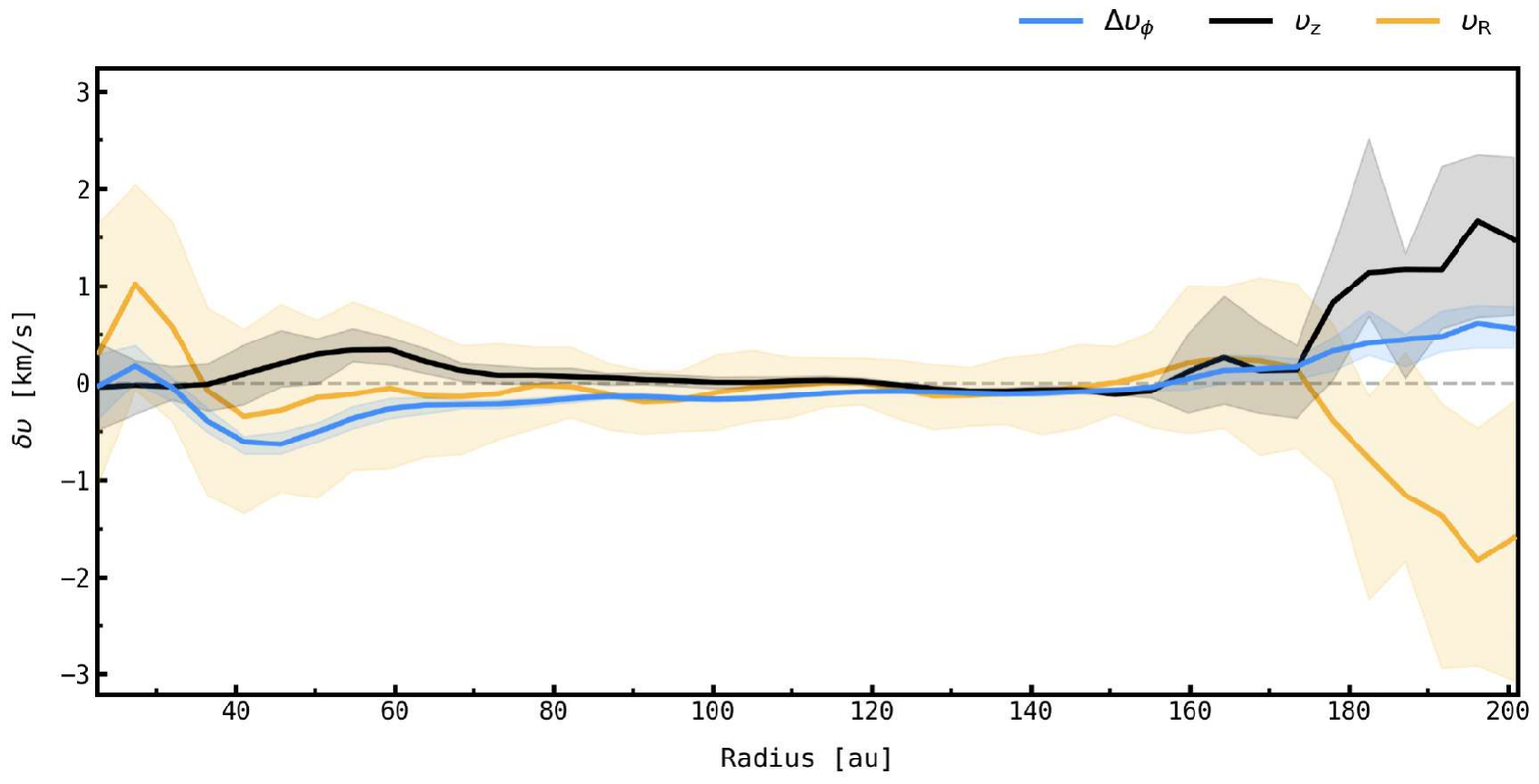}
    \caption{\mdy{Radial velocity profiles of the $^{12}$CO (2-1) emission of V892~Tau fitted by {\sc Discminer}. \textit{Top} : rotation curve. \textit{Bottom} : Offset to Keplerian rotation (blue), radial (yellow) and vertical (black) velocities as a function of radius.}}
    \label{fig:discminer_rotation_curve}
\end{center}
\end{figure}

{\sc Discminer} allows velocity radial profiles from the results of the fitting procedure \lang{to be computed}. The rotation curve of the data can be compared to the best-fit model velocity profile. From the resulting residuals and and their projection in the skyplane, the velocity deviation to a Keplerian profile $\Delta v_{\phi}$, the velocity of radial flows $v_r$ and the velocity of vertical flows $v_z$ can all be computed as a function of radius.
In Figure \ref{fig:discminer_rotation_curve}, we show these quantities from our best-fit model to the $^{12}$CO (2-1) emission of V892~Tau. The disc is dominated by Keplerian rotation, although slight offsets can be observed in the innermost and outermost parts of the disc. In the inner parts, this mainly translates into a rise of $v_r$, potentially triggered by interactions with the inner binary. The outer disc is characterised by inward radial flows, vertical motions of the gas and slight deviations from the keplerian rotation speed. These patterns are consistent with interactions with V892~Tau~NE and support the hypothesis of a bound triple system.

%
\clearpage
\section{Manual rotation of the disc before post-processing}
\label{app : manual move back}

\begin{minipage}{\textwidth}
\centering
\captionof{table}{\mdy{Inclination and position angle (PA) of the observed disc and of the disc at the end of the different simulations.}} 
 \begin{tabular}{c c c c c c} 
\multicolumn{6}{c}{Inclination [$\deg$]} \\
\hline 
   Observations & \textit{ref} & \textit{e05} & \textit{i30} & \textit{i60} & \textit{ei60} \\ 
 \hline \hline

 $54.6\pm1.3$ & $54.6\pm0.4$ & $54.6\pm0.3$ & $52.6\pm0.3$ & $47.2\pm0.5$ & $50.8\pm0.8$ \\
 
  & & \\

\multicolumn{6}{c}{Position Angle [$\deg$]} \\
\hline 
   Observations & \textit{ref} & \textit{e05} & \textit{i30} & \textit{i60} & \textit{ei60} \\ 
 \hline \hline

  $53.0\pm0.7$ & $52.7\pm1.0$ & $53.1\pm0.6$ & $43.4\pm0.3$ & $30.2\pm0.9$ & $34.2\pm1.5$ \\

\end{tabular}
\tablefoot{Each simulated disc was manually moved back to the inclination and PA of the observed disc before post-processing into synthetic observations. }
\label{table:disc_rotation}
\end{minipage}

\vspace{1cm}
\mdy{Discs in multiple systems are expected to precess due to the gravitational torques applied by misaligned stars \citep{PapaloizouTerquem1995}. We observe that precession in our hydrodynamical models of the disc of V892~Tau : the disc is initialised at its observed inclination and position angle (PA), but these values evolve during the simulation because of the gravitational perturbations induced by the inner binary and the external companion. Values of inclination and PA of the disc at the end of each simulation can be found in Table \ref{table:disc_rotation} (see Fig \ref{fig:simgrid_rho} for a visual representation). However, for a proper comparison with the observations, the simulated discs should have the same orientation as the observed disc. But, it would be too computationally expensive to simulate the disc during a whole precession timescale, at the end of which the disc would have come back to its original PA while the inclination would have slightly changed. We tackle this issue by manually rotating the simulated discs back to their observed values of inclination and PA before going into their post-processing into synthetic observations. This assumes that the disc properties other than inclination and PA, and the interactions with the stars are independent of the disc orientation. We are aware of the discrepancy between the observed orientation of the disc and the orientation at the end of the simulations. But, we argue that this discrepancy is reasonably small, which allows us to work in the context of the previous hypothesis. This way, rotating the disc to the observed orientation will allow for an adequate comparison of the observations and the synthetic data.}

\clearpage
\section{Inclination sign of the disc}
\label{app : incl sign}

\begin{figure*}[b]
\centering
\begin{center}
    \includegraphics[width=1.0\textwidth]{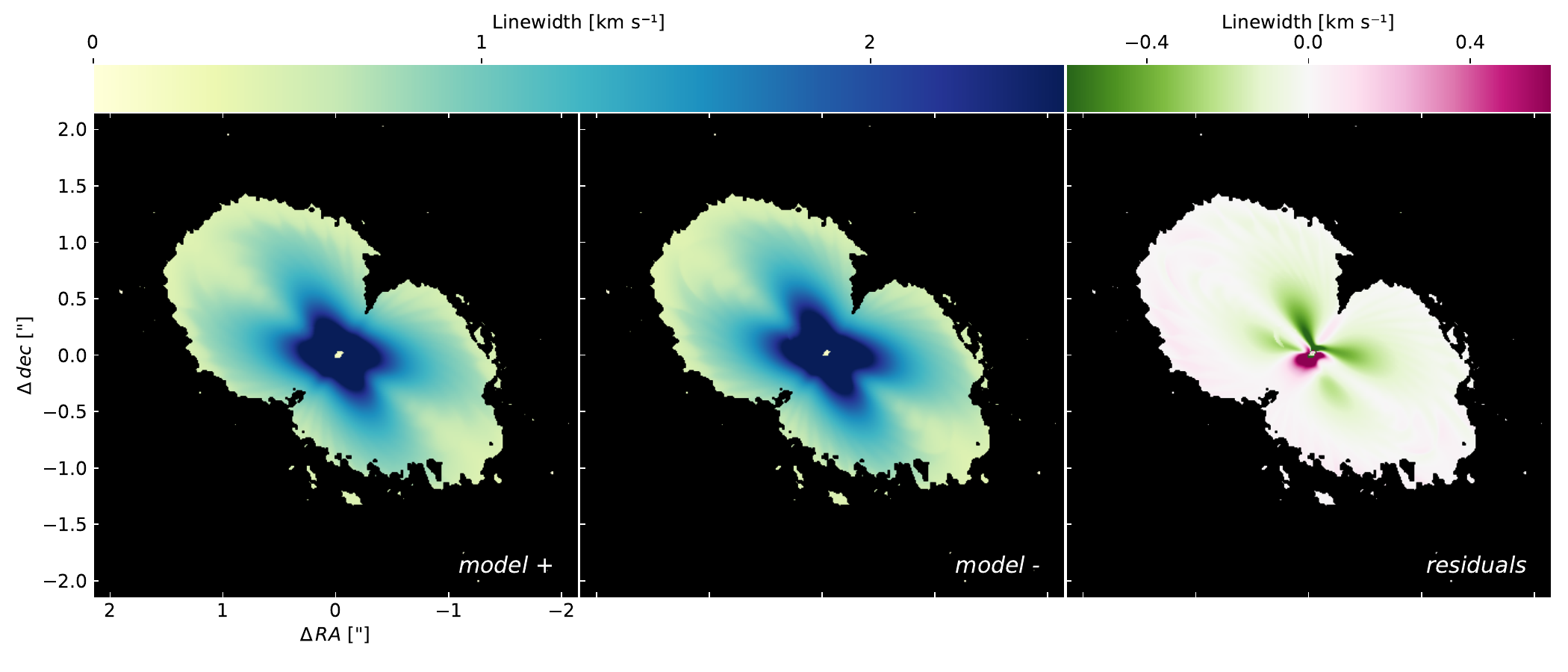}
    \captionof{figure}{Comparison of the fitted linewidth models from {\sc Discminer}. The original model is represented in the left panel, the model with a flipped inclination sign in the middle panel and the residuals between the two are plotted in the right panel. \mdy{A mask defined by the $3\sigma$ contour of the $^{12}$CO (2-1) emission has been applied to the models.}}
    \label{fig:lw_res}
\end{center}
\end{figure*}

\mdy{Usually, it is straightforward to determine the closest side of a disc when its upper and lower emission surfaces are both detected (e.g. IM Lup, \cite{Pinte+2018}). Here and in previous studies of the system, the lower surface of the V892~Tau disc is not clearly detected, making the sign of the inclination uncertain. According to the fitted linewidths maps of the disc displayed on Figure \ref{fig:obs_model_panel} last row, the Keplerian model is able to reproduce the data well. Nonetheless, the linewidths seems larger in the NW side of the disc than in the SE side. When compared to the model, the NW side shows an excess in linewidth (see residual map on Figure \ref{fig:obs_model_panel} last row right panel). Lines are expected to be seen broader in the closest side of the disc to the observer due to the greater contribution of the lower emission surface (e.g. MWC480, \cite{DiscminerII}). Applied to V892~Tau, this could indicate that the NW side of the disc is the closest side to the observer and that the disc inclination is negative.
Trying to confirm that result, we ran an MCMC sampling to fit the disc observations with a model with a negative inclination sign with the same method as described in Sections \ref{subsec:discminer} and \ref{app : Discminer}. In that model, the closest side to the observer is the NW side of the disc. The resulting fitted values were similar to the original model apart from the inclination sign. It lead to a disc model quite identical to the original one in the outer disc, as showed by the linewidth map of each model on Figure \ref{fig:lw_res} below. This could be explained by the disc being significantly flat and its lower emission surface not being detected, which doesn't allow to conclude on the true orientation of the disc. Differences still can be seen between the two models in the central parts of the disc, for which our data are not reliable due to cloud absorption.}

\newpage
\mdy{In general, the far side of the disc is observed with a larger brightness temperature than the near side \citep{Law+2023}, which is in line with our synthetic observations resulting from the simulations where the closest side to the observer is assumed to be the SE side.
According to our $^{12}$CO observations, the NW side is brighter in average than the SE side by approximately 13$\%$. This agreement between the simulations and the observations is in favour of the SE side being the nearside of the CBD.
Moreover, the temperature asymmetry plotted on Figure \ref{fig:sketch_obs} and heading toward the NW cloud also trace the flared far side of the disc, since similar patterns have already been observed in discs with a known orientation (see RXJ~1615 in \cite{Wolfer+2023}, LkCa~15 and HD~34282 in \cite{Law+2023}).
Considering the azimuthal asymmetry in the continuum disc arises from warm dust emission, this emission may come from the visible inner rim of the disc far side \citep{Ribas+2024}. Such a scenario would indicate that the disc near side is the SE side of the disc.}

\mdy{The disc thickness and faint lower surface make the conclusion on the disc true inclination difficult. Nonetheless, we present the previous results as tentative evidences for the SE side of the disc being the closest side to the observer. 
Future scattered light observations or high resolution polarisation maps could help to characterise the dust scattering properties which would bring additional constrains on the disc true orientation.}

\clearpage
\section{Binary-only simulations}
\label{app : binary sims}

We performed additional hydrodynamical SPH simulations of the V892~Tau system. In these simulations, we model the system as a binary system with a disc. The disc parameters are the same as described in Section \ref{subsec:hydro_sim}. Then, we consider different setups where the companion star is ignored, or where the inner binary is merged into a single star with an equivalent mass. These setups are :
\begin{itemize}
    \item the inner binary surrounded by the disc (\textit{IB}).
    
    \item the outer binary initially misaligned of $30\degree$ with respect to the disc (\textit{OB30}).
    
    \item the outer binary initially misaligned of $60\degree$ with respect to the disc (\textit{OB60}).
\end{itemize}

The orbital parameters of the binary star in \textit{IB} are the same as the orbital parameters of the inner binary in the triple system simulations.
The orbital parameters of the binary star in \textit{OB30} are the same as the orbital parameters of the outer binary in \textit{i30}.
The orbital parameters of the binary star in \textit{OB60} are the same as the orbital parameters of the outer binary in \textit{i60}.
These simulations counted $10^6$ SPH particles and ran for a total time of $t=25 $ P$_{out}$, where P$_{out}$ is the outer binary period in the triple system simulations. 

The oscillations of the modelled CBD are triggered both by the inner binary and the outer binary. Yet, the contribution of each component to the dynamics of the disc is unclear : the simulated CBD is found in an intermediate plane between the inner and outer binary orbital planes, its global inclination oscillates without reaching a coplanar nor polar configuration and it precesses non-linearly.
Simulations of the V892~Tau system modelled as a binary star could help to disentangle the dynamical effects of the inner and of the outer binaries on the disc.

\begin{figure*}[t]
\centering
\begin{center}
     \includegraphics[width=0.4\textwidth]{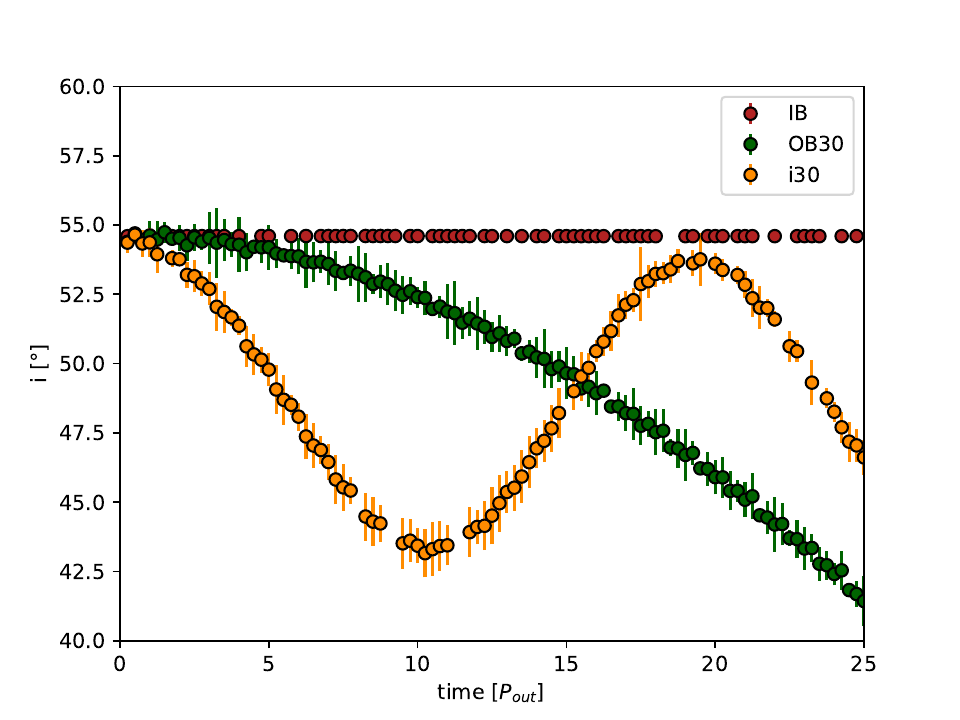}
     \includegraphics[width=0.4\textwidth]{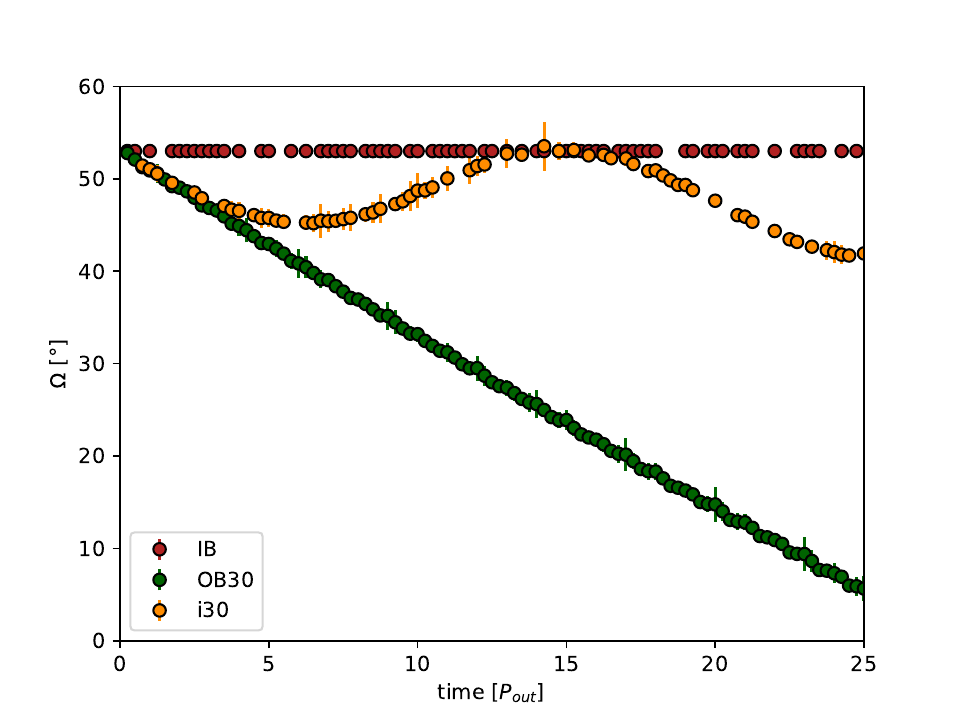}
     \includegraphics[width=0.4\textwidth]{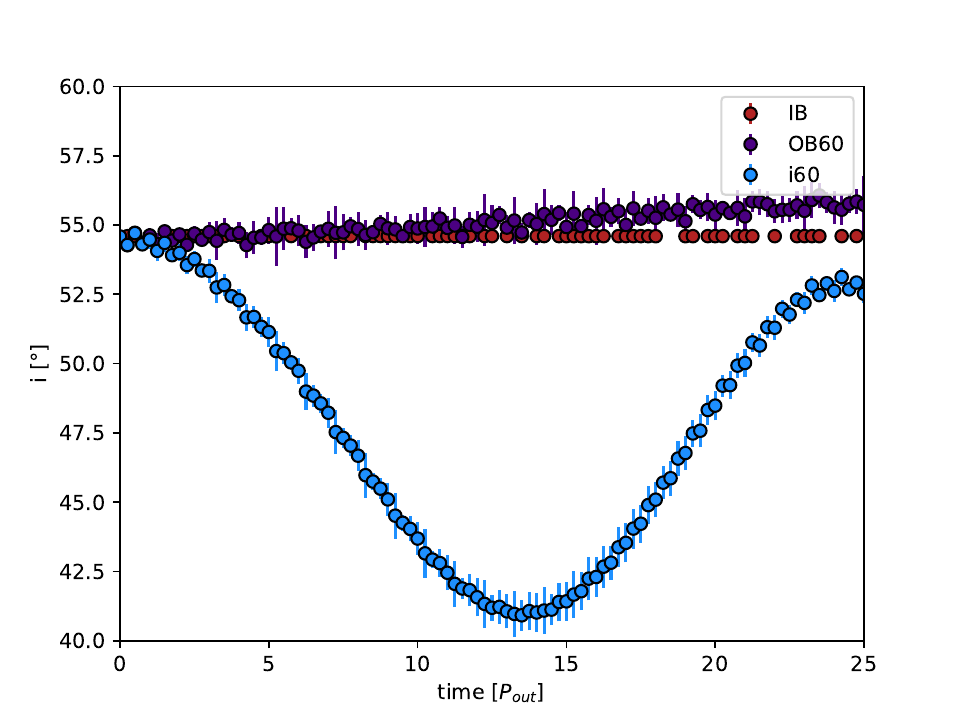}
     \includegraphics[width=0.4\textwidth]{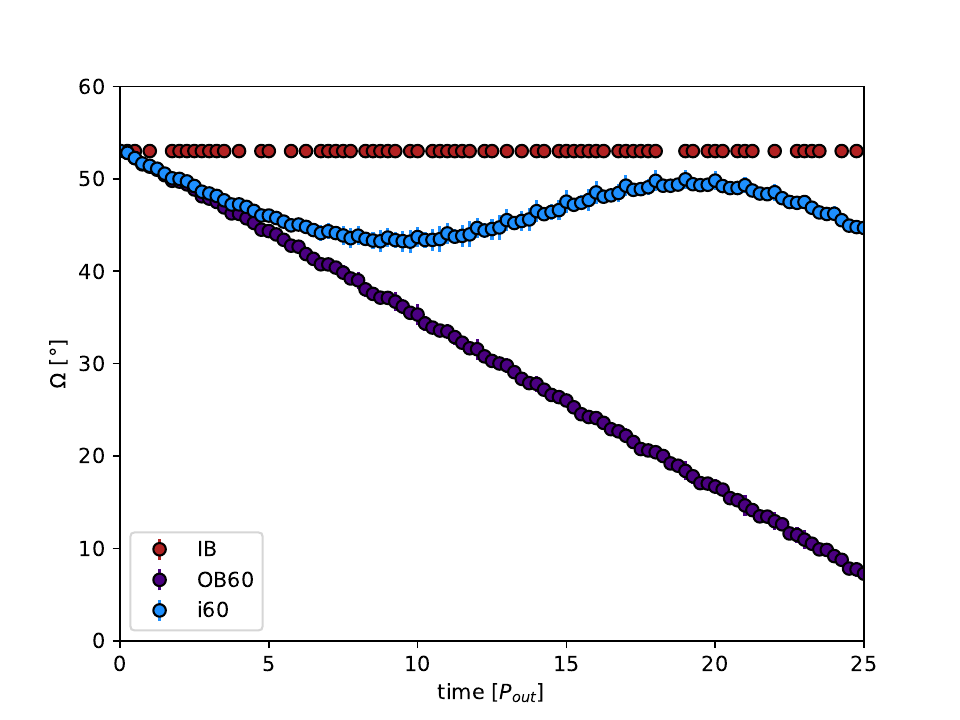}
     \caption{Mean inclination (left) and mean position angle (right) of the disc as a function of time for the binary star setups and the triple systems \textit{i30} and \textit{i60}. The top row shows the discs in \textit{IB, OB30}, and \textit{i30} and the bottom row shows the discs in \textit{IB, OB60}, and \textit{i60}.}
     \label{fig:SPH_comp_binaries}
\end{center}
\end{figure*}

Figure \ref{fig:SPH_comp_binaries} shows the inclination $i$ and the PA $\Omega$ of the disc as a function of time in the simulations \textit{i30}, \textit{i60}, \textit{OB30}, \textit{OB60}, and \textit{IB}.

In the inner binary only simulation, the CBD remains coplanar with the inner binary and its PA remains at the initialised value.
The PA of the disc follows a decreasing trend in all the other simulations. During the very first orbits, the disc has the same linear precession rate in the setups \textit{i30} and \textit{OB30}, and \textit{i60} and \textit{OB60} respectively. Then, the PA of the disc starts to oscillates in the triple system simulations, while the precession stays linear in the outer binary systems \textit{OB30} and \textit{OB60}. The precession slope in those setups is steeper than the average slope in the triple setups, but asymptotically reproduces the precession rate of the disc in triple systems during the first orbits of the companion. The precession rate is similar in the \textit{OB30} and the \textit{OB60} setups.
We expect that kind of linear precession for binary systems (e.g. \cite{PapaloizouTerquem1995}). From dynamical considerations, we expect the precession rate of the disc to be $t_{prec, 0}= 144$ P$_{out}$ in the setup \textit{OB30} and $t_{prec, 0}= 251$ P$_{out}$ in \textit{OB60} \citep{Terquem+1998, Bate+2000}. We measure a precession rate of $t_{prec}= 189\pm1$ P$_{out}$ and $t_{prec}= 195\pm1$ P$_{out}$ respectively, which is in rough agreement with that prediction. 

We already saw that the disc inclination follows damped oscillations in the triple system setups \textit{i30} and \textit{i60}. In the \textit{OB30} setup, the inclination of the disc drops with time, but that drops is not linear nonetheless. The \textit{OB30} disc could then align with the orbital plane of the binary, but the simulation stopped too early to clarify this statement. 
In the \textit{OB60} model, the inclination of the disc rises steadily in a linear way with a slope of $5.6\times10^{-2}\pm 0.2\times10^{-2}$ P$_{out}$. That particular behaviour could be explained by the disc starting to align with the binary orbital plane. The timescale of this alignment is larger that in the \textit{OB30} case due to the higher mutual inclination of the disc to the orbital plane, which results in a smaller torque applied by the companion on the disc \citep{Bate+2000}. 

The three additional simulations \textit{IB}, \textit{OB30} and \textit{OB60} show that the dynamical behaviour of the V892~Tau disc can not be reproduced by a binary system only. The triplicity of the system, which means both the inner binary and outer binary, is responsible for the inclination oscillations and the peculiar precession of the CBD.

\newpage
\begin{minipage}{\textwidth}
\section{Other N-Body simulations}
\label{app : nbody}
\end{minipage}

\begin{figure*}[!h]
\centering
         \includegraphics[width=0.3\textwidth]{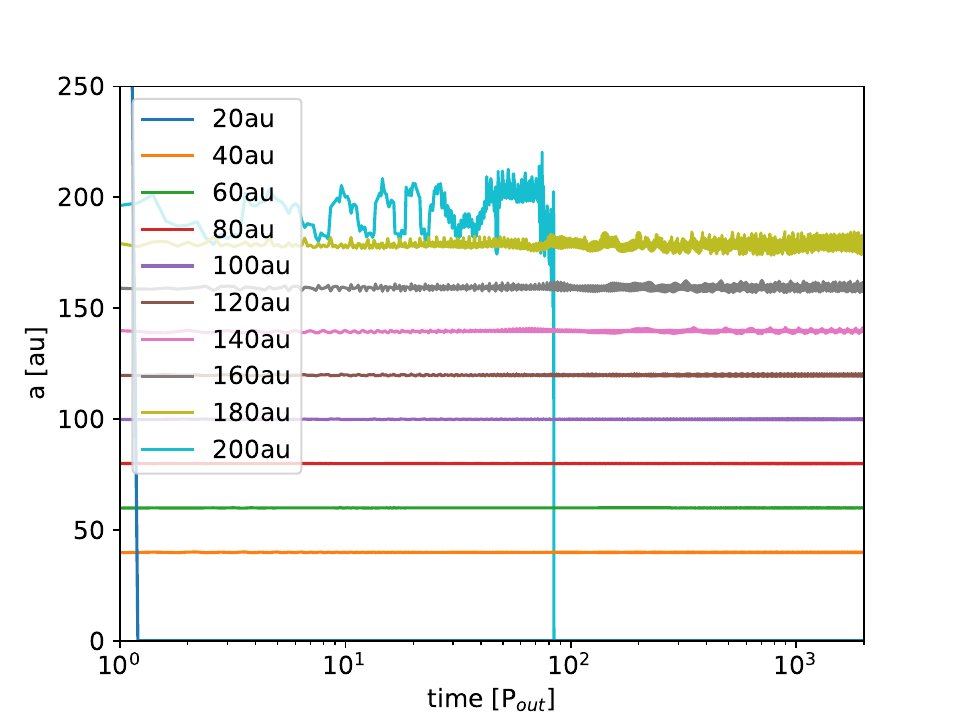}  
        \includegraphics[width=0.3\textwidth]{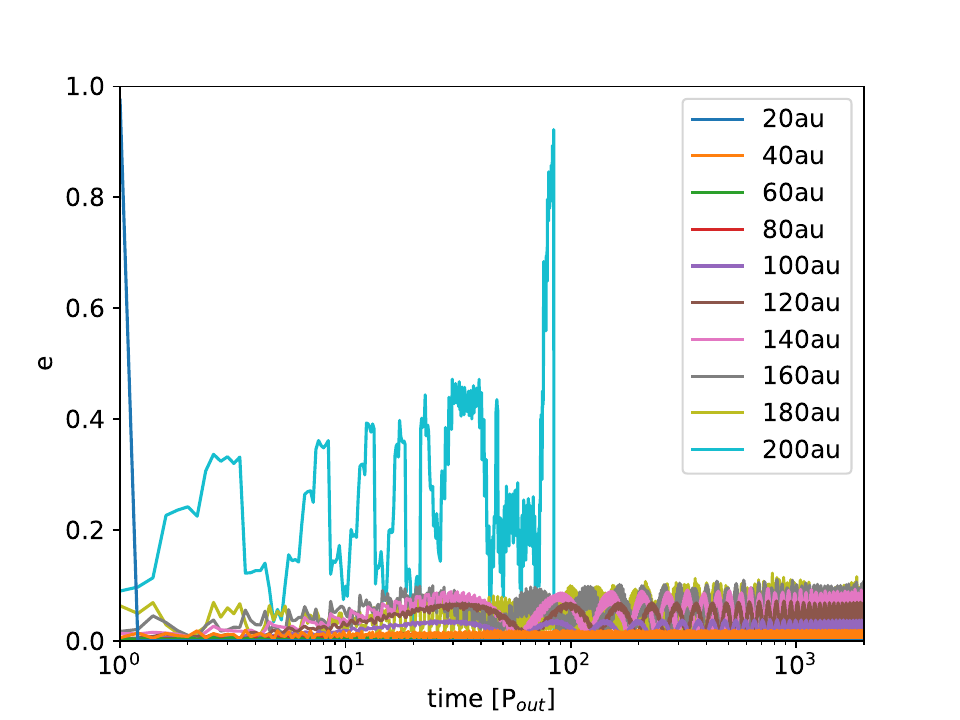}  
        \includegraphics[width=0.3\textwidth]{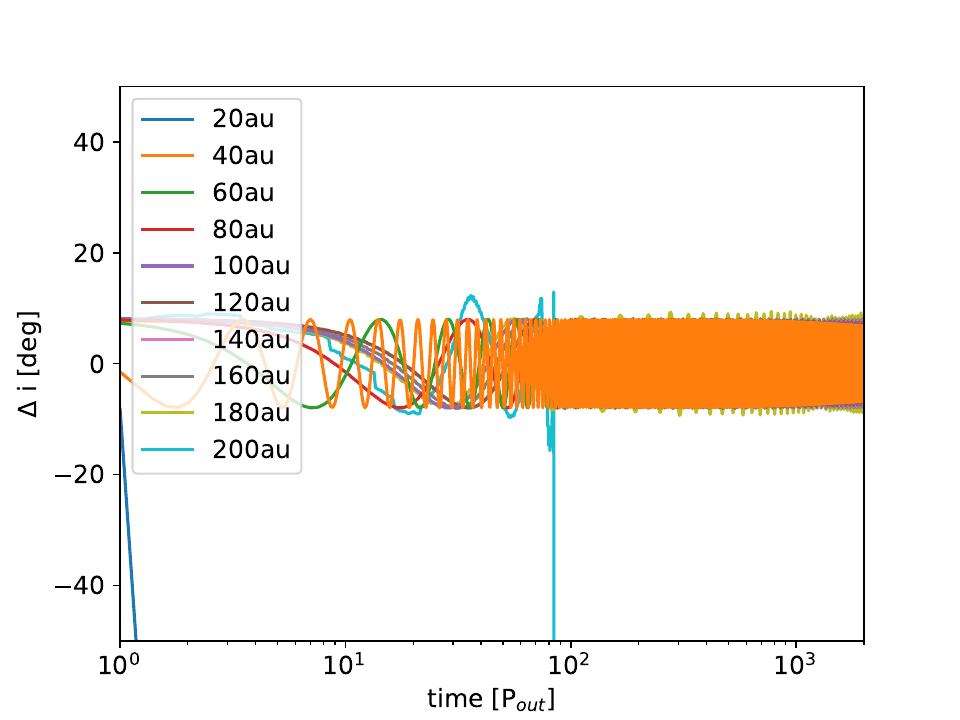}  
    \caption{Same as Figure \ref{fig:nbody} \lang{but} for the \textit{ref} orbital configuration.}
    \label{fig:nbody_ref}
\end{figure*}

\begin{figure*}[!h]
\centering
\begin{center}
         \includegraphics[width=0.3\textwidth]{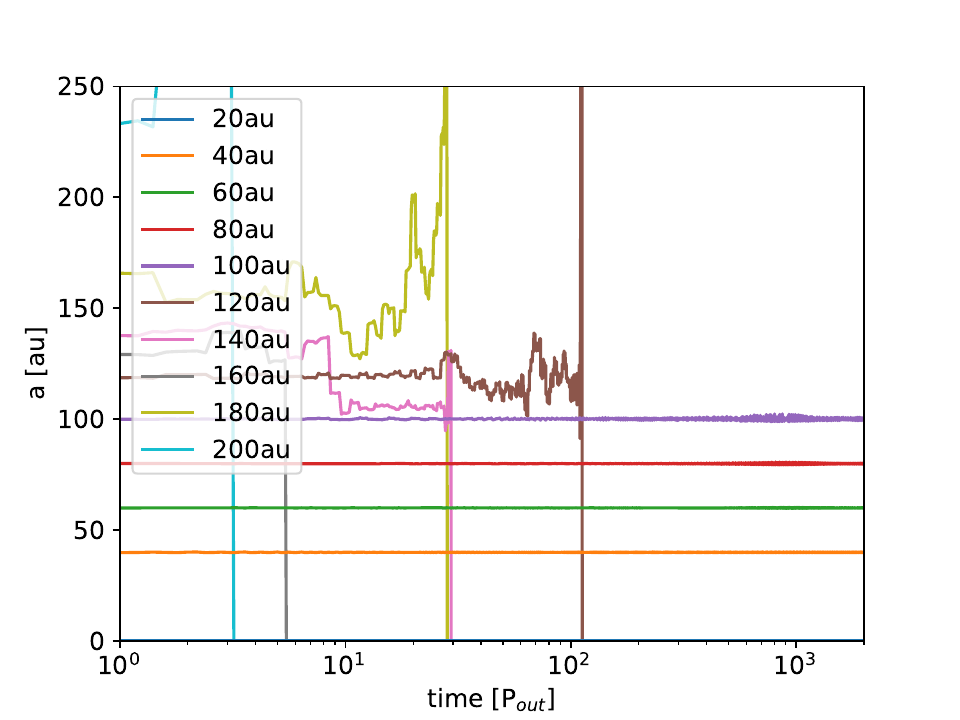}  
        \includegraphics[width=0.3\textwidth]{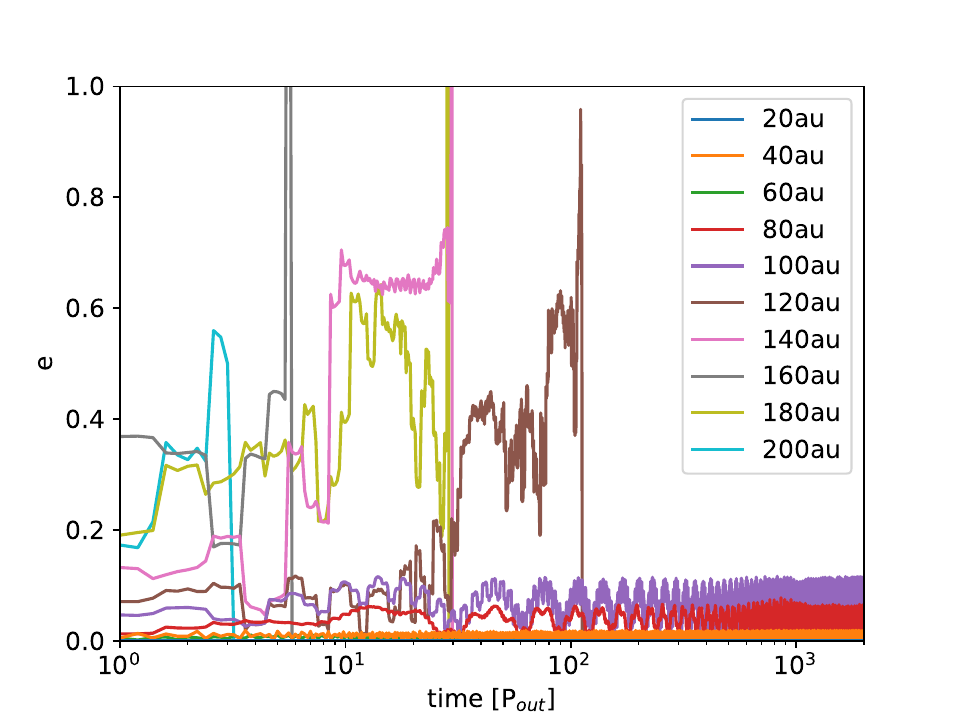}  
        \includegraphics[width=0.3\textwidth]{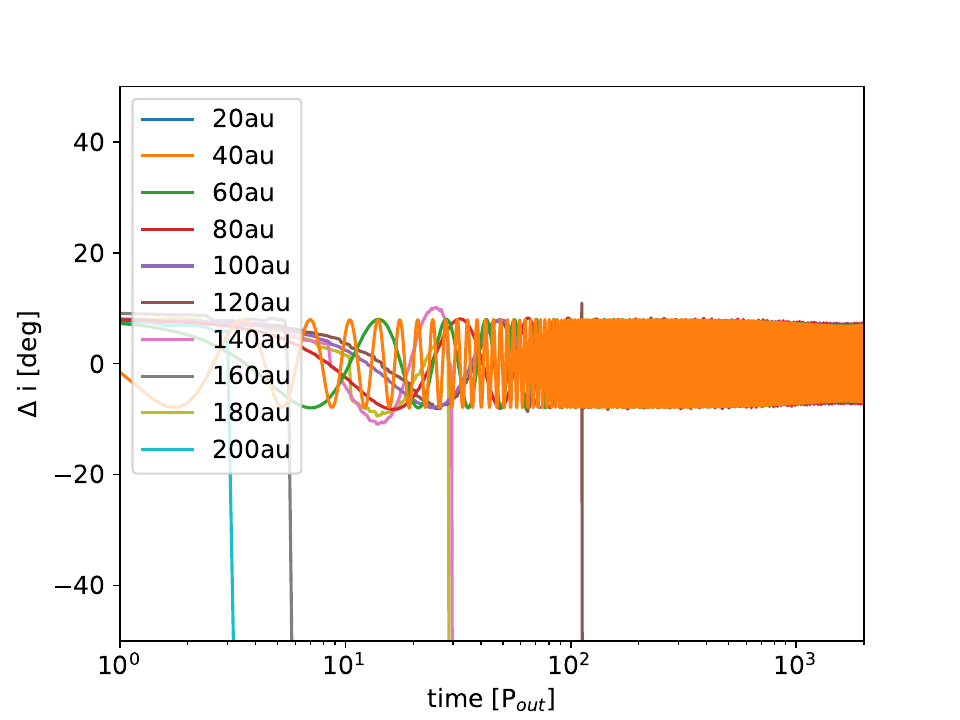}  
    \caption{Same as Figure \ref{fig:nbody} \lang{but} for the \textit{e05} orbital configuration.}
    \label{fig:nbody_e05}
\end{center}
\end{figure*}

\begin{figure*}[!h]
\centering
\begin{center}
         \includegraphics[width=0.3\textwidth]{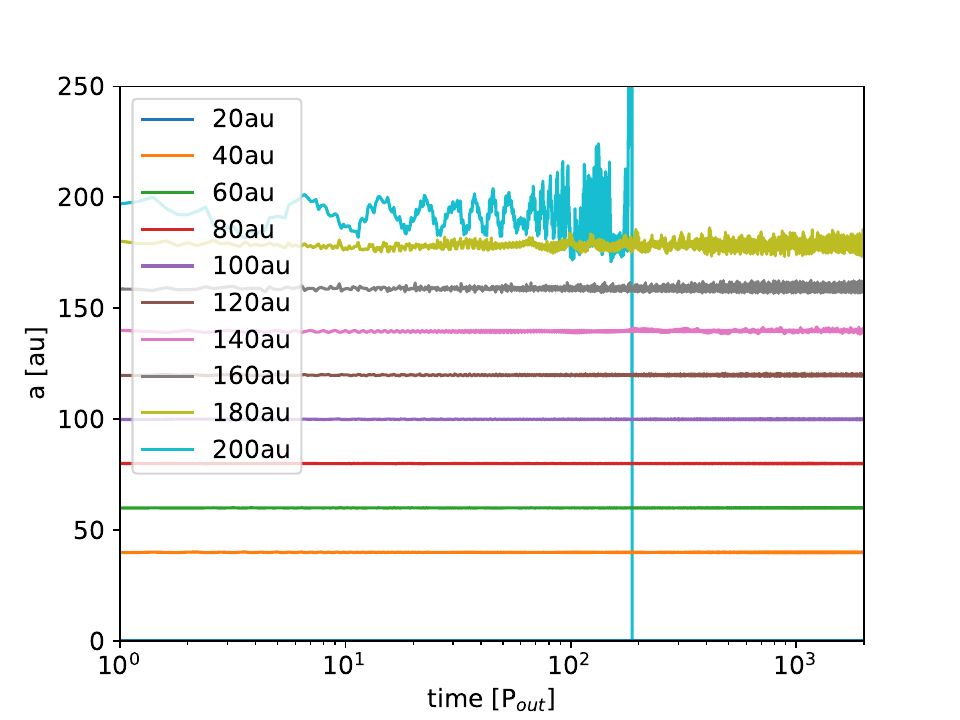}  
        \includegraphics[width=0.3\textwidth]{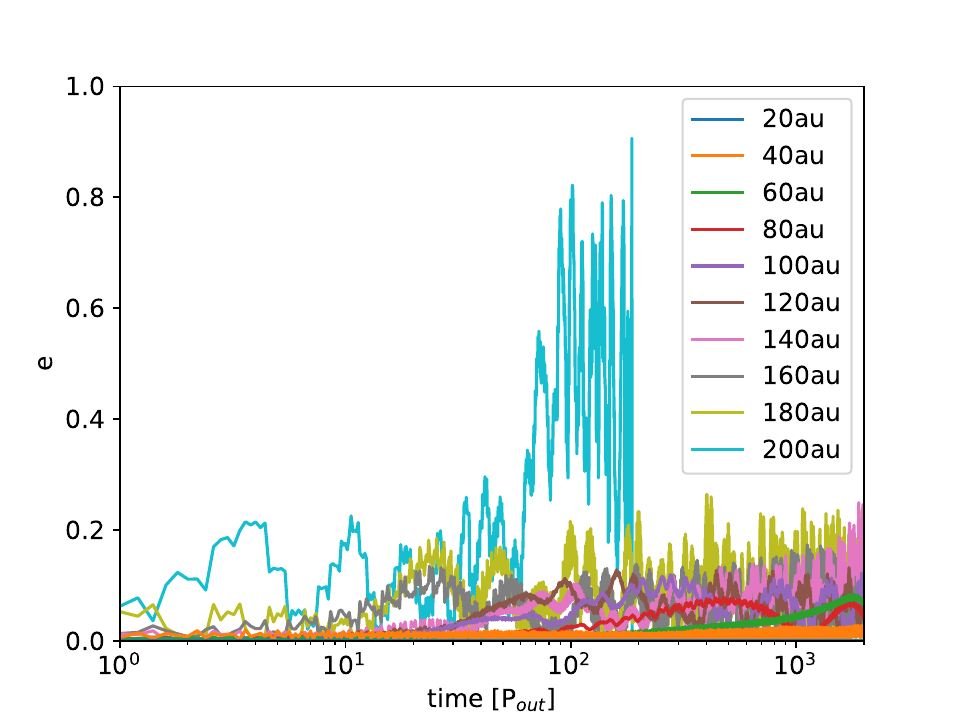}  
        \includegraphics[width=0.3\textwidth]{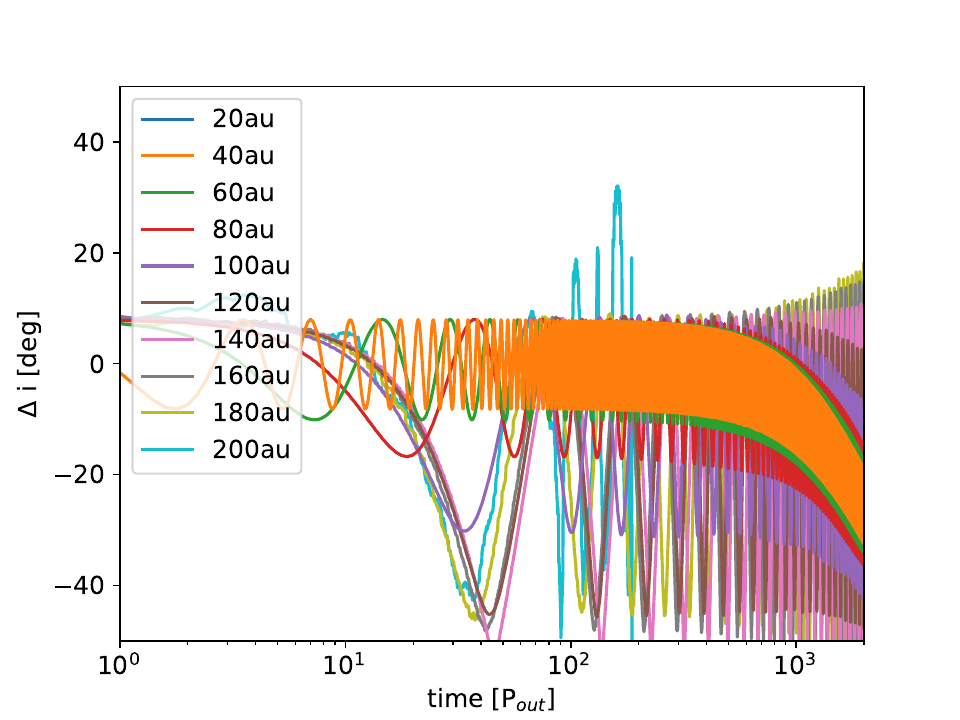}  
    \caption{Same as Figure \ref{fig:nbody} \lang{but} for the \textit{i30} orbital configuration.}
    \label{fig:nbody_i30}
\end{center}
\end{figure*}

\begin{figure*}[!h]
\centering
\begin{center}
         \includegraphics[width=0.3\textwidth]{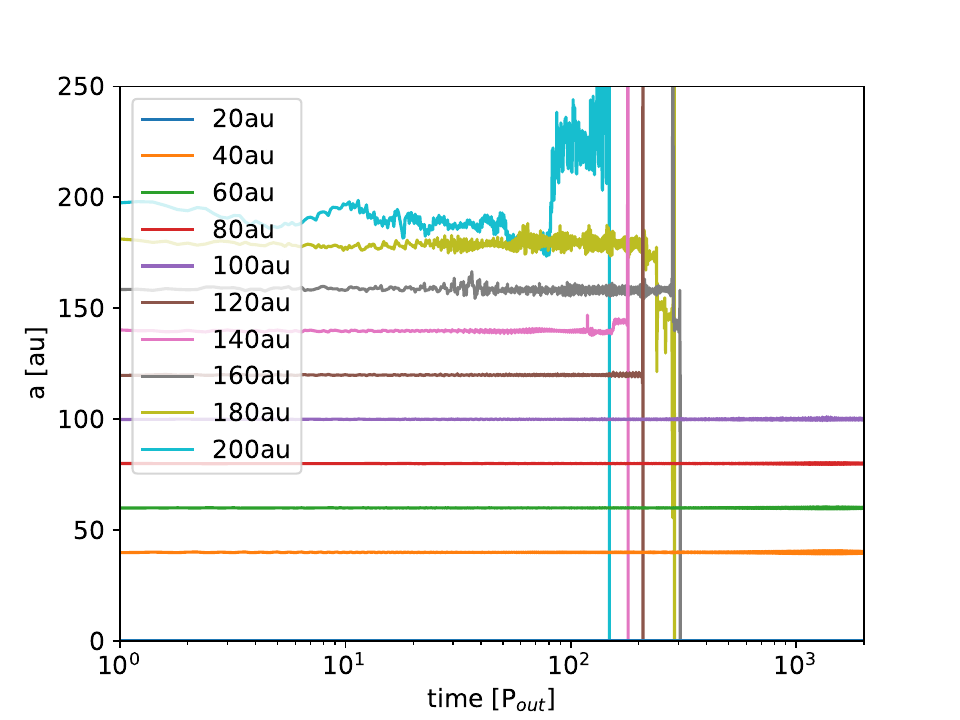}  
        \includegraphics[width=0.3\textwidth]{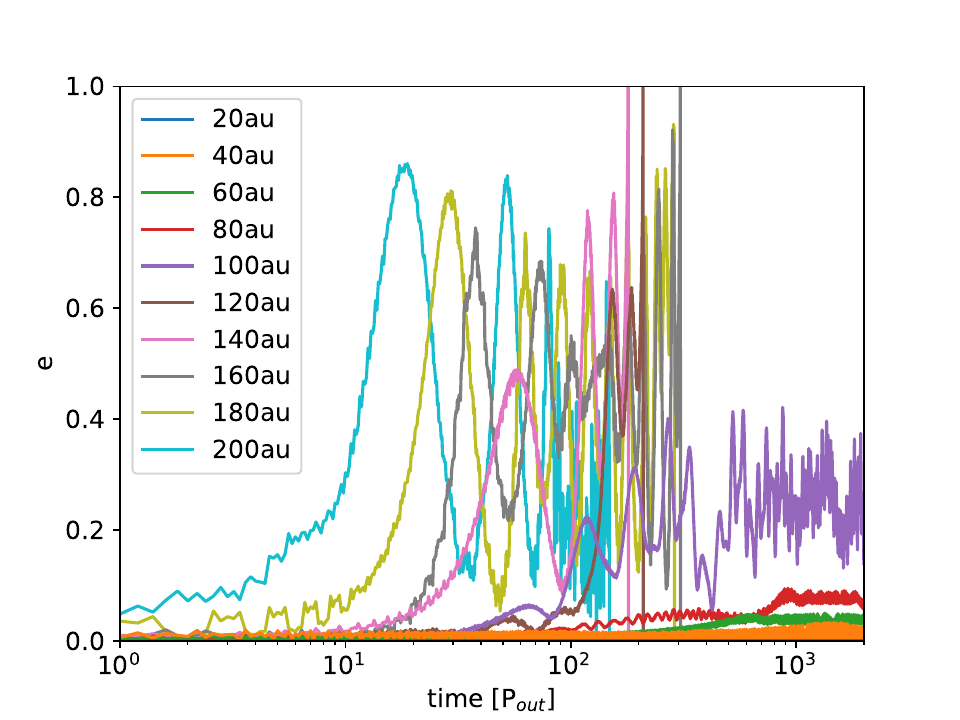}  
        \includegraphics[width=0.3\textwidth]{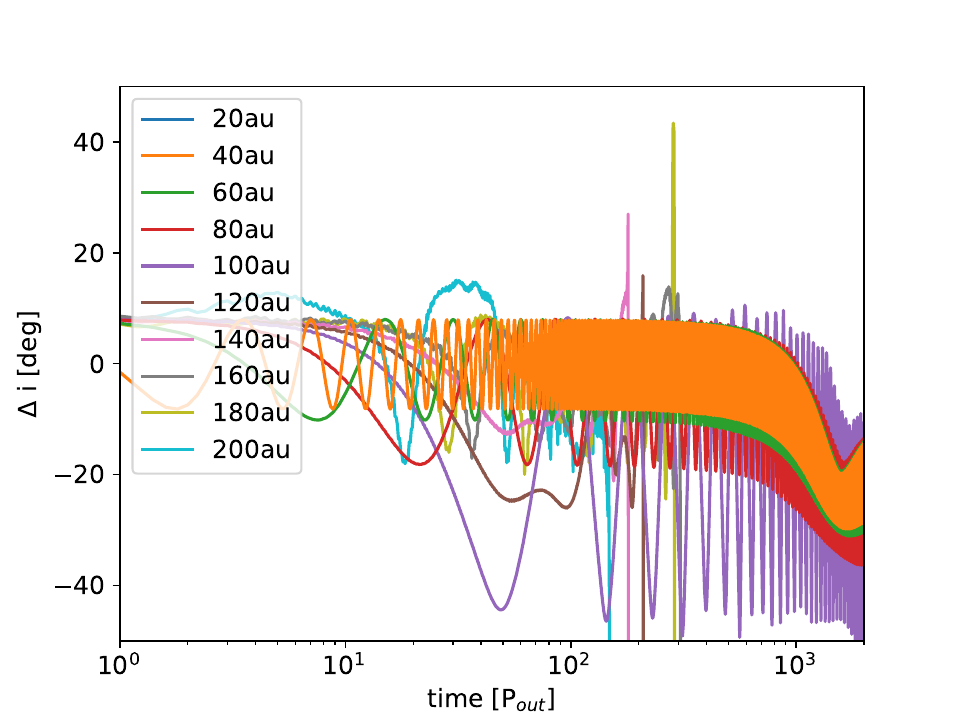}  
    \caption{Same as Figure \ref{fig:nbody} \lang{but} for the \textit{i60} orbital configuration.}
    \label{fig:nbody_i60}
\end{center}
\end{figure*}

\end{appendix}

\end{document}